# Metallicity Distribution Functions of 13 Ultra-faint Dwarf Galaxy Candidates from Hubble Space Telescope Narrowband Imaging

Sal Wanying Fu[1], Daniel R. Weisz[1], Else Starkenburg[2], Nicolas Martin[3], Alessandro Savino[1], Michael Boylan-Kolchin[4], Patrick Côté[5], Andrew E. Dolphin[6,7], Alexander P. Ji[8,9], Nicolas Longeard[10], Mario L. Mateo[11], Ekta Patel[1,12], and Nathan R. Sandford[1]

[1] Department of Astronomy, University of California Berkeley, Berkeley, CA 94720, USA; swfu@berkeley.edu
[2] Kapteyn Astronomical Institute, University of Groningen, Postbus 800, 9700 AV, Groningen, The Netherlands
[3] Université de Strasbourg, Observatoire astronomique de Strasbourg, UMR 7550, F-67000 Strasbourg, France
[4] Department of Astronomy, The University of Texas at Austin, 2515 Speedway, Stop C1400, Austin, TX 78712-1205, USA
[5] National Research Council of Canada, Herzberg Astronomy and Astrophysics Research Centre, Victoria, BC V9E 2E7, Canada
[6] Raytheon Technologies, 1151 E. Hermans Road, Tucson, AZ 85756, USA
[7] Steward Observatory, University of Arizona, 933 N. Cherry Avenue, Tucson, AZ 85721-0065 USA
[8] Department of Astronomy & Astrophysics, University of Chicago, 5640 S. Ellis Avenue, Chicago, IL 60637, USA
[9] Kavli Institute for Cosmological Physics, University of Chicago, Chicago, IL 60637, USA
[10] Laboratoire d'astrophysique, École Polytechnique Fédérale de Lausanne (EPFL), Observatoire, 1290 Versoix, Switzerland
[11] Department of Astronomy, University of Michigan, 311 West Hall, 1085 S. University Avenue, Ann Arbor, MI 48109, USA
[12] Miller Institute for Basic Research in Science, 468 Donner Lab, Berkeley, CA 94720, USA

## Abstract

We present uniformly measured stellar metallicities of 463 stars in 13 Milky Way (MW) ultra-faint dwarf galaxies (UFDs; $M_V = -7.1$ to $-0.8$) using narrowband CaHK (F395N) imaging taken with the Hubble Space Telescope. This represents the largest homogeneous set of stellar metallicities in UFDs, increasing the number of metallicities in these 13 galaxies by a factor of 5 and doubling the number of metallicities in all known MW UFDs. We provide the first well-populated MDFs for all galaxies in this sample, with $\langle[\mathrm{Fe/H}]\rangle$ ranging from $-3.0$ to $-2.0$ dex, and $\sigma_{[\mathrm{Fe/H}]}$ ranging from 0.3–0.7 dex. We find a nearly constant $[\mathrm{Fe/H}]\sim -2.6$ over 3 decades in luminosity ($\sim 10^2$–$10^5\,L_\odot$), suggesting that the mass–metallicity relationship does not hold for such faint systems. We find a larger fraction (24%) of extremely metal-poor ($[\mathrm{Fe/H}] < -3$) stars across our sample compared to the literature (14%), but note that uncertainties in our most metal-poor measurements make this an upper limit. We find 19% of stars in our UFD sample to be metal-rich ($[\mathrm{Fe/H}] > -2$), consistent with the sum of literature spectroscopic studies. MW UFDs are known to be predominantly >13 Gyr old, meaning that all stars in our sample are truly ancient, unlike metal-poor stars in the MW, which have a range of possible ages. Our UFD metallicities are not well matched to known streams in the MW, providing further evidence that known MW substructures are not related to UFDs. We include a catalog of our stars to encourage community follow-up studies, including priority targets for ELT-era observations.

## 1. Introduction

The development of wide-field photometric surveys and deep imaging capacities has revolutionized the discovery of ultra-faint dwarf galaxies (UFDs) around the Milky Way (MW) and within the Local Group (LG; e.g., Belokurov et al. 2007; Laevens et al. 2015; Bechtol et al. 2015; Drlica-Wagner et al. 2015; Homma 2018; Cerny et al. 2023). With luminosities fainter than $10^5\,L_\odot$, these galaxies occupy the faintest-known end of the galaxy luminosity function (Simon 2019). Hubble Space Telescope (HST) star formation history (SFH) studies of these galaxies reveal stellar populations that appear uniformly old, suggesting that they may be some of the faintest and earliest galaxies to have formed in the universe (Brown et al. 2014; Weisz et al. 2014; Gallart et al. 2021; Simon et al. 2021; Sacchi et al. 2021). As high-redshift UFDs are expected to be beyond the observational reach of even the recently launched JWST due to their intrinsic faintness (e.g., Boylan-Kolchin et al. 2016; Weisz & Boylan-Kolchin 2017; Jeon & Bromm 2019), resolved stellar population studies of UFDs in our local neighborhood remain our only window into understanding galaxy formation at the faintest-known scales and the earliest epochs of the universe.

Paramount to these efforts is the characterization of stellar chemical abundances in these systems. At the broadest level, the presence of internal [Fe/H] dispersion is understood to be the distinguishing characteristic between a galaxy and a star cluster, as it has become difficult to categorize many of the newly discovered satellites on the basis of size and luminosity alone (e.g., Kirby et al. 2015; Laevens et al. 2015; Simon et al. 2020).[13] On a more detailed level, the stellar chemistry in a galaxy encodes the astrophysical circumstances of their formation, providing constraints on parameters such as supernovae (SNe) yields, gas inflow/outflow, star formation efficiency, timescales, and burstiness (e.g., Andrews et al. 2017; Weinberg et al. 2017). Additionally, the low masses of UFDs make them particularly sensitive to the baryonic physics implemented in cosmological simulations, and reproducing their internal chemistry remains a key theoretical challenge for the community (Jeon et al. 2017; Revaz & Jablonka 2018; Wheeler et al. 2019; Agertz et al. 2020; Prgomet et al. 2022).

[13] Historically, the defining difference between a galaxy and a star cluster is that the former is embedded in a dark matter halo. An internal metallicity dispersion has been understood as observational evidence for multiple epochs of star formation, which is only possible with a gravitational potential well deep enough to retain SNe ejecta for successive star formation episodes (Willman & Strader 2012).

For the classical LG dwarf galaxies, the astrophysical circumstances of their star formation have been inferred using well-sampled stellar metallicity distribution functions (MDFs; e.g., Carigi et al. 2002; Lanfranchi et al. 2008; Tolstoy et al. 2009; Kirby et al. 2011). However, similar observations have historically been challenging to make in the UFD regime. Simon (2019) highlights the paucity of metallicity information in UFDs. Nearly half of known MW UFDs lack any metallicity information, while the remainder only have a handful of stars bright enough to observe with current ground-based facilities. Next-generation photometric surveys are already delivering on their promise to discover more faint satellites and out to larger distances (e.g., Homma et al. 2019; Mutlu-Pakdil et al. 2021; Cerny et al. 2021; Smith et al. 2023), and even observations with next-generation spectrographs on extremely large telescopes (ELTs) will not be able to observe adequate numbers of stars in these galaxies to sufficiently populate the galaxy's MDF. For example, Figure 9 from Simon (2019) projects that medium-resolution spectroscopy on forthcoming ELTs may only be able to reach $\lesssim$10 stars at best in a Ret II equivalent UFD at 250 kpc.

An alternative solution is through photometric metallicities. Though optical broadband photometry is largely insensitive to metallicity,[14] medium bands and narrowbands that target specific absorption features in cool stars, e.g., red giant branch stars (RGB), have a long history of metallicity measurements in the MW. Building on a long legacy of CaHK imaging surveys of stars in the MW, the CFHT/Pristine survey (Starkenburg et al. 2017a) has shown that the combination of CaHK, $g$- and $i$-band imaging can be used to measure metallicities of individual stars to a precision of 0.2–0.3 dex for stars as metal-poor as $[\mathrm{Fe/H}] = -3.0$ enabling the detection of extremely metal-poor stars ($[\mathrm{Fe/H}] < -3.0$) candidates in the MW (e.g., Youakim et al. 2017; Venn et al. 2020).

Motivated by the success of the PRISTINE survey, we designed an HST program (GO-15901; PI: Weisz) that leverages its excellent blue-optical sensitivity and its underused CaHK filter (UVIS/F395N) to measure metallicities of faint stars in a sizable sample of UFD satellites of the MW. The first published results from this program analyzed Eridanus II (Eri II), the brightest galaxy observed by this program. Specifically, Fu et al. (2022) demonstrated that the HST CaHK photometric metallicities are in good agreement with the calcium triplet (CaT) calibration (Li et al. 2017) that is the current community standard for ground-based spectroscopic studies characterizing UFDs (Kirby et al. 2015; Li et al. 2018; Longeard et al. 2018; Fritz et al. 2019; Simon et al. 2020; Chiti et al. 2022), as well as its agreement with less-often used photometric metallicity calibrations such as those from RR Lyrae stars (Martínez-Vázquez et al. 2021). The well-populated HST-based MDF served as a basis for demonstrating that the star formation of Eri II was characterized by strong outflows and low star formation efficiency (Sandford et al. 2022), which are in good agreement with theoretical expectations (Muratov et al. 2015). These results demonstrate HST's unique ability to provide insight into the baryon cycle of the faintest galaxies in the Universe.

[14] An exception is the metallicity-sensitive SDSS $u$ band, e.g., Ivezić et al. (2008); An et al. (2013).

In this paper, we present MDF measurements for the 13 UFDs observed by our program. This work represents the largest homogeneous set of stellar metallicity measurements in UFDs to date and will enable a wide range of science in future papers from this program. The goals of this paper are to detail the metallicity determinations and provide qualitatively new insight into our knowledge of the MDFs of UFDs.

This paper is organized as follows. We describe our observations and photometry in Section 2 and detail our methodology for measuring metallicities in Section 3.1. We apply our method to a few detailed examples in Section 4 and discuss caveats and systematics. We discuss the MDFs for each galaxy in the sample individually in Section 5 and place our results into a broader context in Section 6. We summarize our work in Section 7.

## 2. Observations and Data Reduction

In this section, we detail the target sample selection process, summarize the observations, and describe the photometric reduction process. The design of this program sought to balance observational efficiency with the need to observe enough UFDs to broadly characterize MDFs across a population of faint galaxies using the HST equivalent of the CaHK color–color space used by the Pristine survey. This typically required acquiring new F395N and F475W (Sloan Digital Sky Survey (SDSS) $g$ band) imaging with HST/UVIS and pairing it with archival Advanced Camera for Surveys (ACS)/F814W and ACS/F606W data. All of the HST data analyzed in this program can be found in MAST at this link:10.17909/cfn9-gt96.

### 2.1. Target Selection

To select our sample, we started from all known UFDs within ∼300 kpc of the MW as listed in Simon (2019). We first eliminated galaxies with existing well-populated MDFs and removed galaxies that were angularly too large on the sky for HST to efficiently observe them in a single pointing (i.e., HST must cover $\gtrsim$70% $r_h$). For observational efficiency, we removed a small number of galaxies that did not have archival F814W imaging. We also eliminated galaxies that would have required large numbers of orbits due to a paucity of sufficiently bright stars, i.e., the galaxy is distant and/or did not have more than ∼20 stars that could be observed in four orbits of HST time. This process resulted in 18 systems that range in distance from $20 \lesssim D \lesssim 300$ kpc and span a factor of ∼1000 in luminosity.

Following the execution of our program (see Section 2.2), some of the systems in the original sample changed status. At the time of the proposal, Sgr II was a UFD candidate (Longeard et al. 2020), but there is now strong evidence it is a globular cluster (GC; Longeard et al. 2021; Baumgardt et al. 2022). Additionally, our analysis of the Sgr II data is also consistent with its GC status. Thus, we omit this system from our study. Dra II is also a UFD candidate whose status is still under debate (e.g., Longeard et al. 2018; Baumgardt et al. 2022); we chose to

**Table 1**
Observational Properties of UFDs

| Galaxy | $M_V$ (mag) | $r_h$ (arcmin) | $(m-M)$ (mag) | $r_h$, WFC3 FoV | Exp. Time F395N (s) | F475W (s) | F606W (s) | F814W (s) | $S/N_{F395N}=5$ F395N (mag) | F475W (mag) |
|---|---|---|---|---|---|---|---|---|---|---|
| Eridanus II | −7.1 | $2.3 \pm 0.12$[a] | $22.8 \pm 0.3$[a] | 1.13 | 5517 | 7644 | 28,580 | 7900 | 26 | 26 |
| Canes Venatici II | −5.2 | $1.83 \pm 0.21$[b] | $21.02 \pm 0.3$[c] | 1.26 | 6584 | 1014 | 20,860 | 20,860 | 26 | 26 |
| Hydra II | −4.9 | $1.7 \pm 0.3$[d] | $20.89 \pm 0.11$[e] | 1.53 | 6584 | 1014 | 4746 | 4746 | 26 | 26 |
| Reticulum II | −4.0 | $6.3 \pm 0.4$[f] | $17.5 \pm 0.1$[f] | 0.41 | 4534 | 780 | 4627 | 4627 | 25.5 | 25 |
| Horologium I | −3.8 | $1.1 \pm 0.15$[g] | $19.18 \pm 0.09$[g] | 2.36 | 4526 | 780 | 4627 | 4627 | 26 | 26 |
| Grus I | −3.5 | $1.77 \pm 0.4$[h] | $20.4 \pm 0.2$[h] | 1.47 | 9481 | 1314 | 4766 | 4766 | 26 | 26 |
| Reticulum III | −3.3 | $2.4 \pm 0.9$[i] | $19.81 \pm 0.31$[i] | 0.92 | 8040 | 919 | 4662 | 4662 | 26 | 26 |
| Willman 1 | −2.9 | $2.52 \pm 0.21$[j] | $17.9 \pm 0.4$[k] | 0.97 | 4534 | 780 | 4627 | 4627 | 26 | 25 |
| Phoenix II | −2.7 | $1.5 \pm 0.3$[f] | $19.6 \pm 0.2$[f] | 1.73 | 4534 | 780 | 4627 | 4627 | 26 | 25.5 |
| Eridanus III | −2.1 | $0.315 \pm 0.03$[l] | $19.8 \pm 0.04$[l] | 8.25 | 4534 | 780 | 4627 | 4627 | 25.5 | 25.5 |
| Tucana V | −1.6 | $1.0 \pm 0.3$[i] | $18.71 \pm 0.34$[i] | 2.6 | 3120 | 730 | 4661 | 4661 | 25 | 25 |
| Segue 1 | −1.3 | 4.5[m] | $16.8 \pm 0.2$[m] | 0.58 | 8776 | 1314 | 4605 | 4605 | 26 | 25 |
| Draco II | −0.8 | $2.7 \pm 1.0$[n] | $16.9 \pm 0.3$[n] | 0.96 | 8040 | 919 | 4662 | 4662 | 26 | 25 |

**Notes.** Observational characteristics of the UFDs analyzed by this program are presented in order of descending luminosity. We provide information on their on-sky size in relation to the HST FoV, their distance modulus, and exposure times of the images used in this study. As a summary of data depth, for each UFD, we report the F475W and F395N magnitudes where F395N $S/N = 5$. In all cases except Eri II and CVn II, the archival F606W and F814W imaging are from GO-14734 (PI: Kallivayalil). For CVn II, the F606W and F814W imaging are from GO-12549 (PI: Brown). For Eri II, all of the broadband imaging is from programs GO-14224 (PI: Gallart) and GO-14234 (PI: Simon). All of the data can be found in MAST at the following doi:10.17909/cfn9-gt96.

**References.**
[a] Crnojević et al. (2016)
[b] Sand et al. (2012)
[c] Greco et al. (2008)
[d] Martin et al. (2015)
[e] Vivas et al. (2016)
[f] Mutlu-Pakdil et al. (2018)
[g] Jerjen et al. (2018)
[h] Koposov et al. (2015a)
[i] Drlica-Wagner et al. (2015)
[j] Muñoz et al. (2018)
[k] Willman et al. (2006)
[l] Conn et al. (2018)
[m] Belokurov et al. (2007)
[n] Laevens et al. (2015)

include it due to this ambiguity. Finally, Indus II was originally classified as an extremely faint UFD, but subsequent studies revealed that it was a spurious detection of noise (Cantu et al. 2021); our reduction of this data also confirms that there is no stellar system at this location. As discussed in Fu et al. (2022), we used existing deep F475W and F814W imaging for Eri II and only added new F395N imaging. Owing to unresolvable alignment issues (i.e., lack of bright stars in the field) in the photometry process for the remaining three UFDs observed by our program (Pisces II, Pictor I, and Segue II), we exclude them from our analysis for this work.

Thus, for this work, we only focus on 13 systems. Table 1 lists the observational properties of our UFD sample.

### 2.2. Observations and Data Reduction

We acquired new F395N and F475W observations for our 18 systems between 2020 January and October. The F395N and F475W filters are equivalent to the Pristine CaHK narrowband and *g*-band filters used for measuring stellar metallicities, and as shown in Fu et al. (2022), are also able to recover MDFs. We required that the UVIS fields spatially overlap archival ACS imaging, but did not place any roll angle constraints in order to maximize schedulability. Visits were 1–2 orbits in duration.

We use DOLPHOT to perform point-spread function (PSF) photometry simultaneously on F395N, F475W, F606W, and F814W `flc` images for each of our galaxies. We then perform a quality cut on the resulting catalog by requiring that every star has a signal-to-noise ratio $(S/N) > 5$, $|\text{sharp}|^2 < 0.3$, and $\text{crowd} < 1$ in F606W and F814W, which are the highest S/N data.

### 2.3. Color–Magnitude Diagrams

Figure 1 shows the gallery of color–magnitude diagrams (CMDs) for our sample in F606W–F814W, and Figure 2 shows the corresponding F475W–F814W CMDs. The stellar population sequence for all of our galaxies is narrower in F606W–F814W than in F475W–F814W because (i) the former color combination is less sensitive to temperature, metallicity, and age and (ii) the F606W photometry has much higher S/N owing to longer integration. We therefore use F606W–F814W CMD for UFD candidate member selection, as described in several sections in the paper.

### 2.4. Member Selection

As most of our stars are too faint for radial velocities or proper motions, we primarily determine the membership of each star by selecting stars that fall on/near the RGB and/or main-sequence turn-off (MSTO) on the F606W–F814W CMDs. For each UFD, we then crossmatch each star to catalogs of radial velocities to remove MW foreground stars. We provide detailed discussions on membership vetting for each UFD in Section 5.

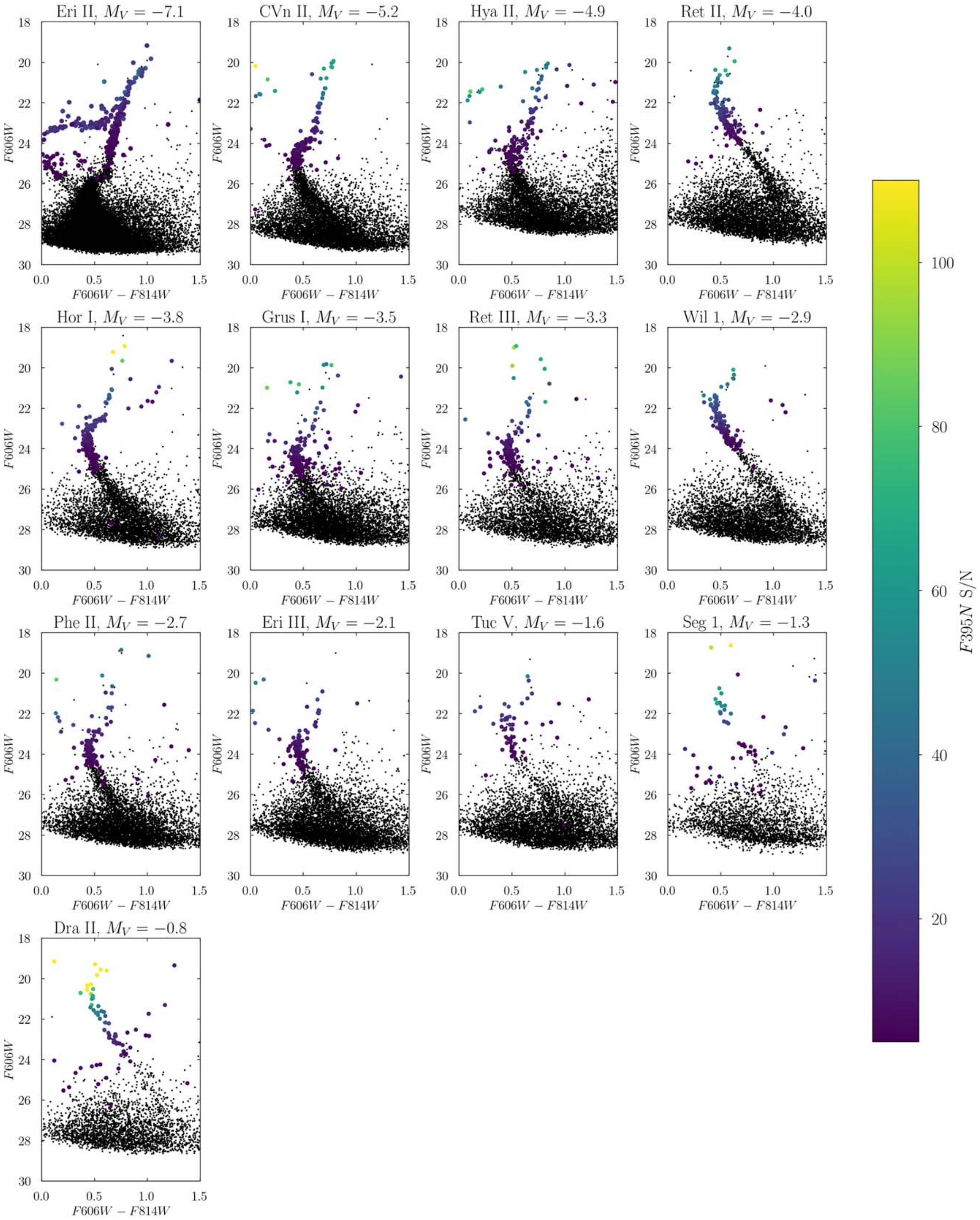

**Figure 1.** A gallery of F606W–F814W CMDs for our galaxies in order of decreasing luminosity from the top left to the bottom right. We color code stars by their F395N S/N for F395N S/N >5 and have not yet applied any membership selection or removed spurious detections.

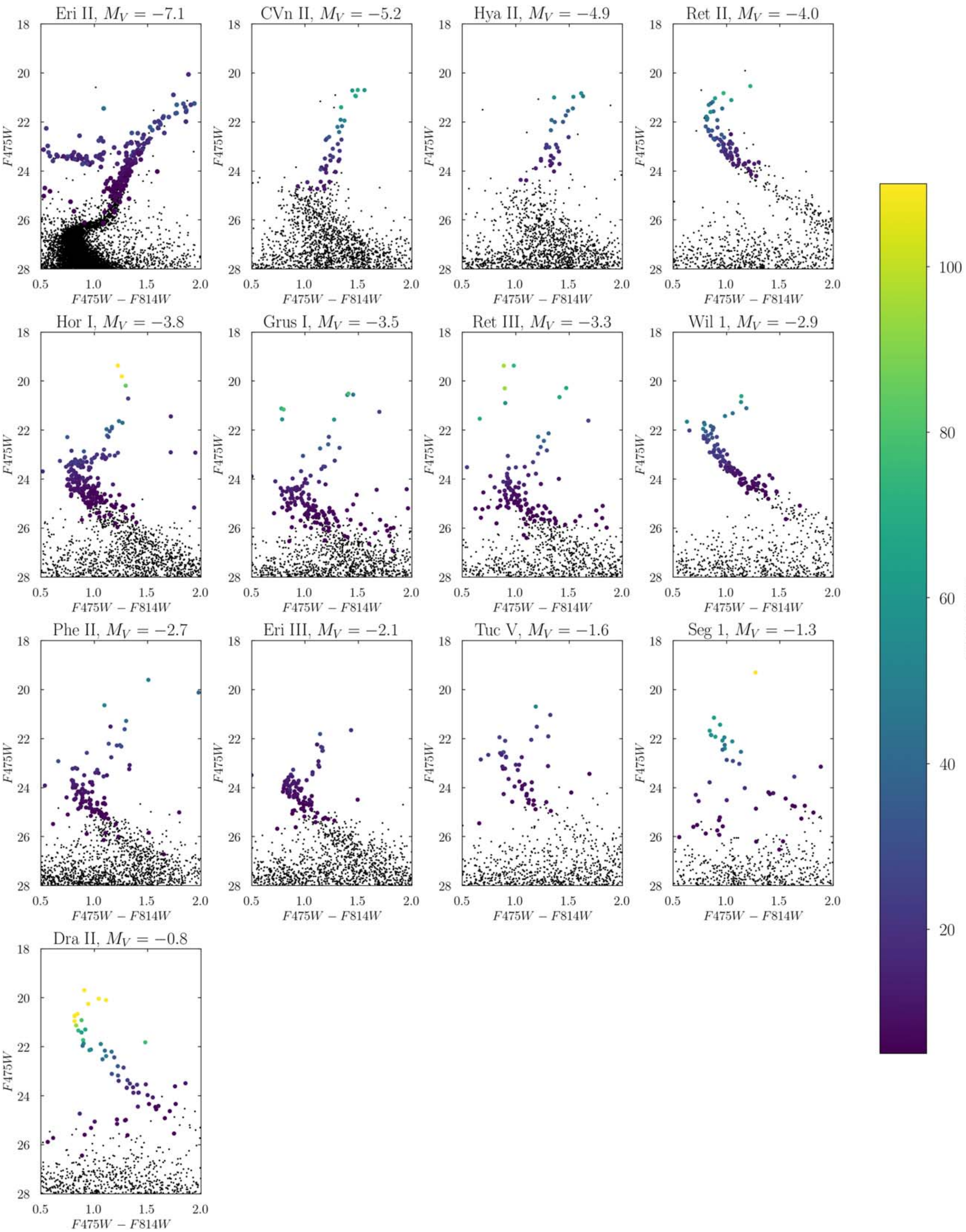

**Figure 2.** The same as Figure 1 only for F475W–F814W CMDs. The broader CMD features are partially driven by the lower S/N of the F475W data and the increased metallicity sensitivity of F475W–F814W vs. F606W–F814W.

We note a few difficulties of relying on spectroscopic velocities for cleaning our entire sample: (i) The on-sky footprint of spectroscopic studies extends beyond our HST field of view and the target density for spectroscopy is smaller due to limitations in slit and fiber placement; (ii) The stars observed spectroscopically are often much brighter than the stars in our sample; and (iii) The brightest stars observed by spectroscopy are often missing from our data due to saturation effects in the archival broadband HST data. Using kinematic information, we were able to remove a total of 8 contaminants: six contaminants in Seg 1 (Simon et al. 2011),[15] one contaminant in Ret II (Simon et al. 2015), and one contaminant in Grus I (Chiti et al. 2022). Some stars in the 13 UFDs we present in this work have been studied via spectroscopy, and 10 of our UFDs have at least some stars in common with spectroscopic studies. For these stars, we compare our metallicities to those in the literature in Section 5 and Appendix C.

Additionally, we manually remove stars which pass the CMD selection criteria, but which, upon closer inspection, have colors that are inconsistent with their presumed astrophysical properties. We perform the following checks: (1) That member stars fall within the $[Fe/H] = -4$ and $[Fe/H] = -1$ isochrones of the $[\alpha/Fe] = +0.4$ MESA Stellar Isochrones and Tracks (MIST) models (Choi et al. 2016; Dotter 2016) on both F606W–F814W and F475W–F814W and CMDs, and (2) that the metallicity of stars inferred from the F395N photometry in subsequent sections are consistent with their positions on the broadband CMD: that metal-poor stars should be on the bluer side of the stellar population sequence, and the inverse for metal-rich stars. The stars removed from this process tend to have low S/N in F395N, around the cutoff threshold of 10. This manual vetting criteria is still quite broad, and in the absence of kinematic and astrometric data for membership studies, we choose to err on the side of inclusivity in determining our sample of members. We provide a table of observed stars in this paper and invite the community to follow up with complementary observations.

Finally, we expect the MW foreground to be a minimal source of contamination for our member samples due to the small HST field of view (FoV). We provide an illustration of this expectation in Appendix A. Where applicable we also discuss the concern of potential foreground impact for individual galaxies in Section 5. We do not expect the foreground to significantly affect our MDFs or the general results of our paper.

### 2.5. Artificial Star Tests

We use artificial star tests (ASTs) to compute uncertainties on photometry and construct an error profile (bias and scatter) for individual star metallicity fitting as described in Section 3.1. ASTs involve inserting a star of a known magnitude, in our case F395N, F475W, F606W, and F814W, into each image, and attempting to recover its magnitudes using the same DOLPHOT procedure that we use for the original photometric reduction. By running many ASTs, we build up the statistics to construct well-sampled error profiles for each of our UFDs.

In advance of the metallicity fitting procedure, we generate ASTs around each member star that we identify in each UFD, with about $10^4$ ASTs run per star. The general idea is to distribute the ASTs such that they (a) cover all model tracks in CaHK space and (b) sample the relevant regions of the CMD. To do this, we center the ASTs for each star within 0.2 mag of its F475W magnitude. We then require the input AST list to satisfy the criteria $0.7 < \mathrm{F475W{-}F814W} < 2.0$ and $-2.0 < \mathrm{CaHK} < -0.4$. This is a departure from the procedure for generating ASTs for the Eri II sample in Fu et al. (2022), as it provides more efficient coverage of the 4-dimensional AST space. We apply this procedure to each galaxy, including Eri II.

We discuss and illustrate the ASTs in greater detail for select galaxies in our sample in Section 4 and only provide a general summary here. Our error profile is based on the difference between the recovered magnitude and its known input magnitude in each filter, (out–in). At a given magnitude, we use the (out–in) quantity at that magnitude to define the scatter. We use the mean of the (out–in) quantity to compute the bias.

In all of our galaxies, the error and bias introduced by F606W and F814W are minimal ($\lesssim$0.01 mag) because the imaging is much deeper. The scatter in F475W is larger (>0.05 mag) at fainter magnitudes, and there is a modest bias (~0.02 mag at F475W = 24.5 mag) in recovered magnitudes versus input magnitudes for fainter ASTs. However, the error and bias in the F395N filter are the chief sources of photometric uncertainty for the metallicity measurements as it has the shallowest imaging of all the data sets (uncertainty ~0.1 mag, bias~0.1 mag for F475W ~ 23.5 mag). We provide concrete examples in Section 4.

## 3. Determining MDFs

### 3.1. Individual Metallicity Measurements

To build MDFs, we first infer metallicities for individual stars by adapting the technique used for CMD-based star formation history (SFH) fitting as described in Dolphin (2002). We first construct the equivalent of a Hess diagram with the $x$-axis as F475W–F814W and on the $y$-axis as CaHK = F395N–F475W–1.5*(F475W–F814W), which is motivated by the Pristine survey. The resulting Hess-like diagram runs from $0.7 < \mathrm{F475W{-}F814W} < 2.0$ and $-2.0 < \mathrm{CaHK} < -0.4$, and bins are 0.025 mag by 0.025 mag. Individual stars are then modeled in this pixelated space. Stellar metallicities are inferred by comparing the overlap of an individual star's Hess-like diagram with that of model CaHK tracks of various metallicities that have been corrected for observational effects. We now describe the process of constructing the metallicity models for fitting each individual star.

We begin with $\alpha$-enhanced ($[\alpha/Fe] = +0.40$), 13 Gyr MIST CaHK color–color tracks. We use the MIST suite because they have the most metal-poor limit ($[Fe/H] = -4.0$) out of all isochrones currently available.[16] Additionally, we choose to use $[\alpha/Fe] = +0.40$ models because observations thus far have demonstrated that stars at the typical UFD metallicity of $[Fe/H] \sim -2.5$ tend to be $\alpha$-enhanced (Vargas et al. 2013) and that UFD stars are uniformly old (Brown et al. 2014).[17] The impact of our choice in $\alpha$-enhancement is within the uncertainties as verified in Appendix B.1. The monometallic tracks we use are 0.05 dex apart. We apply dust corrections to the MIST CaHK tracks using the filter-appropriate extinction values from Schlegel et al. (1998); for most of our galaxies, the extinction is minimal ($A_V \lesssim 0.05$). We use the extinction coefficient for F395N provided by the MIST models for our corrections.

For every point in the model CaHK tracks, we select ASTs that match its color in CaHK color space, and the magnitude of the star in F475W. We use the results of those ASTs to calculate the expected bias to apply to that point. As a result of the bias effects discussed in Section 2, the model tracks become redder in the CaHK color index and F475W–F814W color. The overall impact of accounting for bias effects is lowering the inferred metallicity. We illustrate these effects in the CaHK color panels in Figures 3, 5, and 6.

[15] Of the contaminants removed using this method, four were MW foreground and two were stars from the 300S stellar stream in the same field.

[16] As an example, the Dartmouth Stellar Evolution Database (Dotter et al. 2008) Bag of Stellar Tracks and Isochrones (BaSTI; Hidalgo et al. 2018) models only extend down to $[Fe/H] = -2.5$ and $[Fe/H] = -3.2$ respectively.

[17] We also note that for these models, the mapping of CaHK color to metallicity in CaHK space is largely invariant with age for old stars.

Next, we pixelate the bias-applied model CaHK tracks into Hess-like diagrams. We use the ASTs to calculate the standard deviation for the ASTs in each pixel, and to convolve it with the number of expected stars from the model tracks. The result at the end of this process are a series of Hess diagrams for monometallic populations that have been applied with the specific observational characteristics of each UFD. We refer to the Hess diagram corresponding to individual metallicities as a basis function. We normalize the counts in each basis function so that it is equal to 1, and infer a star's metallicity measurement by comparing the overlap of its Hess-like diagram with that of basis functions of various metallicities. Because the number counts can be low in many cases, we adopt a Poisson likelihood function of

$$\log L = \sum_{m_i \neq 0} d \ln m_i - m_i - \ln(d_i!), \quad (1)$$

where $m_i$ are the number of counts in the model bin, and $d_i$ is the data in each bin.

We adopt uniform priors on the metallicity, ranging between the limits of our metallicity grid: −4.0 and +0.0 and then sample the posterior distribution using `emcee` (Foreman-Mackey et al. 2013) by initializing 50 walkers and running the Markov chain Monte Carlo sampler for 10,000 steps, with a burn-in time of about 50 steps per star. We assess convergence using the Gelman–Rubin (GR) statistic (Gelman & Rubin 1992). Compared to Fu et al. (2022), this approach allows us to account for uncertainties in both CaHK and F475W–F814W. The result of this process is a posterior distribution for the metallicity of each star. We determine metallicity measurements and statistical and systematic uncertainties in the following ways:

For stars with well-constrained posterior distributions, we report the median measurement and their uncertainties corresponding to the 68% confidence interval. Following the investigation into systematics in Appendix B.5, we assign a systematic uncertainty of either 0.2 or 0.3 dex depending on if the star is on the RGB or MSTO/(main sequence (MS), respectively.

For stars with posterior distributions that have a well-defined peak but truncation at the metal-poor end, we also report the median measurement and their uncertainties corresponding to the 68% confidence interval. We assign a systematic error of 0.5 dex if their median measurement is below −3.0; if their median measurement is above that, then we follow the schema based on the star's evolutionary phase.

For stars with posterior distributions that only show an upper limit (i.e., no clear peak), we report that upper limit. For stars that fall outside the metal-poor end of the grid, with undefined posterior distributions, we assign them an upper limit of −4. For stars that are unconstrained altogether, we remove them from the sample; since they are low S/N to begin with, their measurements should not significantly change the nature of our inferred MDFs.

We discuss our measurement reporting procedure in greater detail by using the example of CVn II in Section 4.1 and present example fits to CVn II stars in Figure 4. We present the table of measurements in Table 4, reporting both the random uncertainties from photometry, and from systematic uncertainties that we determined following our procedure in Appendix B.5.

### *3.2. Fitting the MDF*

Following standard practices in the field, we model the MDF of each UFD assuming a Gaussian distribution, for which the parameters of interest are the mean and dispersion of the MDF. For the MDF of each galaxy, we adopt the two-parameter Gaussian likelihood function from Walker et al. (2006):

$$\log L = -\frac{1}{2}\sum_{i=1}^{N} \log(\sigma^2_{\rm [Fe/H]} + \sigma^2_{{\rm [Fe/H]},i}) - \frac{1}{2}\sum_{i=1}^{N} \frac{([{\rm Fe/H}]_i - \langle[{\rm Fe/H}]\rangle)^2}{\sigma^2_{\rm [Fe/H]} + \sigma^2_{{\rm [Fe/H]},i}}, \quad (2)$$

where $\langle[\mathrm{Fe/H}]\rangle$ and $\sigma_{\rm [Fe/H]}$ are the mean metallicity and metallicity dispersion of the galaxy, and $[\mathrm{Fe/H}]_i$ and $\sigma_{{\rm [Fe/H]},i}$ are the metallicity and metallicity uncertainties for each star. In this procedure, we assume Gaussian uncertainties on the individual metallicity measurements, and discuss their derivation later in this section. We adopt a uniform prior on the mean and require that it remain within the range set by the most metal-poor and metal-rich stars for a galaxy. We also require that $\sigma_{\rm [Fe/H]} \geqslant 0$. We use `emcee` to sample the posterior distribution, initializing 50 walkers for 10,000 steps. The autocorrelation time for each galaxy is about 50 steps, and the corresponding GR statistic indicates the chains have likely converged.

The above likelihood function assumes symmetric uncertainties on the individual star metallicity measurements. Due to the uneven spacing between monometallic CaHK tracks, that is not the case for the vast majority of stars. We thus make the following adjustments:

For stars with posterior distributions that are well constrained enough where we can report a median and 68% confidence interval uncertainties, we average the asymmetric uncertainties and then add them in quadrature with their corresponding systematic uncertainty (see Appendix B.5 for more detail) in order to arrive at the final uncertainty used for the MDF measurement.

For stars whose posterior distributions allow us to constrain an upper limit, we adopt a point measurement that is the median of the posterior distribution. We adopt a Gaussian uncertainty by averaging the uncertainties from the 68% confidence interval. If the upper limit of the star is below −3 (i.e., an extremely metal-poor candidate), then we add the uncertainties in quadrature with a systematic error of 0.5 dex. If the upper limit of the star is above −3, then we add the uncertainties in quadrature with their corresponding systematic uncertainty.

The above schematic accounts for the vast majority of stars analyzed in our study. Finally, there are a few stars whose CaHK color places them beyond the metal-poor end of the grid. These stars are particularly low S/N ($\lesssim$15). We adopt a point measurement of −4.0 with an uncertainty floor of 1.0 dex, which reflects our low confidence in our estimate.

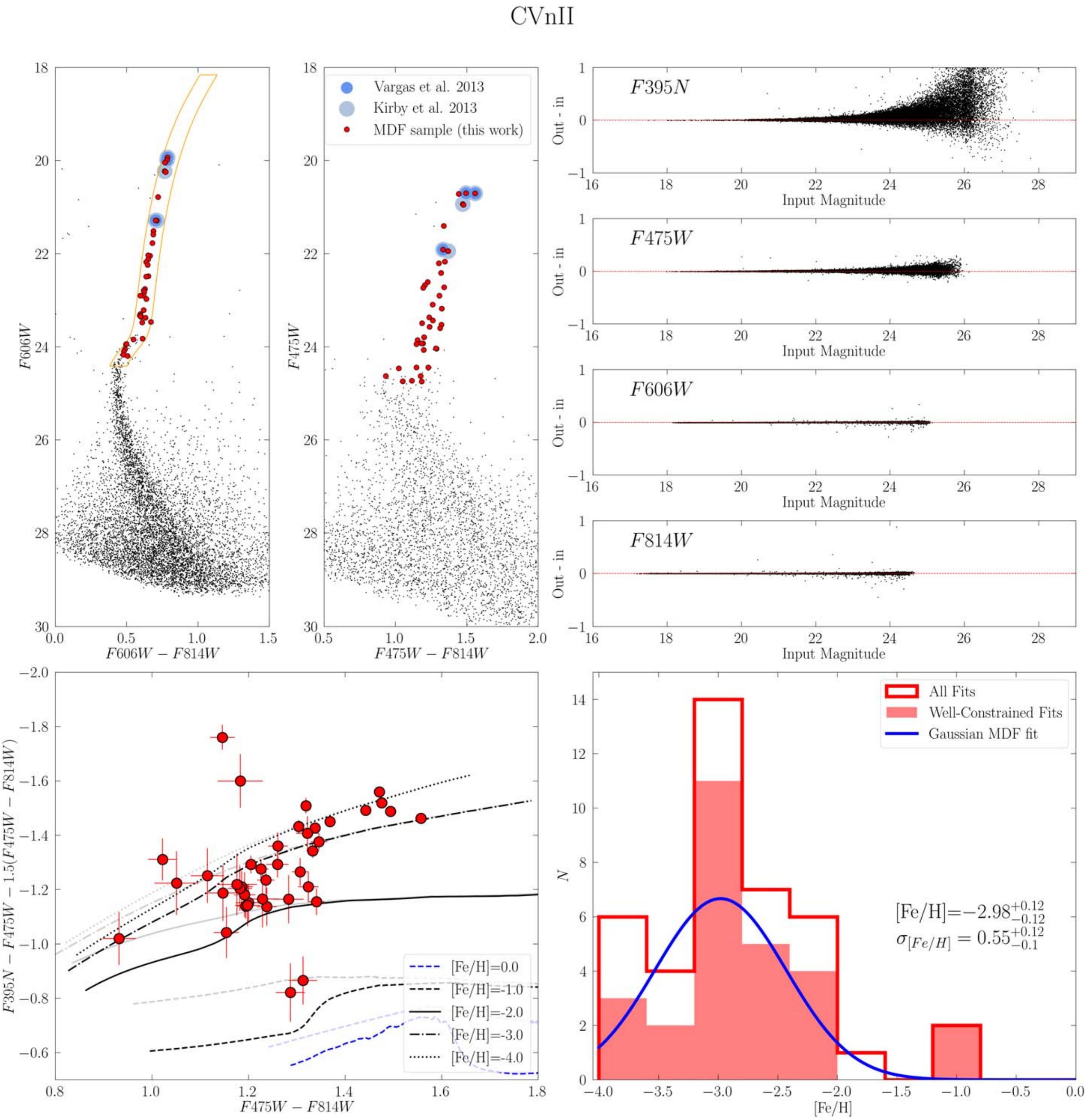

**Figure 3.** An illustrative example of our process for measuring a galaxy's MDF from HST imaging. This case study is for CVn II. The upper left panels show the CMDs of CVn II, with members shown in red and stars in common with other studies highlighted in blue. The upper right panels show the scatter and bias in the ASTs for each filter. The bottom left panel shows the CaHK diagram with member stars plotted in red and select metallicity tracks as lines. The high-opacity lines are the convolution of the intrinsic models low-opacity with the ASTs. The lower right panel shows the MDF of CVn II. The shaded red regions are well-constrained metallicity fits for individual stars, while the unshaded regions reflect stars with poorly constrained fits (e.g., off the metallicity grid, truncated probability density functions (PDFs)). The blue line shows a Gaussian fit to all stars on the histogram.

## 4. Illustrative Examples of MDF Measurements

To illustrate the process of measuring MDFs, we provide detailed examples for three systems that represent the range of data quality and galaxy type across our sample. The systems are CVn II (Section 4.1), Grus I (Section 4.2), and Dra II (Section 4.3). CVn II is a bright UFD with a fairly well-populated RGB. Grus I is an intermediate-luminosity UFD with an RGB present. Dra II is a faint UFD with no RGB at all.

### 4.1. CVn II

Figure 3 illustrates the MDF derivation process for CVn II. Using the F606W broadband CMD, we select 34 candidate member stars (red points) along the RGB that have F395N S/N > 10. We highlight stars in common with Kirby et al. (2013) (circled in light blue) and Vargas et al. (2013) (circled in dark blue), two other spectroscopic studies of CVn II. We compare our metallicities to these literature values in Section 4.4.

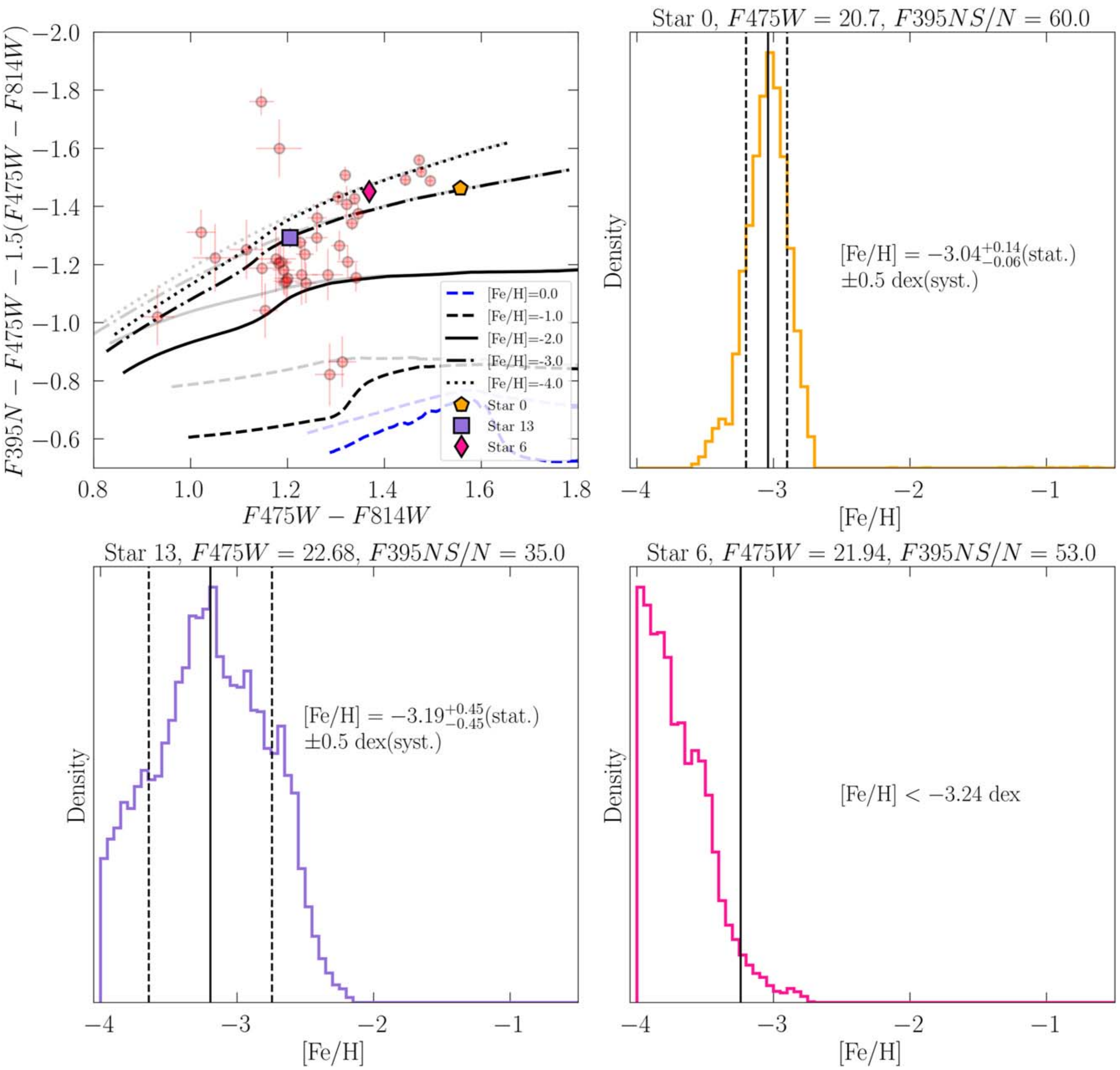

**Figure 4.** Example posterior distribution functions of stars in CVn II. (Upper left) Position of example stars in CaHK space, plotted against the MIST monometallic tracks used to infer metallicities in this work. (Upper right) Example of a star with a well-constrained PDF. (Lower left) Example of a star with a well-constrained PDF peak that is truncated at the metal-poor end; we also designate this star as an extremely metal-poor candidate. (Lower right) Example of a star for which we only constrain an upper limit is also an extremely metal-poor candidate.

The four panels in the top-right corner of Figure 3 show the ASTs (i.e., the difference in input and recovered magnitude versus input magnitude). The uncertainties in photometry are dominated by F395N and F475W. The typical scatter in F395N and F475W toward the faint magnitude limit of F475W ∼ 24 mag are 0.15 and 0.02 mag, respectively. Toward fainter magnitudes, there is also a bias in (out–in) in both filters. Characteristic to all the UFDs that we analyze, the bias effect is largest in F395N (∼0.1 mag).

The bottom left panel of Figure 3 shows the CVn II RGB stars in CaHK color space, with monometallic MIST isochrone tracks for $[\alpha/\mathrm{Fe}] = +0.4$ overplotted. The low-opacity lines are the monometallic models without ASTs applied, and the high-opacity models are the same models with the AST noise model applied. While the bias effect is less prominent for brighter and redder stars, it does become significant for stars bluer than F475W–F814W = 1.4. Without accounting for this bias effect, a star's inferred metallicity would be larger than it actually is. For example, a star at F475W = 23.5 and with F475W–F814W = 1.20, CaHK = −1.1 would have an inferred metallicity of −2.0 and −2.3 before and after applying the bias effect, respectively.

Plotted in CaHK space, it is apparent by eye that the stars in CVn II span a range of metallicities because they do not fall along a single monometallic track. The placement of stars in CVn II suggests that there are stars as metal-rich as [Fe/H] = −1.2 and at least as metal-poor as [Fe/H] = −4, with the bulk of them centered at [Fe/H] = −3.0.

We now discuss the nature of our individual metallicity measurements by describing the broad categories of individual posterior distributions and presenting examples in Figure 4. Some posterior distributions are well within the metallicity grid and have well-defined peaks; this is usually the case for stars at intermediate metallicities ($\gtrsim -3.0$) and intermediate-to-high S/N in F395N (Figure 4, top right). Others are truncated at the metal-poor end, corresponding to the metal-poor limit of the metallicity grid, but with well-defined peaks; these are often stars with [Fe/H] $< -3$ of intermediate or high S/N (Figure 4, bottom left) There are also stars that fall outside of the metallicity grid, so their metallicities are not constrained by the fitting process, and we can only obtain an upper limit (Figure 4, bottom right). This includes stars across a range of S/Ns.

The bottom-right panel of 3 shows the MDF of CVn II (red) and our Gaussian fit to the MDF (blue line). We infer $\langle[\mathrm{Fe/H}]\rangle = -2.98^{+0.12}_{-0.12}$ dex and $\sigma_{[\mathrm{Fe/H}]} = 0.55^{+0.12}_{-0.10}$ dex. We identify 15 stars as extremely metal-poor candidates for spectroscopic follow-up. We also find two stars with metallicity [Fe/H]$\sim -1.2$, albeit with uncertainties of about 0.5 dex. As we discuss in Section 4.4, spectroscopic studies of CVn II find similarly metal-rich member stars at larger radii, supporting the notion that our metal-rich stars may be bona fide members of CVn II. In Section 4.4, we compare our metallicities to literature values.

### 4.2. Grus I

Figure 5 illustrates the MDF process for Grus I. Grus I has a sparsely populated RGB and the S/N of our F395N data is sufficiently high that we are able to include MSTO stars in our analysis (i.e., the F395N S/N is $> 10$ down to the MSTO). The layout of Figure 5 is identical to the CVn II example (Figure 3). Stars in common with the literature study of Chiti et al. (2022) are indicated in light blue. One of the stars that passed our isochrone selection was ruled by Chiti et al. (2022) to be a kinematic nonmember, so it is represented in the CMD panel plots as a purple point. The ASTs for Grus I also reveal a systematic bias for F395N and F475W for fainter stars that results in the reddening of the monometallic tracks in CaHK space (e.g., at F475W = 24, a bias of 0.17 mag in F395N and 0.02 mag in F475W).

In the CaHK color space (bottom left), the highest S/N stars show a clear scatter, indicative of a metallicity spread. This is also present for the lower S/N stars, though it is more difficult to visually discern. We identified five extremely metal-poor candidates that would be compelling for spectroscopic follow-up studies. On the metal-rich end, we also find stars with [Fe/H] up to $-1.0$. From this sample, we infer $\langle[\mathrm{Fe/H}]\rangle = -2.62^{+0.14}_{-0.15}$ dex and $\sigma_{[\mathrm{Fe/H}]} = 0.61^{+0.12}_{-0.11}$ dex.

### 4.3. Dra II

Figure 6 shows our detailed MDF derivation for Draco II. Dra II has no RGB at all and thus the metallicities all come from lower MS stars. Our narrowband data for Dra II reaches S/N = 10 at F475W $\sim 24$. The layout of Figure 6 is the same as the previous two examples. We highlight stars in common with the ground-based CaHK study of Dra II by Longeard et al. (2018) in light blue.

The bottom left panel of Figure 6, shows the Dra II stars on the CaHK color space. Overplotted are two versions of the monometallic MIST isochrone tracks for [$\alpha$/Fe]-enhanced lower MS star models, which illustrate the bias profile computed from the ASTs. The effect of the AST bias is to lower the inferred metallicity of the star, similar to the case of RGB stars. For the case of Dra II, the highest S/N stars are also the stars that are bluest in F475W–F814W.

The MDF of Dra II (bottom-right panel) spans a metallicity range of $-4.0$ to $-1.5$. From our Gaussian MDF fitting, we infer $\langle[\mathrm{Fe/H}]\rangle = -2.72^{+0.10}_{-0.11}$ dex and $\sigma_{[\mathrm{Fe/H}]} = 0.40^{+0.12}_{-0.12}$ dex.

### 4.4. Comparison to the Literature

Figure 7 compares the metallicities and MDFs derived from our data to what is available in the literature for each galaxy. For CVn II, our sample includes five stars that were spectroscopically studied by Kirby et al. (2013), and of those five stars, three were also analyzed by Vargas et al. (2013). We color code stars with [$\alpha$/Fe] measurements by the value of their alpha. For this limited sample, we generally find that point estimate metallicities from Kirby et al. (2013) are systematically 0.6 dex more metal-rich than our findings. Including uncertainties, this level of disagreement is $\sim 1.5\sigma$. Given the use of different models, spectral features, and broad approaches, we are encouraged by the similarity of our findings. The stars that have alpha measurements from Vargas et al. (2013) are also color coded by the literature point estimates, and typical [$\alpha$/Fe] uncertainties from that study are 0.2 dex. Their point estimates are indicated in the color coding. As discussed in Fu et al. (2022), increasing $\alpha$-enhancements in our modeling leads to lower inferred metallicities. For the single star that has a slightly lower value of $\alpha$ from Vargas et al. (2013) than we assume, re-inferring the metallicity with our $\alpha$-value would bring the measurements into closer agreement. For the other two stars which are $\alpha$-enhanced, differences in $\alpha$-enhancements alone cannot reconcile the differences.

We also compare our MDF against those of Kirby et al. (2013) and Vargas et al. (2014). Kirby et al. (2013) infer $\langle[\mathrm{Fe/H}]\rangle = -2.12 \pm 0.05$ dex and $\sigma_{[\mathrm{Fe/H}]} = 0.59$ dex from 14 stars in CVn II. While our $\sigma_{[\mathrm{Fe/H}]}$ measurement is in agreement with Kirby et al. (2013), they infer a mean that is higher than ours by $\sim 0.8$ dex. This is because our MDF includes more stars that are below [Fe/H]$= -3.0$. Given the agreement shown in the 1:1 comparisons, it is unlikely that all of our extremely metal-poor stars are systematically too metal-poor. Additionally, it is likely that we are finding more extremely metal-poor stars in CVn II simply by virtue of an expanded sample size of 40. On the metal-rich end, the stars that we detect between $-1.5$ and $-1.0$ are not the same as those from Kirby et al. (2013), but that there are spectroscopically confirmed metal-rich stars affirms our confidence that the ones in our sample are also bona fide members of CVn II. Finally, we run a Kolmogorov–Smirnov (KS) test using `scipy.stats.ks_2samp` to test the null hypothesis that the MDF we measure for CVn II and the MDF from Kirby et al. (2013) are drawn from the same underlying distribution. The resulting $p$-value for the test is 0.11, suggesting that there are insufficient grounds to reject this null hypothesis, and that it is possible for these MDFs to share the same underlying distribution.

The center column compares our metallicities for Grus I against those in the literature. Chiti et al. (2022) identified eight members of Grus I using Magellan/IMACS spectroscopy and measured radial velocities and metallicities from the CaT feature. Of the three stars that we have in common with that study, two of them have CaT metallicity measurements. They are shown in the top panel. One star agrees within $<1.5\sigma$ ([Fe/H]$= -2.5$), while the other is $\sim$1 dex more metal-poor in Chiti et al. (2022). C-enhancement could contribute to this discrepancy by adding absorption into the F395N filter that would make our measurement of this star more metal-rich, but there are currently no known C-enhanced stars in Grus I. Ji et al. (2019) studied two of the brightest stars in Grus I using high-resolution Magellan/MIKE spectroscopy. However, these stars are saturated in the archival HST broadband imaging, and cannot be compared.

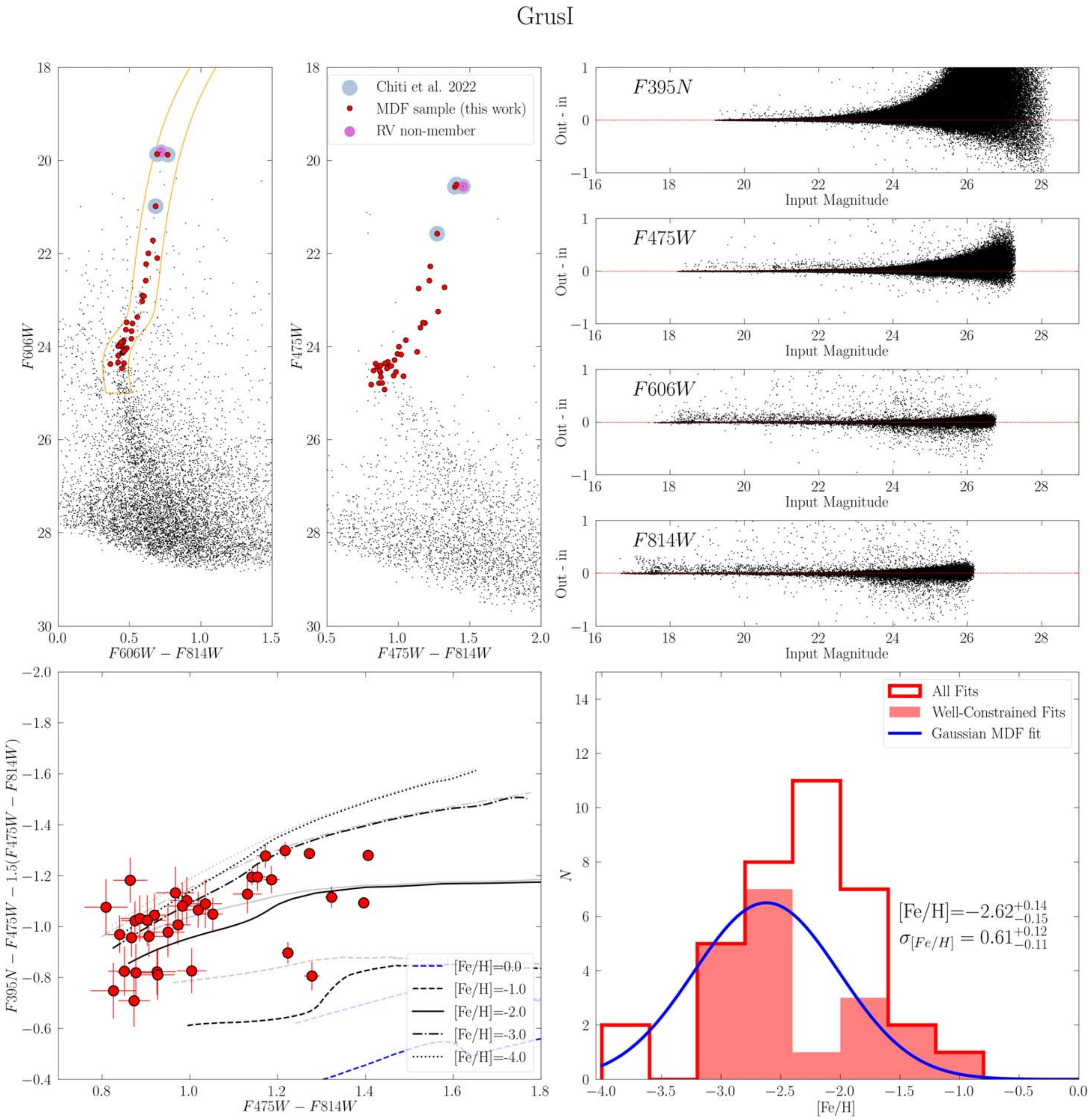

**Figure 5.** Same as Figure 3 only for the fainter, less populated system, Grus I.

In the bottom panel, we compare our MDF against those of Ji et al. (2019) to Chiti et al. (2022). Chiti et al. (2022) infer $\langle$[Fe/H]$\rangle = -2.62 \pm 0.11$ and were only able to place an upper limit on $\sigma_{\rm [Fe/H]}$ of 0.44 dex. We measure $\langle$[Fe/H]$\rangle = -2.62^{+0.14}_{-0.15}$, which is in 1$\sigma$ agreement with Chiti et al. (2022). Our $\sigma_{\rm [Fe/H]}$ measurement of $0.61^{+0.12}_{-0.11}$ is larger than what Chiti et al. (2022) found, though our sample is also significantly larger and has better-populated tails. Taking both distributions at face value, the KS test produces a $p$-value of 0.06, suggesting that these MDFs may share the same underlying distribution.

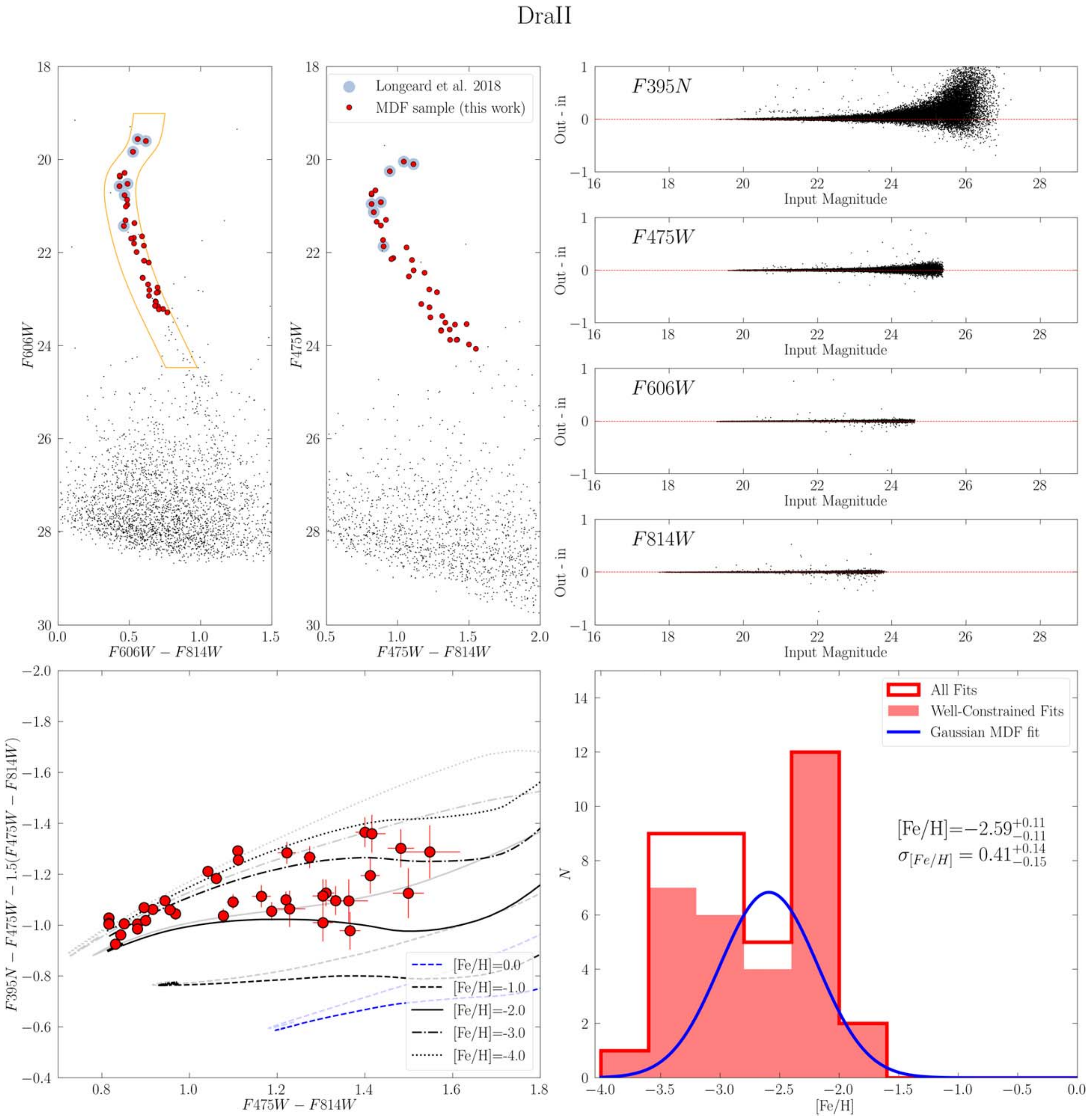

**Figure 6.** Same as Figure 3 only for Dra II, which has no RGB stars. The monometallic tracks shown in the bottom left panel are for MSTO and MS stars. Accordingly, the MDF is entirely based on stars from these evolutionary phases.

In the right column, we compare our findings for Dra II against those from Longeard et al. (2018). Longeard et al. (2018) studied 12 Dra II stars using the Pristine narrowband CaHK filter. We have seven stars in common with those in Longeard et al. (2018).

The top panel shows that our well-constrained metallicity measurements are in better than $1\sigma$ agreement with the measurements from Longeard et al. (2018). There are three stars for which we can only provide an upper limit, and which are more metal-rich than $-3.0$ in the Longeard et al. (2018) study. We note that Longeard et al. (2018) impose a metallicity floor on their measurements at [Fe/H]$=-3.0$, limiting this comparison.

In the bottom panel of the Dra II column, we compare our MDF to the one derived by Longeard et al. (2018). Longeard et al. (2018) measure $\langle$[Fe/H]$\rangle = -2.7 \pm 0.1$ dex and place an upper limit on $\sigma_{\rm [Fe/H]}$ of 0.24. Their $\langle$[Fe/H]$\rangle$ measurement is in $1\sigma$ agreement with ours. Additionally, we also resolve $\sigma_{\rm [Fe/H]}$ because our MDF spans a wider range. The floor of $-3.0$ adopted by Longeard et al. (2018) complicates their ability to resolve $\sigma_{\rm [Fe/H]}$.

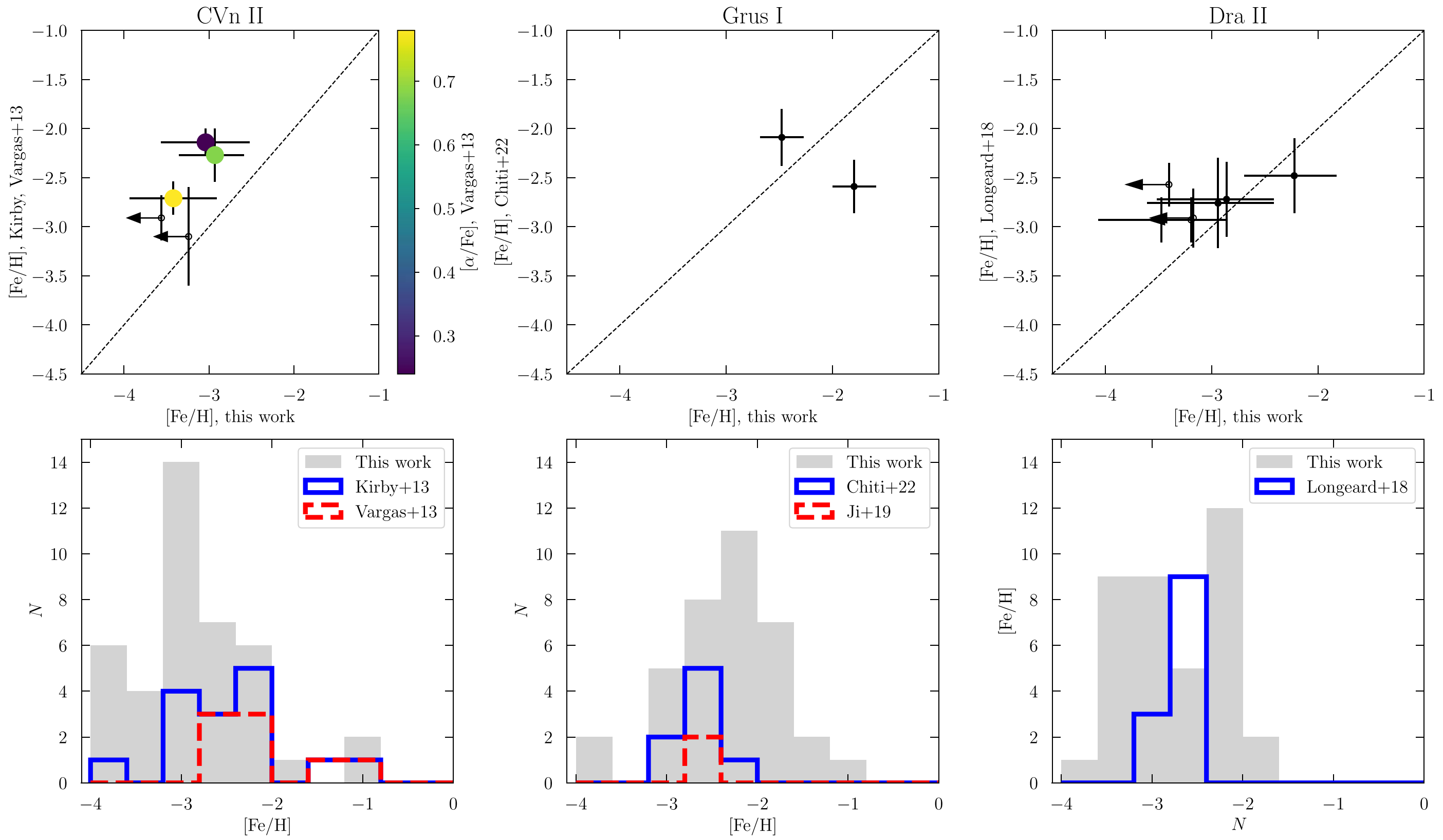

**Figure 7.** A comparison of metallicities for stars in common with our example galaxies CVn II, Grus I, and Dra II. The top row shows a 1:1 comparison of stars common in both studies. Our metallicities and those in the literature generally agree to $\sim 1.5\sigma$. The lower panels show a comparison between our MDFs (gray) and all MDFs in the literature for each galaxy. The histogram bins are 0.4 dex wide.

The corresponding KS test yields a $p$-value of 0.09, suggesting a similarity between our MDF for Dra II and the one from the literature

We refer the reader to Appendix C for the full set of 1:1 comparisons between our measurements and those from the literature across all the UFDs in this study.

## 5. Results

We undertake an exhaustive comparison of our MDFs to those previously published in the literature for the same galaxies. For completeness, here we list the papers used for each galaxy's literature values and refer to the face value results from these papers in the following sections: Eri II (Li et al. 2017; Martínez-Vázquez et al. 2021; CVn II (Kirby et al. 2013; Hya II (Kirby et al. 2015); Ret II (Simon et al. 2015; Koposov et al. 2015b; Walker et al. 2016; Ji et al. 2016, 2023); Hor I (Koposov et al. 2015b; Nagasawa et al. 2018); Grus I (Ji et al. 2019; Chiti et al. 2022); Ret III (Fritz et al. 2019); Wil 1 (Willman et al. 2011); Phe II (Fritz et al. 2019); Eri III (none); Tuc V (Simon et al. 2020); Seg 1 (Geha et al. 2009; Norris et al. 2010; Simon et al. 2011; Frebel et al. 2014); and Dra II (Longeard et al. 2018). We present a summary of our direct comparisons to the literature in Appendix C.

### 5.1. MDFs for Individual Galaxies

#### 5.1.1. Eri II

Eri II ($M_V = -7.1,\ L = 10^{4.8}\,L_\odot$) was first discovered in Bechtol et al. (2015) and Koposov et al. (2015a). Since its discovery, Li et al. (2017) identified 28 RGB members within 8′ of Eri II using Magellan/IMACS spectroscopy and measured metallicities for 16 of them using the CaT equivalent width calibration. They measure $\langle[\mathrm{Fe/H}]\rangle = -2.38 \pm 0.13$ and $\sigma_{[\mathrm{Fe/H}]} = 0.47^{+0.12}_{-0.09}$. Martínez-Vázquez et al. (2021) derived metallicities for 46 RR Lyrae stars in Eri II; their metallicity inference method is calibrated to the $\langle[\mathrm{Fe/H}]\rangle$ measured by Li et al. (2017), and they report $\sigma_{[\mathrm{Fe/H}]}$ of 0.2 dex. As the authors remark, their smaller inferred $\sigma_{[\mathrm{Fe/H}]}$ is expected because the most metal-poor and metal-rich stars in a galaxy's MDF would not end up in the instability strip for RR Lyrae.

Due to its brightness and the abundance of available literature references, we analyzed this galaxy in the first paper of our program in Fu et al. (2022) to verify the efficacy of CaHK for recovering the MDFs of UFDs. In Fu et al. (2022), we report metallicities for 60 resolved RGB stars in Eri II, measure $\langle[\mathrm{Fe/H}]\rangle = -2.50 \pm 0.07$ and $\sigma_{[\mathrm{Fe/H}]}=0.42 \pm 0.06$. In this work, we reanalyze Eri II following the procedure outlined in previous sections, which also newly includes a treatment of systematic uncertainties. Selecting along the RGB of Eri II, we obtained a sample of 75 stars for this analysis. We ended up with a larger sample size because the F606W–F814W CMD in which we made our member selection is a higher S/N than the F475W–F814W CMD used for member selection in Fu et al. (2022). The resulting MDF spans a similar range to that from our previous study.

From these stars, we measure $\langle[\mathrm{Fe/H}]\rangle = -2.63^{+0.06}_{-0.06}$ and $\sigma_{[\mathrm{Fe/H}]} = 0.26^{+0.08}_{-0.09}$. Both the mean metallicity and the metallicity dispersion are lower than the values inferred from Fu et al. (2022) by over $1\sigma$. This is because the MIST models used for the analysis in this paper are scaled to the Grevesse & Sauval (1998) solar abundances, while the MIST models used in Fu et al. (2022) are scaled to the Asplund et al. (2009) solar abundances.[18] The metallicity measurements made using the Grevesse & Sauval (1998) scaled models are on average more metal-poor by $0.1 \sim 0.2$ dex. This difference accounts for the change in $\langle[\mathrm{Fe/H}]\rangle$. The lower $\sigma_{[\mathrm{Fe/H}]}$ is due to a larger fraction of extremely metal-poor stars in our sample, which contributes to a smaller dispersion measurement because of the 0.5 dex systematic uncertainty floor.

#### 5.1.2. CVn II

CVn II ($M_V = -5.1$, $L = 10^{4.0}\,L_\odot$) was originally discovered in SDSS by Belokurov et al. (2007) and subsequently studied spectroscopically by Simon & Geha (2007), Kirby et al. (2013), and Vargas et al. (2013). From a sample of 14 stars, Kirby et al. (2013) measure $\langle[\mathrm{Fe/H}]\rangle = -2.2 \pm 0.05$ dex and $\sigma_{[\mathrm{Fe/H}]} = 0.59$ dex.

From a sample of 40, we measure a metallicity of $-2.98^{+0.12}_{-0.12}$ and $\sigma_{[\mathrm{Fe/H}]} = 0.55^{+0.12}_{-0.10}$. Our measurement of $\langle[\mathrm{Fe/H}]\rangle$ places CVn II as one of the most metal-poor UFDs known to date.

As detailed in Section 4.1, our $\sigma_{[\mathrm{Fe/H}]}$ measurement is in good agreement with the spectroscopic study, but we measure a lower $\langle[\mathrm{Fe/H}]\rangle$ due to more low-metallicity stars in our sample. Differences in $\langle[\mathrm{Fe/H}]\rangle$ appear to be driven by our larger fraction of EMPs: 38% of our stars are extremely metal-poor, versus 20% in the Kirby et al. (2013) sample. One possibility is that the small sample of Kirby et al. is missing EMPs, leading to a higher $\langle[\mathrm{Fe/H}]\rangle$ estimate. Alternatively, if half of our EMPs are actually more metal-rich by 0.5 dex, it would bring our mean metallicities into better agreement.

#### 5.1.3. Hya II

Hya II ($M_V = -4.9$, $L = 10^{3.9}\,L_\odot$) was discovered by Martin et al. (2015) in the Survey of the Magellanic Stellar History conducted using the DECam instrument on the Blanco telescope. Kirby et al. (2015) followed up on this discovery by observing Hya II using Keck/DEIMOS, targeting the CaT lines. They identified 13 members of Hya II, and among that subset were able to measure metallicities for five of those stars. They measured for Hya II $\langle[\mathrm{Fe/H}]\rangle = -2.02 \pm 0.08$, and $\sigma_{[\mathrm{Fe/H}]} = 0.40^{+0.48}_{-0.26}$.

We present the MDF of Hya II, constructed from 31 RGB stars, in Figure 10. The stars span a metallicity from $-4.0$ to $-1.5$, and the bulk of them are at around a metallicity of $-3.0$. From this sample, we measure $\langle[\mathrm{Fe/H}]\rangle = -3.05^{+0.12}_{-0.13}$ and $\sigma_{[\mathrm{Fe/H}]} = 0.47^{+0.13}_{-0.12}$. Compared to the measurements from Kirby et al. (2015), our $\sigma_{[\mathrm{Fe/H}]}$ are in $1\sigma$ agreement, but our $\langle[\mathrm{Fe/H}]\rangle$ is lower than that study by about $\sim$0.7 dex because our sample is dominated by stars below $-2.5$. We have seven stars in common with the sample studied by Kirby et al. (2015), and none of them are foreground interlopers.

We also note the star in our sample that is at $[\mathrm{Fe/H}] \sim -1.5$, setting it apart from the rest of the stars in our sample. We do not have data to verify conclusively whether it may be an outlier, so instead we recompute the Gaussian MDF excluding that star to infer $\langle[\mathrm{Fe/H}]\rangle = -3.08^{+0.11}_{-0.12}$ and $\sigma_{[\mathrm{Fe/H}]} = 0.33^{+0.12}_{-0.13}$. As expected, this new calculation results in a lower $\sigma_{[\mathrm{Fe/H}]}$, but this new value is still in $1\sigma$ agreement with the value computed from using the full sample. Since the uncertainty on $\sigma_{[\mathrm{Fe/H}]}$ from Kirby et al. (2015) is nearly 0.5 dex on the upper end, this new calculation also does not represent a significant improvement in agreement from the calculation made using the full sample. Additionally, the removal of the most metal-rich star in our sample worsens the discrepancies between the $\langle[\mathrm{Fe/H}]\rangle$ measurements. Most notably, we measured the MDF from 30 stars whereas Kirby et al. (2015) measured the MDF from five, and our expanded sampling may explain the bulk of these discrepancies.

#### 5.1.4. Ret II

Ret II ($M_V = -4.0$, $L = 10^{3.5}\,L_\odot$) was discovered in the DES by Bechtol et al. (2015) and Koposov et al. (2015a). Since then, it has been a dwarf galaxy of great interest to the community because its stars contain evidence of a rare *r*-process enrichment event (Ji et al. 2016; Roederer et al. 2016). The most recent study on Ret II (Ji et al. 2023) identified 32 member stars of the satellite using Very Large Telescope (VLT)/GIRAFFE and Magellan/M2FS spectroscopy and aimed to measure the abundances of the *r*-process element barium. As part of this study, they provided constraints on the metallicities of 29 stars in that sample. From the 13 stars for which they were able to constrain metallicities beyond an upper limit, they measure $\langle[\mathrm{Fe/H}]\rangle = -2.64 \pm 0.11$ and $\sigma_{[\mathrm{Fe/H}]} = 0.32^{+0.10}_{-0.07}$.

We identify 77 candidate members in Ret II along the MSTO and MS. Using the catalog of Simon et al. (2015), we remove one star whose velocity is inconsistent with Ret II membership. From the remaining 76 stars, which span a wide range in metallicity, we measure $\langle[\mathrm{Fe/H}]\rangle = -2.64^{+0.1}_{-0.11}$ and $\sigma_{[\mathrm{Fe/H}]} = 0.72^{+0.09}_{-0.08}$. Our $\langle[\mathrm{Fe/H}]\rangle$ measurement is in agreement with that in Ji et al. (2023) and earlier works on the dwarf (Koposov et al. 2015b; Simon et al. 2015; Walker et al. 2015). However, we resolve a larger $\sigma_{[\mathrm{Fe/H}]}$.

Direct comparison with existing metallicities is challenging. Due to saturation effects in our photometry, our sample lacks substantial overlap with the Ji et al. (2023) sample. We have three stars in common with that study, with two stars of that subset constrained only by an upper limit in Ji et al. (2023). The star for which the measurement is constrained is in agreement to $2\sigma$, and the upper limits point in the correct direction toward agreement.

Our large metallicity dispersion in Ret II is driven by the larger number of stars in the tails of the MDF than what Ji et al. (2023) report. Although the uncertainties on the metallicities of these stars are large (i.e., we adopt a 0.3 dex systematic uncertainty, see Appendix B.5), these stars still contribute to a higher metallicity dispersion measurement as compared to the literature. Because this dispersion is so much larger than what is reported by spectroscopic data, we undertake further scrutiny to see if any systematic effects in our analysis, or to Ret II in particular, can reconcile the difference.

First, we consider the possibility of foreground contamination. The MW halo has very few extremely metal-poor ($[\mathrm{Fe/H}] < -3.0$) stars (e.g., Conroy et al. 2019), which suggests the foreground is unlikely the source of a large metal-poor component.

Instead, we consider the possibility that our large dispersion may result from substantial contamination from higher metallicity MW foreground stars. Figure 15 illustrates this possibility. Here, we plot the simulated TRILEGAL MW foreground in the direction of Ret II. We find that, on average, one foreground star passes our isochrone and spatial cuts. A single star does not significantly alter the dispersion measurement.

[18] C. Conroy, A. Dotter (private communication)

The TRILEGAL model also suggests that if the foreground is significant, it should be present in other areas of the CMD, i.e., not just those that pass our isochrone cut. For Ret II, the CMDs do not exhibit large populations of objects outside the RGB and MS of Ret II, further suggesting that foreground contamination is not significant. A clear example of a galaxy with a larger degree of foreground contamination is Seg 1, for which there are many objects on either side of the actual galaxy RGB and MS.

Second, we consider the possibility that there may be some unknown MW substructure along the line of sight that is not accounted for in the smooth MW halo models of TRILEGAL. Such a contaminating substructure would have to be able to account for a large fraction of the ∼20 metal-rich and/or metal-poor stars observed in the tails of the MDF. Additionally, it would have to be concentrated almost entirely along the MS of Ret II. To date, there are no currently known substructures in the vicinity of Ret II from wider-field spectroscopic studies of Ret II (Walker et al. 2015; Simon et al. 2015; Ji et al. 2023), as well as from photometric imaging (Drlica-Wagner et al. 2015; Mutlu-Pakdil et al. 2018). It is unlikely that a previously undiscovered stellar substructure would be first detected in a narrow HST pointing compared to other observations.

Third, we consider the possibility that our dispersion measurements are inflated by lower S/N stars. That is, we allow for the possibility that some unaccounted for systematic effects in the photometry of the faint stars have biased our metallicity determinations.[19] To explore this effect, we remove faint stars in our sample with $m_{\rm F475W} > 22.5$ ($M_{\rm F475W} > 5$) and refit our MDF with the remaining 24 stars, finding $\langle[\rm Fe/H]\rangle = -3.07 \pm 0.19$ and $\sigma_{[\rm Fe/H]} = 0.70^{+0.17}_{-0.15}$. This cut disproportionately removes the metal-rich stars, which, in part is due to a known effect of magnitude-limited cuts (i.e., metal-rich stars are fainter at a fixed magnitude owing to increased atmospheric opacity; Manning & Cole 2017). As a result, we measure a lower mean for this revised sample. The resulting dispersion is still large and in agreement with the dispersion inferred using our full sample.

Finally, we consider that unaccounted for impacts of binarity are inflating our dispersion measurements. Our investigation of the impact of binarity in Appendix B.3 suggests that we may inflate the metallicity of an unresolved binary star by up to 0.2 dex by fitting them with single stellar models. We consider the upper limit of binarity on our MDF by measurement by assuming that all of our stars that are more metal-rich than −2.0 have been affected by the maximum possible impact of binarity, i.e., all 20 of these stars have an $\sim 0.5\,M_\odot$ solar mass companion (see Figure 16). We subtract 0.2 dex from their metallicity measurements and recompute our Gaussian fit, finding $\langle[\rm Fe/H]\rangle = -2.69 \pm 0.1$ and $\sigma_{[\rm Fe/H]} = 0.63 \pm 0.1$. The resulting mean and dispersion remain in agreement with the values inferred using our full sample.

In summary, all of our tests retain the large dispersion inferred using the original sample. At worst, they implausibly shift the mean out of agreement with the mean measured in the original sample and from the literature. We do not claim to have resolved the tension with the literature, but at this time there is no obvious solution to resolving these discrepancies. We welcome spectroscopic follow-up studies by the community to improve membership selection and refine these measurements, and provide coordinates and magnitudes for all the stars in Ret II used to derive its MDF in Table 4.

[19] See Appendix D for further discussion on this context.

#### 5.1.5. Hor I

Hor I ($M_V = -3.8$, $L = 10^{3.4}\,L_\odot$) was discovered in DES data by Bechtol et al. (2015) and first followed up spectroscopically by Koposov et al. (2015b) using VLT/GIRAFFE. In that study, they identified five candidate members. From those stars, they report $\langle[\rm Fe/H]\rangle = -2.76 \pm 0.1$ and $\sigma_{[\rm Fe/H]} = 0.17^{+0.20}_{-0.03}$, and detect $\alpha$-enhancement of $[\alpha/\rm Fe] = 0.30 \pm 0.07$. Nagasawa et al. (2018) later observed three of these stars using Magellan/MIKE and performed a detailed chemical abundance analysis. The stars they observed have a low average metallicity of [Fe/H]∼ −2.6, and low α-enhancement at $[\alpha/\rm Fe] \sim 0.0$.

We present the MDF of Hor I, measured from 27 RGB stars, in Figure 10. The stars in the MDFs span a wide range of metallicity, with some potentially being extremely metal-poor stars, to stars as metal-rich as ∼ −1.0 dex. The bulk of the stars in Hor I are between −3.0 and −2.0.

The stars observed in the aforementioned studies fall beyond our HST footprint, so we cannot make direct comparisons of the measurements. Instead, we compare the broad features of metallicity measurement results. Similar to the aforementioned studies, we find a low $\langle[\rm Fe/H]\rangle$ for Hor I of $-2.79^{+0.12}_{-0.13}$. We also resolve $\sigma_{[\rm Fe/H]} = 0.56^{+0.11}_{-0.09}$. We note that there is a star in the MDF of Hor I more metal-rich than ∼ −1.5 dex that could be an interloper, but absent detailed kinematic and astrometric data, it is difficult to ascertain for sure that it is not a member of Hor I. We instead recompute the Gaussian MDF excluding these stars to infer $\langle[\rm Fe/H]\rangle = -2.84^{+0.11}_{-0.12}$ dex and $\sigma_{[\rm Fe/H]} = 0.41^{+0.11}_{-0.10}$ dex. As expected, the removal of this star preserves the metal-poor nature of Hor I. $\sigma_{[\rm Fe/H]}$ also decreases as a result, but we still clearly resolve a metallicity spread in the dwarf.

#### 5.1.6. Grus I

Grus I ($M_V = -3.5$, $L = 10^{3.3}\,L_\odot$) was discovered by Koposov et al. (2015a) in the DES survey. At the time of discovery, its classification was unknown. Ji et al. (2019) studied the chemical abundances of two stars in Grus I and used the deficiency in neutron capture elements in those two stars to classify Grus I as a dwarf galaxy. Chiti et al. (2022) were able to observe additional members of Grus I using Magellan/IMACS spectroscopy to measure its metallicity and velocity dispersion. They were unable to resolve $\sigma_{[\rm Fe/H]}$, but the measured velocity dispersion of $\sigma_{rv} = 2.5^{+1.3}_{-0.8}$ km s$^{-1}$ translates to a large dynamical mass-to-light ratio for Grus I and informs its classification as a dwarf galaxy.

In Section 4.2, we presented a detailed comparison between our Grus I MDF measurement and those from Chiti et al. (2022). We measure $\langle[\rm Fe/H]\rangle = -2.62^{+0.14}_{-0.15}$ and $\sigma_{[\rm Fe/H]} = 0.61^{+0.12}_{-0.11}$. Our $\langle[\rm Fe/H]\rangle$ measurement is consistent with the Chiti et al. (2022) study, but we were able to resolve $\sigma_{[\rm Fe/H]}$, whereas that study could only provide an upper limit. Our results further support the conclusion that Grus I is a dwarf galaxy.

### 5.1.7. Ret III

Ret III ($M_V = -3.3$, $L = 10^{3.2}\,L_{\odot}$) was discovered in DES by Drlica-Wagner et al. (2015), and followed up spectroscopically by Fritz et al. (2019) using VLT/FLAMES targeting the CaT features. From that study, Fritz et al. (2019) found three likely members in Ret III, and used them to measure $\langle[\mathrm{Fe/H}]\rangle = -2.81 \pm 0.09$ and $\sigma_{[\mathrm{Fe/H}]} = 0.35^{+0.21}_{-0.09}$. They were unable to resolve a velocity dispersion, putting an upper limit on the value at 31.2 km s$^{-1}$.

We identify 18 RGB stars in Ret III. From this sample, we measure $\langle[\mathrm{Fe/H}]\rangle = -2.46^{+0.12}_{-0.15}$ and $\sigma_{[\mathrm{Fe/H}]} = 0.31^{+0.2}_{-0.18}$. Our $\sigma_{[\mathrm{Fe/H}]}$ measurement is in good agreement with the Fritz et al. (2019) measurement, but we infer a larger $\langle[\mathrm{Fe/H}]\rangle$.

We also have two stars in common with the Fritz et al. (2019) study, all of which were deemed spectroscopic members. For star 0 in our sample, we measure a metallicity of $[\mathrm{Fe/H}] = -2.38^{+0.04}_{-0.03}$(stat.) $\pm$ 0.2(syst.), compared to their measurement of $-3.24 \pm 0.15$. For star 1 in our sample, we measure $[\mathrm{Fe/H}] = -2.07^{+0.04}_{-0.04}$(stat.) $\pm$ 0.2(syst.), compared to their measurement of $-2.32 \pm 0.15$. Our measurements for star 0 and star 1 are larger than the Fritz et al. (2019) measurements by $>3\sigma$ and $\sim 2\sigma$, respectively. The Fritz et al. (2019) measurements were derived from spectral synthesis methods that assume $[\alpha/\mathrm{Fe}] = +0.5$, which is more enhanced than our assumption, but this is not enough to fully account for discrepancies. We discuss these comparisons further in Appendix C.

### 5.1.8. Wil 1

Wil 1 ($M_V = -2.9$, $L = 10^{3.1}\,L_{\odot}$) was discovered by Willman et al. (2005) as one of the first faint MW satellites known to the community. Similar to Seg 1, Wil 1 was one of the stellar associations whose classification as either a star cluster or a dwarf galaxy has been the subject of ongoing debate.

The most comprehensive spectroscopic study of Wil 1 done to date was by Willman et al. (2011) using Keck/DEIMOS spectroscopy. They identified 45 candidate members of Wil 1 and 40 of those are high confidence. Their stars span the bright RGB of Wil 1 down to the MS. Although they find an irregular kinematic distribution for Wil 1, they detect evidence of a large metallicity spread in Wil 1 from metallicity measurements of the two RGB stars in their sample, with one star at $[\mathrm{Fe/H}] = -1.73 \pm 0.12$, and the other at $[\mathrm{Fe/H}] = -2.65 \pm 0.12$.

We crossmatch our sample against that from Willman et al. (2011) and find that we have 10 stars in common. Of the 10 stars we have in common with their study, one was identified as a probable nonmember in Wil 1.

We identify 68 stars in Wil 1 on the MSTO and MS. Our stars span a wide range of metallicity from $-4.0$ to $-1.0$. We measure $\langle[\mathrm{Fe/H}]\rangle = -2.53^{+0.11}_{-0.11}$, and $\sigma_{[\mathrm{Fe/H}]} = 0.65^{+0.10}_{-0.09}$. A significant fraction of the stars we find are at metallicities above $-2.0$ (31%). With a sample size of over an order magnitude larger, we confirm the large metallicity range suggested by measurements of the two stars from Willman et al. (2011). The large metallicity dispersion we recover supports results from mass-segregation studies that Wil 1 is a dwarf galaxy (Baumgardt et al. 2022).

### 5.1.9. Phe II

Phe II ($M_V = -2.7$, $L = 10^{3.0}\,L_{\odot}$) was discovered in the DES footprint by Bechtol et al. (2015) and Koposov et al. (2015a), and subsequently followed up spectroscopically by Fritz et al. (2019) using VLT/FLAMES spectroscopy targeting the CaT features. From that study, the authors identify six likely members, and five whose membership they are certain of. From the five members, they measure $\langle[\mathrm{Fe/H}]\rangle = -2.51^{+0.19}_{-0.17}$ and $\sigma_{[\mathrm{Fe/H}]} = 0.33^{+0.29}_{-0.16}$. They also measure a radial velocity dispersion of $\sigma_v = 11.0^{+9.4}_{-5.3}$ km s$^{-1}$. From these measurements resolving both a metallicity dispersion as well as a large velocity dispersion (and therefore a large mass-to-light ratio), the authors conclude that Phe II is a dwarf galaxy.

We identify 10 RGB stars in Phe II, one of which we have in common with Fritz et al. (2019). The star we have in common with that study is a kinematic member of Phe II. Our measurement for that star is $[\mathrm{Fe/H}] = -2.23 \pm 0.04$ (stat.) $\pm$ 0.2(sys.), while Fritz et al. (2019) measure for it a metallicity of $[\mathrm{Fe/H}] = -2.65 \pm 0.15$. Our measurements for this star agree within $2\sigma$.

From our overall sample, we measure $\langle[\mathrm{Fe/H}]\rangle = -2.36^{+0.18}_{-0.16}$ and $\sigma_{[\mathrm{Fe/H}]} = 0.41^{+0.22}_{-0.17}$. These measurements agree with those from Fritz et al. (2019) at the $1\sigma$ level. Alongside the absence of mass segregation detected for Phe II stars (Baumgardt et al. 2022), our results support the conclusion that Phe II is a dwarf galaxy.

### 5.1.10. Eri III

Eri III ($M_V = -2.1$, $L = 10^{2.8}\,L_{\odot}$) was simultaneously discovered by Bechtol et al. (2015) and Koposov et al. (2015a) in the DES survey. To date, it has not been observed by spectroscopy, but Conn et al. (2018) targeted it for follow-up Gemini imaging to derive its structural properties. With a half-light radius of $r_h = 8.3^{+0.9}_{-0.8}$ pc, and a magnitude of $M_V = -2.1$, the categorization of Eri III, like many recently discovered satellites, is ambiguous from structural parameters alone.

We identify 13 candidate RGB members in Eri III and present the resulting MDF in Figure 10. The stars in Eri III span a wide range of metallicities, from $-3.0$ to $-1.2$. Using these measurements, we infer $\langle[\mathrm{Fe/H}]\rangle = -2.03^{+0.16}_{-0.18}$ and $\sigma_{[\mathrm{Fe/H}]} = 0.49^{+0.22}_{-0.18}$. Since we resolve a nonzero $\sigma_{[\mathrm{Fe/H}]}$ for this satellite, we classify it as a dwarf galaxy. Baumgardt et al. (2022) are unable to resolve mass segregation in Eri III above the $3\sigma$ level, supporting our classification of Eri III from the metallicity dispersion. Spectroscopic follow-up to determine kinematic properties and membership for this satellite would help confirm or refute this categorization.

### 5.1.11. Tuc V

Tuc V ($M_V = -1.6$, $L = 10^{2.6}\,L_{\odot}$) was first discovered by Drlica-Wagner et al. (2015) in the DES footprint. Conn et al. (2018) targeted Tuc V for follow-up imaging, but did not find evidence in the vicinity of Tuc V of a bound stellar association. From this imaging, they suggest that Tuc V may be a chance overdensity in the SMC halo due to its proximity, or a dissolving star cluster.

Since then, Simon et al. (2020) studied Tuc V using Magellan/IMACS spectroscopy in the CaT region and identified three candidate members of Tuc V. They were able to measure CaT metallicities for two of the candidate members, finding $\langle[\mathrm{Fe/H}]\rangle = -2.16 \pm 0.23$. They were unable to resolve $\sigma_{[\mathrm{Fe/H}]}$.

As shown in the CMD figures (Figures 1 and 2), we clearly see a stellar sequence in our imaging of the Tuc V field,

including a defined MSTO. We identify six stars along the RGB of Tuc V and from these six stars, measure $\langle[\mathrm{Fe/H}]\rangle = -2.41^{+0.26}_{-0.34}$ and $\sigma_{[\mathrm{Fe/H}]} = 0.61^{+0.44}_{-0.28}$, thereby resolving a nonzero metallicity dispersion at the $2\sigma$ level. We therefore classify Tuc V as a dwarf galaxy, supporting the conclusion from the absence of mass segregation in this satellite as measured by Baumgardt et al. (2022).

### 5.1.12. Seg 1

Seg 1 ($M_V = -1.3$, $L = 10^{2.5} L_\odot$) was one of the first-discovered UFDs, uncovered by Belokurov et al. (2007) in SDSS data. At the time of discovery, it was among the faint satellites that ushered in a new paradigm in dwarf galaxy studies, where structural parameters alone became insufficient to determine a stellar association's classification as either star cluster or galaxy. Follow-up spectroscopic studies to determine its classification are Geha et al. (2009), Norris et al. (2010), Simon et al. (2011), and Frebel et al. (2014). Simon et al. (2011) conducted a complete spectroscopic study of stars in the field of Seg 1 down to 22 mag in the SDSS $r$ band, encompassing the RGB and the MSTO. Frebel et al. (2014) performed high-resolution chemical abundance analysis of six RGB stars in Seg 1, adding their sample to the one other star analyzed in the same manner by Norris et al. (2010), and analyzed this data in the context of Seg 1 history. The metallicities of the seven stars studied and compiled by Frebel et al. (2014) range from −3.8 to −1.4.

We crossmatched our sample against the Simon et al. (2011) sample and found that we have 13 stars in common from our initial CMD selection among the MS. Of the 13 stars, six were ruled out as kinematic nonmembers by Simon et al. (2011). Of the six nonmembers, four have characteristic velocities of MW stars, and two belonged to the 300S stellar stream in the vicinity of the dwarf (Grillmair 2014; Fu et al. 2018). Since the observed Seg 1 stars are on the MS where the CaT calibration does not extend, we do not have metallicity measurements from this study for comparison.

From 12 stars, we measure for Seg 1 $\langle[\mathrm{Fe/H}]\rangle = -2.36^{+0.23}_{-0.25}$ and $\sigma_{[\mathrm{Fe/H}]} = 0.65^{+0.26}_{-0.21}$. The stars in our sample also span the same metallicity range as the stars from Frebel et al. (2014). We also have one star in common with Frebel et al. (2014), which we discuss in Appendix B.2 in the context of assessing the impact of carbon enhancements. Frebel et al. (2014) identified that star as the most metal-rich known in Seg 1, with $[\mathrm{Fe/H}] = -1.42 \pm 0.24$ and also carbon enhancement with $[\mathrm{C/Fe}] = 1.44 \pm 0.2$. In our work, we find that this star has $[\mathrm{Fe/H}] = -1.86^{+0.11}_{-0.12}$(stat.) $\pm$ 0.3(syst.); these measurements are in agreement within $\sim 1.5\sigma$.

### 5.1.13. Dra II

Dra II ($M_V = -0.8$, $L = 10^{2.3} L_\odot$) was discovered by Laevens et al. (2015) in the Pan-STARRS survey. It is the faintest galaxy in our sample and has no RGB stars. We describe its MDF inference in detail in Section 4.3.

At the time of its discovery, its classification as either a star cluster or a galaxy was uncertain from structural parameters alone. To resolve this uncertainty, Longeard et al. (2018) observed Dra II using the Pristine narrowband photometry and Keck/DEIMOS spectroscopy to obtain metallicity and kinematic information. They measure a low $\langle[\mathrm{Fe/H}]\rangle$ for Dra II (−2.7 dex), a low $\sigma_{[\mathrm{Fe/H}]}$ (<0.24 dex), and place an upper limit on the velocity dispersion of Dra II at $\sigma_v < 5.9$ km s$^{-1}$. Coupled with an orbital history for Dra II that takes it within 25 kpc of the Galactic Center, Longeard et al. (2018) suggest that Dra II is a potentially disrupting dwarf galaxy.

More recently, as part of a broader study, Baumgardt et al. (2022) measured the degree of stellar mass segregation in Dra II. They find $R_{\mathrm{bright}}/R_{\mathrm{faint}} = 0.78 \pm 0.07$, meaning that brighter, more massive stars in the satellite are more centrally concentrated than fainter, less massive ones.[20] Since mass segregation is an expected feature in star clusters, they conclude that Dra II is a star cluster.

Our analysis reveals that Dra II is unambiguously a dwarf galaxy, based on its large metallicity spread. Specifically, we find $\langle[\mathrm{Fe/H}]\rangle = -2.72^{+0.10}_{-0.11}$ and $\sigma_{[\mathrm{Fe/H}]} = 0.40^{+0.12}_{-0.12}$. Compared to Longeard et al. (2018), our $\langle[\mathrm{Fe/H}]\rangle$ is similar, but our dispersion is much larger. We discuss the reasons for this difference in Section 4.3. There are no clear systematics in our data that would lead to such a large spread. Conversely, we do not see any plausible way to reconcile the observed CaHK data with the lack of $\sigma_{[\mathrm{Fe/H}]}$ that is expected for GCs. The remaining mystery is how to interpret the inferred mass segregation with a large metallicity spread. Additional kinematic information to (1) refine velocity dispersion measurements and (2) confirm dwarf membership of the metal-poor and metal-rich stars in Dra II, may help to bring clarity to the true nature of this enigmatic object.

## 5.2. Broad Characterization of MDFs across the Sample

Figure 8 shows broadband CMDs of our entire sample, with the stars that we use for our MDF determinations color coded by their F395N S/N. The sample displayed in this figure has been cleaned for contamination using available kinematic data, which we detail in Section 5. We verify that all stars that we select in F606W–F814W also fall along the stellar population sequence in F475W–F814W. For stars that fall on the edge of the selection box, we find that their metallicities are not outliers compared to the rest of the distribution, and therefore retain them in the sample. The galaxies with the largest sample sizes are Ret II and Wil 1, which respectively have 76 and 68 stars from the lower RGB and MSTO. The galaxy with the smallest sample sizes are Phe II and Tuc V at 10 and six stars, respectively.

Figure 9 shows the stars for each galaxy in CaHK color space. For each galaxy, we overplot $[\alpha/\mathrm{Fe}] = +0.4$ MIST monometallic isochrones for either RGB or MS, appropriately selected for each galaxy. For most of our galaxies, the sample of members fall along the RGB and lower RGB, so those corresponding tracks are presented in their respective panels. The exceptions to this are Ret II, Wil 1, Seg 1, and Dra II, where the sample is dominated by MSTO and lower MS stars. For bluer F475W–F814W colors ($\leqslant$1.4), the CaHK tracks for RGB and MS evolutionary phases look similar and using either set of tracks would produce comparable MDF measurements. On the redder end, the CaHK tracks for MS stars are more closely spaced, but few of our stars fall in that color regime. Overall, visual inspection alone shows that each galaxy hosts stars of multiple, distinct metallicities.

[20] $R_{\mathrm{bright}}$ and $R_{\mathrm{faint}}$ respectively refer to the radii of bright and faint stars in the satellite. A system with no mass segregation, e.g., dwarf galaxies, would have $R_{\mathrm{bright}}/R_{\mathrm{faint}}$ consistent with 1.

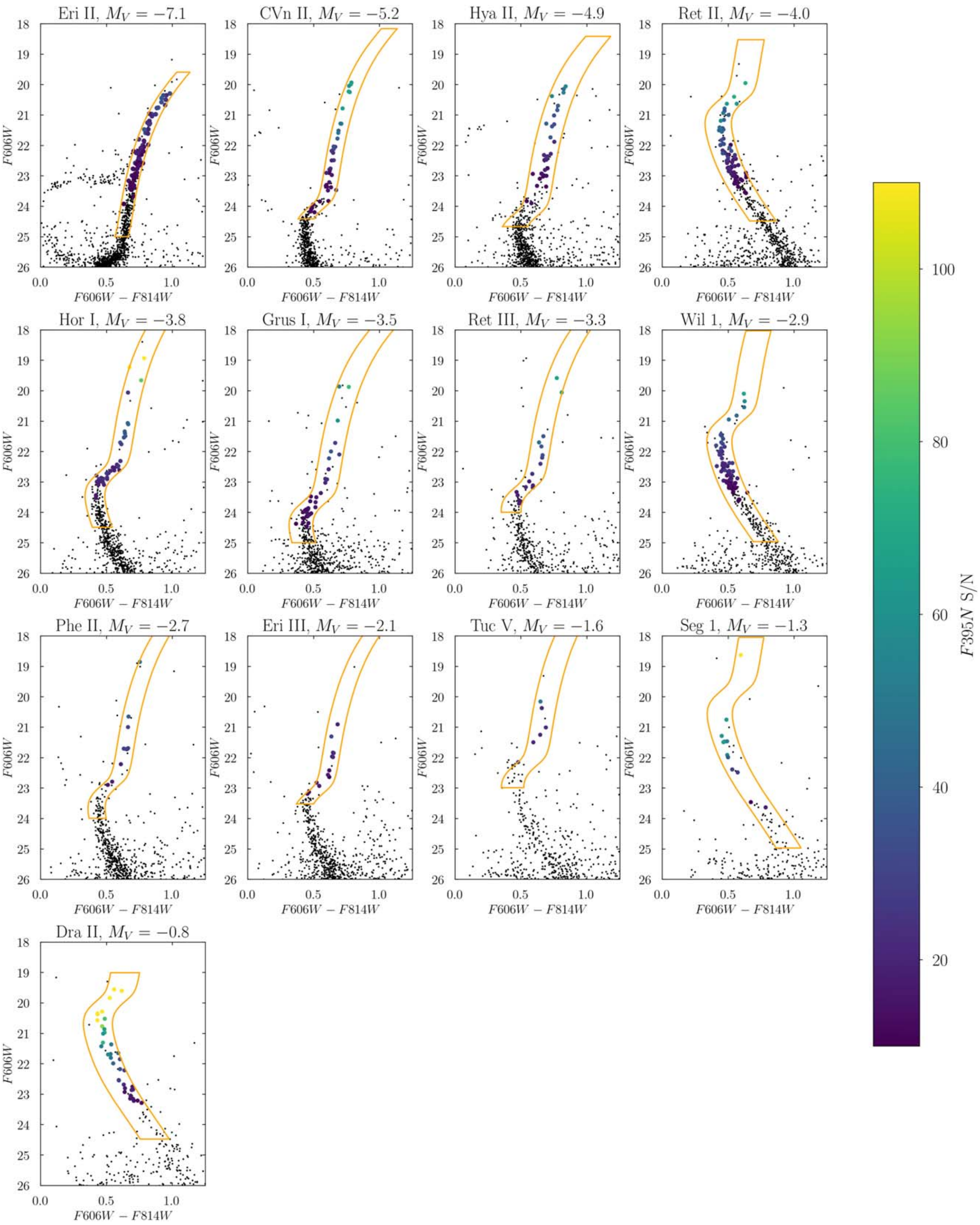

**Figure 8.** A gallery of F606W–F814W CMDs with membership selection indicated by the orange boxes drawn around the RGB. After making this initial selection, we crossmatch against literature radial velocity data where available to remove foreground interlopers. Stars used to infer MDFs are color coded with F395N S/N >10.

Figure 10 shows MDFs for each galaxy resulting from applying the metallicity and MDF fitting methodologies described in Section 3.1. In alignment with the intuition guided by the stars' positions on the CaHK color plots in Figure 9, all UFDs have a wide range of metallicities, with stars for which we indicate upper limits of $\lesssim -3.0$ up to stars as metal-rich as $-1$. Some MDFs are broad and flat (Hya II, Ret II, Grus I, Seg 1), while others are clearly peaked (Eri II,

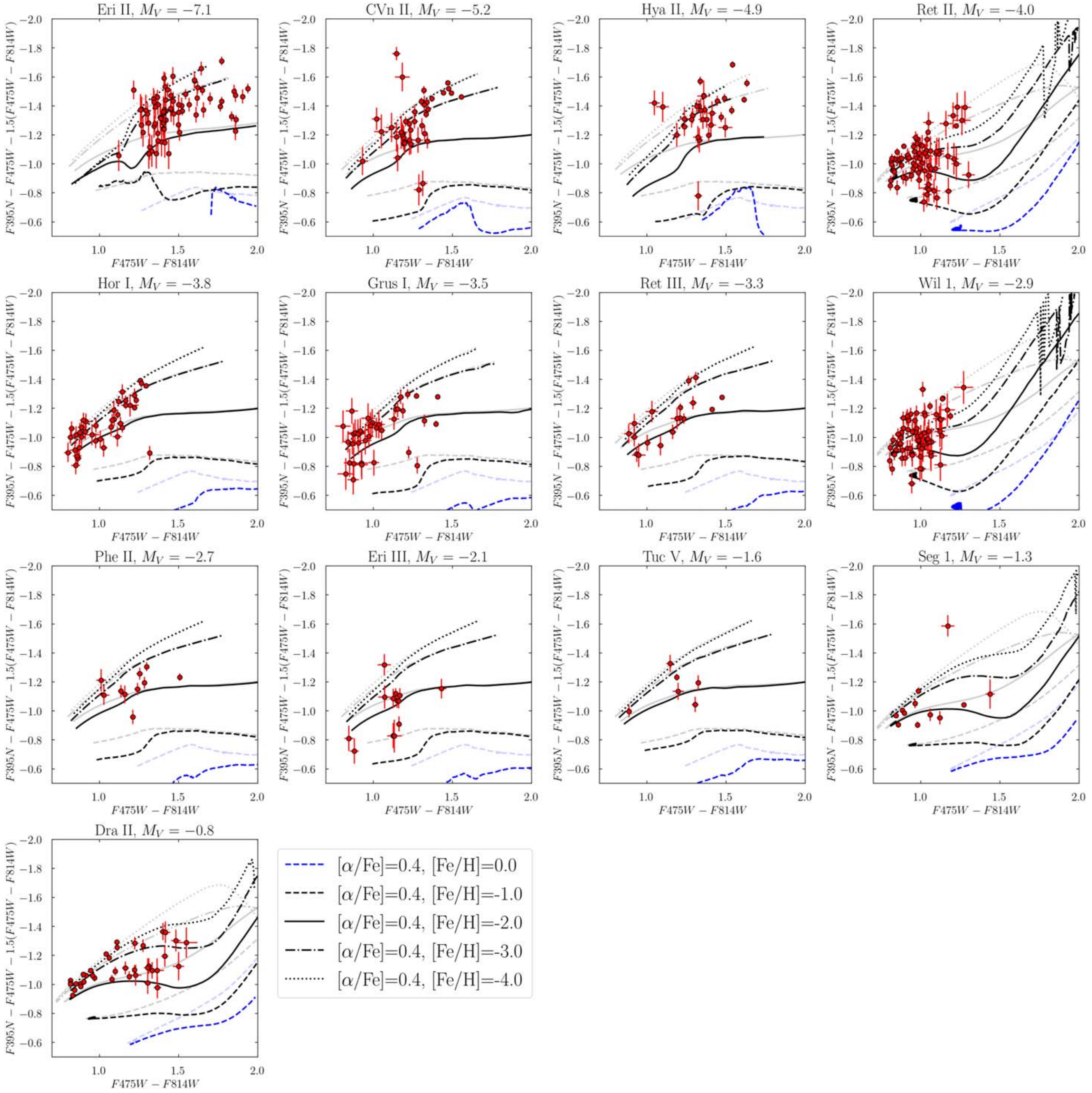

**Figure 9.** A gallery of CaHK diagrams for each UFD, ordered by decreasing luminosity. We have overplotted the MIST α-enhanced ($[\alpha/\mathrm{Fe}] = +0.40$) CaHK tracks for RGB stars in most cases and MS tracks for Ret II, Wil 1, Seg 1, and Dra II, which are dominated by MS and MSTO stars. The tracks have been convolved with ASTs run for each galaxy. These tracks are solely for illustrative purposes to demonstrate the impact of ASTs, and actual tracks used for fitting individual stars may look different due to the process described in Section 3.1. It is clear from the distribution of stars that each galaxy hosts stars over a wide range of metallicities.

Hor I, Phe II, Dra II). The majority of the MDFs display evidence of metal-poor tails, although the characterization of their full extent is limited by the edge of our metallicity grid. We also compute additional summary statistics to quantify deviations from Gaussianity, and are unable to resolve significant departures from Gaussianity for the majority of our MDFs. We present these results in Appendix E.

The left panel of Figure 11 shows the composite MDF from our study, derived by stacking the MDFs of all the individual UFDs. We measure metallicities for 463 stars across 13 UFDs, and disaggregate the composite MDF by contributions from each galaxy. For all of these measurements, we measure a mean of $-2.66 \pm 0.04$ and a sigma of $0.56 \pm 0.03$. We discuss comparisons with composite measurements from the literature in Section 6.1, and summarize our MDF measurements for individual UFDs in Table 2.

As part of validating our measurements, we compare our measurements with those in the literature where available. We discuss comparisons within individual UFDs in Section 5. In Appendix C, we provide a summary figure of direct literature comparisons as Figure 19, and discuss comparisons between specific methods in greater detail. In summary, our measurements are in agreement with the literature measurements to within $\sim 1.5\sigma$. Given the heterogeneity of literature measurements, we consider this an affirming result of the fidelity of CaHK metallicities.

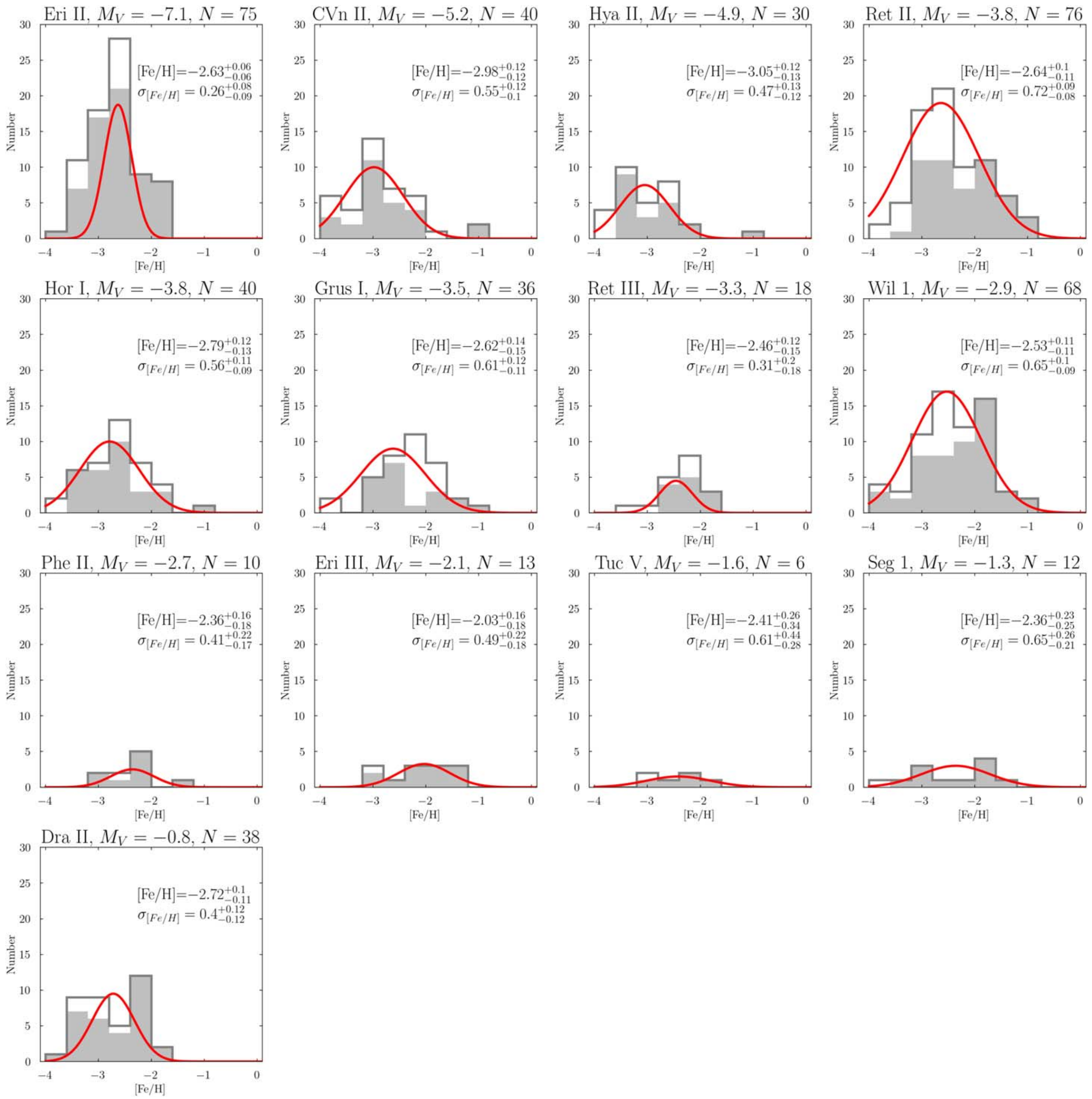

**Figure 10.** A gallery of MDFs for each galaxy based on our CaHK fitting. Metallicity bin sizes are 0.4 dex wide, which is comparable to typical stellar metallicity uncertainties. Well-constrained fits are indicated by shaded gray regions, whereas poorly constrained fits—largely upper limits—are shown as open histograms. We overplot the MDFs with the best-fit Gaussian (red). With the exception of Ret III, we resolve $\sigma_{\rm [Fe/H]}$ above at least the $2\sigma$ level for our UFD sample.

In Appendix F, we present the table of measurements in Table 4, reporting both the random uncertainties from photometry, and from systematic uncertainties that we determined following our procedure in Appendix B.5. We also identify 112 extremely metal-poor ([Fe/H]$< -3.0$) star candidates, with five of them being stars that are low S/N, but whose photometry place them blueward of the CaHK grid. On the other end of the MDF, we identify 86 stars that are more metal-rich than [Fe/H]$= -2.0$. We provide these stars in Tables 5 and 6, respectively, and also in Appendix F.

## 6. Discussion

### 6.1. Comparison to the Literature

The right panel of Figure 11, shows a composite MDF (red) from the 463 stars collected from all UFDs in our sample. For

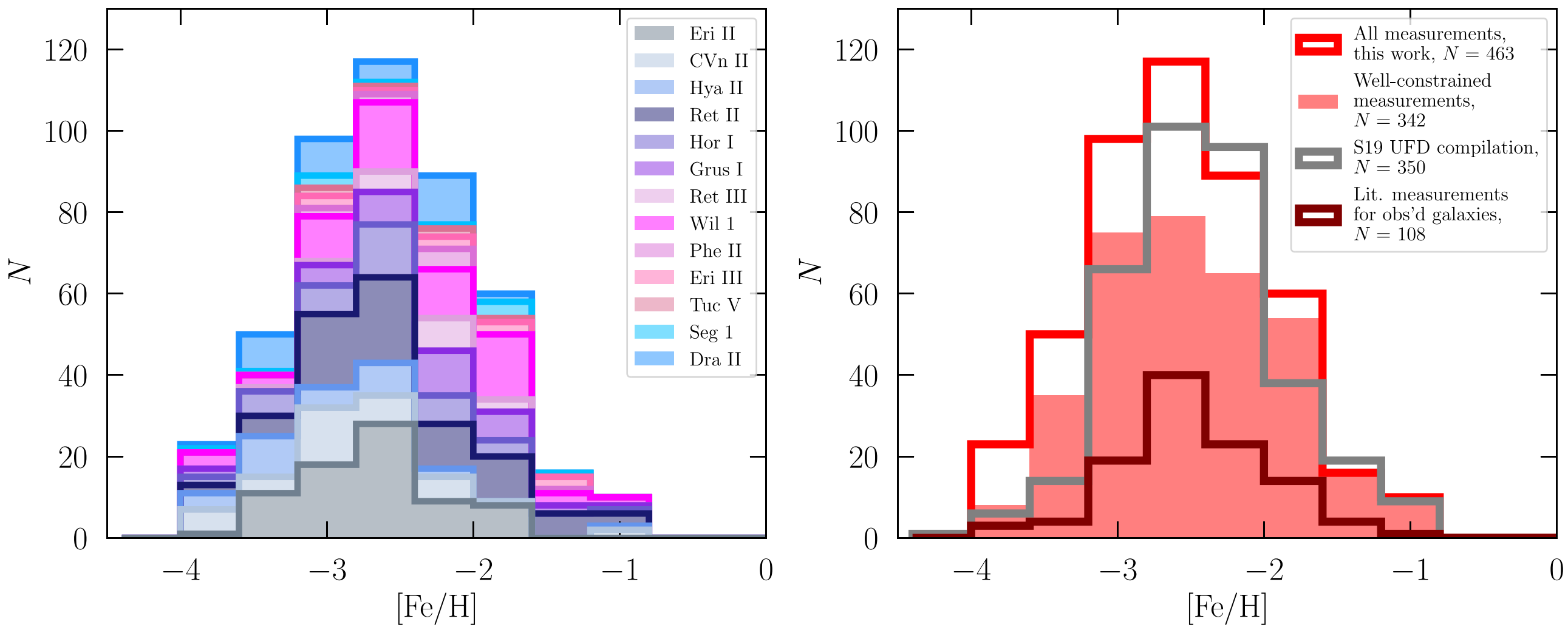

**Figure 11.** The composite MDF for all UFDs in our sample, shown against select UDF MDFs from the literature. (Left) A breakdown of the composite MDF by contribution from each UFD in our sample. (Right) Our MDF (red) compared to those in the literature. The maroon line represents the composite MDF made from all literature measurements available for the same galaxies as those in our sample, including stars not observed by our program. Our program increases the number of stellar metallicities in these galaxies by nearly a factor of 5. The gray line represents the composite UFD MDF made from all available UFD stellar metallicity measurements compiled by Simon (2019). Our work more than doubles the number of metallicities in all UFDs. There is a significant increase in the number of extremely metal-poor star candidates in these systems compared with previous studies. In all cases, our work demonstrates the excellent ability for space-based CaHK narrowband imaging to significantly expand the number of UFD stars with metallicity measurements.

**Table 2**
Summary of MDF Measurements

| Galaxy | $M_v$ (mag) | ⟨[Fe/H]⟩ (dex) | $\sigma_{\rm [Fe/H]}$ (dex) | $N_{\rm stars}$ | $N_{\rm [Fe/H]< -3.0}$ | $N_{\rm [Fe/H]> -2.0}$ |
|---|---|---|---|---|---|---|
| Eridanus II | −7.1 | $-2.63^{+0.06}_{-0.06}$ | $0.26^{+0.08}_{-0.09}$ | 75 | 17 | 8 |
| Canes Venatici II | −5.2 | $-2.98^{+0.12}_{-0.12}$ | $0.55^{+0.12}_{-0.10}$ | 40 | 15 | 3 |
| Hydra II | −4.9 | $-3.05^{+0.12}_{-0.13}$ | $0.47^{+0.13}_{-0.12}$ | 30 | 15 | 1 |
| Reticulum II | −4.0 | $-2.64^{+0.1}_{-0.11}$ | $0.72^{+0.09}_{-0.08}$ | 76 | 17 | 20 |
| Horologium I | −3.8 | $-2.79^{+0.12}_{-0.13}$ | $0.56^{+0.11}_{-0.09}$ | 40 | 10 | 5 |
| Grus I | −3.5 | $-2.62^{+0.14}_{-0.15}$ | $0.61^{+0.12}_{-0.11}$ | 36 | 5 | 10 |
| Reticulum III | −3.3 | $-2.46^{+0.12}_{-0.15}$ | $0.31^{+0.2}_{-0.18}$ | 18 | 2 | 3 |
| Willman 1 | −2.9 | $-2.53^{+0.11}_{-0.11}$ | $0.65^{+0.10}_{-0.09}$ | 68 | 13 | 21 |
| Phoenix II | −2.7 | $-2.36^{+0.18}_{-0.16}$ | $0.41^{+0.22}_{-0.17}$ | 10 | 1 | 1 |
| Eridanus III | −2.1 | $-2.03^{+0.16}_{-0.18}$ | $0.49^{+0.22}_{-0.18}$ | 13 | 0 | 6 |
| Tucana V | −1.6 | $-2.41^{+0.26}_{-0.34}$ | $0.61^{+0.44}_{-0.28}$ | 6 | 1 | 1 |
| Segue 1 | −1.3 | $-2.36^{+0.23}_{-0.25}$ | $0.65^{+0.26}_{-0.21}$ | 12 | 2 | 5 |
| Draco II | −0.8 | $-2.72^{+0.10}_{-0.11}$ | $0.40^{+0.12}_{-0.12}$ | 38 | 14 | 2 |

**Note.** A summary of our MDF measurements. We list ⟨[Fe/H]⟩ and $\sigma_{\rm [Fe/H]}$ inferred from the UFDs in our sample, the number of stars used to make the measurement, and the number of stars of interest at the extreme ends of the UFD's MDF. We refer the reader to Section 5 for detailed discussions on the determination of ⟨[Fe/H]⟩ and $\sigma_{\rm [Fe/H]}$, and comparisons where available to previous studies.

comparison, we overplot with the maroon line the composite MDFs for *all* existing literature stellar metallicity measurements in the same UFDs. The literature sample consists of ∼110 stars and includes stars not in our sample (e.g., at larger radii than the HST footprint). In the same panel, we also show with a gray line literature metallicity measurements for *all* UFDs around the MW, including those for galaxies not in our sample.

There are several key takeaways from Figure 11. The first is the drastic increase in sample size. Compared to the literature, in our sample of 13 UFDs, we increase the number of stellar metallicity measurements by nearly a factor of ∼5. In almost all of our galaxies, we at least double the number of stars with metallicity measurements, with the exception of Seg 1. For Ret III, Wil 1, Eri III, and Tuc V in which previous efforts yielded less than three stars with metallicity measurements per galaxy, our work provides large enough samples to measure robust metallicity spreads. The largest gain is for Wil 1: the 68 stars in our study represent a dramatic improvement over the Willman et al. (2011) study, in which only two of its RGB stars had metallicity measurements.

Second, we significantly increase the total number of stars in all UFDs with reliable metallicity determinations. Compared to the literature MDF, drawn from 26 galaxies (Simon 2019), we double the total number of stars with metallicities; only a small fraction of our data overlaps with existing measurements, as discussed in previous sections.

This larger sample has some implications for MDF interpretation. For example, there are some modest differences in the mean and scatter in our MDF versus the previous composite literature MDF. For all of our measurements, we find a mean of $-2.66 \pm 0.04$ and a sigma of $0.56 \pm 0.03$. If we only include well-constrained fits, which disproportionately exclude our extremely metal-poor star candidates, we find a higher mean of $-2.38 \pm 0.03$ and a smaller scatter of $0.40 \pm 0.03$. In comparison, the literature values are a mean of $-2.41 \pm 0.03$ and a sigma of $0.49 \pm 0.02$.

These differences are driven by pronounced differences in the tails of the composite MDFs. On the metal-poor end, we identify 112 stars as extremely metal-poor ([Fe/H]$< -3.0$) candidates. These candidates make up 24% of all stars in our sample. This is a much larger fraction than the literature values both in the same galaxies (12%) and across all UFDs (14%, Simon 2019). Due to limitations of the current CaHK methodology (e.g., our grid sharply ends at [Fe/H] $= -4.0$), it is plausible that not all of these candidates are bona fide extremely metal-poor stars. Removing the five low S/N stars that fall beyond our metallicity grid decreases our extremely metal-poor star fraction to 23%. Within the Galaxy, spectroscopic follow-up of the Pristine-identified EMPs had only a 20% success rate (i.e., 80% were not EMPs; Youakim et al. 2017; Aguado et al. 2019). Our observational situation is markedly different than a broad survey of the MW: we have targeted the central regions of UFDs, coherent objects with known distances, and use CMD selection to limit interlopers. While there may be possible foreground contaminants in excess of our estimates (see Appendix A), the contamination rate is likely to be small. Thus, the main source of uncertainty is in the coarse performance of CaHK techniques for EMPs. Spectroscopic follow-up will be important, but may not be possible until ELTs are online owing to the faintness of most stars in our sample.

At the other extreme, we find that 19% of our sample (86 stars) are more metal-rich than [Fe/H]$= -2.0$. The fraction of similarly metal-rich stars in the literature MDFs is 18% for the same UFDs as in our sample and 19% for all UFDs (Simon 2019). Thus, our metal-rich star fraction is in agreement with the literature. We have confidence in our metal-rich star measurements because the performance of CaHK is more precise for more enriched stars, and we expect a small number of contaminants given our observational strategy. There is no broad consensus on whether UFDs should host such metal-rich stars, with some suggesting that they are unlikely to be actual member stars (e.g., Fritz et al. 2019). However, there are enough clear examples of spectroscopically confirmed metal-rich stars in UFDs (e.g., CVn II, Wil I, Seg 1) to reinforce the reasonability of our findings.

A final, and particularly important point, is that of homogeneity. The literature composite MDFs shown in Figure 11 are drawn from over a dozen different studies of which rely on many different spectroscopic observations (e.g., wavelength range, S/N, resolution), metallicity inference techniques (e.g., full spectral synthesis, CaT), and underlying assumptions (e.g., line lists). As highlighted in Sandford et al. (2023), these types of differences lead to $\sim$0.3 dex variations in the [Fe/H] values for RGB stars in the MW GC M15, underscoring the importance of homogeneous measurements. Thus, a key product of this work is that all of the metallicity determinations are self-consistent, on the same scale, and have uniformly determined uncertainties. The net result is that not only have we greatly increased the sample of stellar metallicities in UFDs, we have also provided clear evidence for a metallicity floor, large internal dispersions, etc. that are free of study-to-study systematics.

### *6.2. Faint End of the Dwarf Galaxy Mass–Metallicity Relation*

Figure 12 shows our results in the context of the known mass–metallicity relation for dwarf galaxies. We find that across three orders of magnitude in luminosity from $10^2$–$10^5\,L_\odot$, there is no mass–metallicity relation for UFDs. Rather, their mean metallicities are scattered about [Fe/H] $= -2.6$. Early spectroscopic studies had already hinted at such an empirical result, but confidence in this feature was uncertain due to the small sample sizes used to make measurements in individual UFDs, as well as hard-to-quantify systematic differences between metallicity measurements made using various techniques (Simon 2019).

Our large sample of homogeneous metallicities over a range of UFDs definitively shows a break in the relation compared to that for larger galaxies. We quantify the nature of this feature by conducting a linear fit to the mean metallicities of our measurements, following the procedure laid out in Hogg et al. (2010). The resulting slope of the line in units of $[\mathrm{Fe/H}]/\mathrm{Log}_{10}L_\odot$ is $-0.17 \pm 0.11$, with an intercept of $-2.06 \pm 0.38$ dex. The value of this slope is consistent with zero, and in any case, is discrepant from the positive slope of the relation for more luminous dwarfs.

While this feature was previously thought of as a *floor* in the relation in early theoretical works attempting to reproduce it (e.g., Wheeler et al. 2019), we instead conceptualize the $\langle$[Fe/H]$\rangle$ of the UFD population as a normal distribution centered around [Fe/H]$= -2.61 \pm 0.08$ with $\sigma_{[\mathrm{Fe/H}]} = 0.24^{+0.09}_{-0.06}$ dex (Figure 12). We also make the same Gaussian fit to the Simon (2019) compilation of mean metallicities for systems with less than $10^5\,L_\odot$, finding a mean of $-2.40 \pm 0.06$ and a dispersion of $0.21 \pm 0.06$. The mean metallicity is in agreement to within $2\sigma$, and the dispersion is in good agreement. Under this framework, galaxies such as Eri III and Hya II are $2\sigma$ deviations from the distribution. The scatter that we observe about this distribution also traces UFDs' known sensitivity to the stochasticity of baryonic processes.

The galaxy mass–metallicity relation is the observational synthesis of the baryonic physics driving galaxy formation across all scales. Previously, the dwarf galaxy mass–metallicity relation established by Kirby et al. (2013) was shown to be universal across galaxies of different morphological types and environments in the LG, and could be somewhat matched with mass–metallicity relations from other techniques that extend to galaxies as massive as $10^{12}\,M_\odot$. The canonical interpretation of this phenomenon is that more massive galaxies are better able to retain SNe enrichment products to form successive generations of stars compared to their fainter counterparts. That UFDs no longer display the same trends suggests that this picture of galaxy evolution requires additional refinement for the faintest end of the luminosity function.

In the right panel of Figure 12, we compare the dwarf galaxy data against simulated UFDs from the independent simulations of Jeon et al. (2017), Wheeler et al. (2019), Agertz et al. (2020), and Sanati et al. (2023). While simulations have begun to reproduce the mean metallicities of UFDs more luminous than $10^4\,L_\odot$, fainter UFDs still remain a theoretical challenge to simulate.

The mean metallicities of UFDs are shown to be particularly sensitive to the details of feedback implementation, where increasingly strong feedback mechanisms suppress subsequent enrichment and star formation and produce low mean metallicities (Agertz et al. 2020). In the case of FIRE-2 simulations, which implement strong feedback, many of the lowest-luminosity UFDs never enrich beyond the initialized metallicity floor (Wheeler et al. 2019), as shown by the lower-luminosity FIRE UFDs in Figure 12 at $[Fe/H] = -4.0$. So far, there have been numerical experiments varying parameters such as star formation, radiative transfer, SNe energies (Agertz et al. 2020; Sanati et al. 2023), metallicity-dependence of Type Ia SNe DTDs (Gandhi et al. 2022), and top heaviness and stochastic sampling of the IMF (Applebaum et al. 2020; Prgomet et al. 2022 respectively), in order to explore the parameter space of physics that elevate enrichment levels in UFDs. Overall, this area of study is still ongoing.

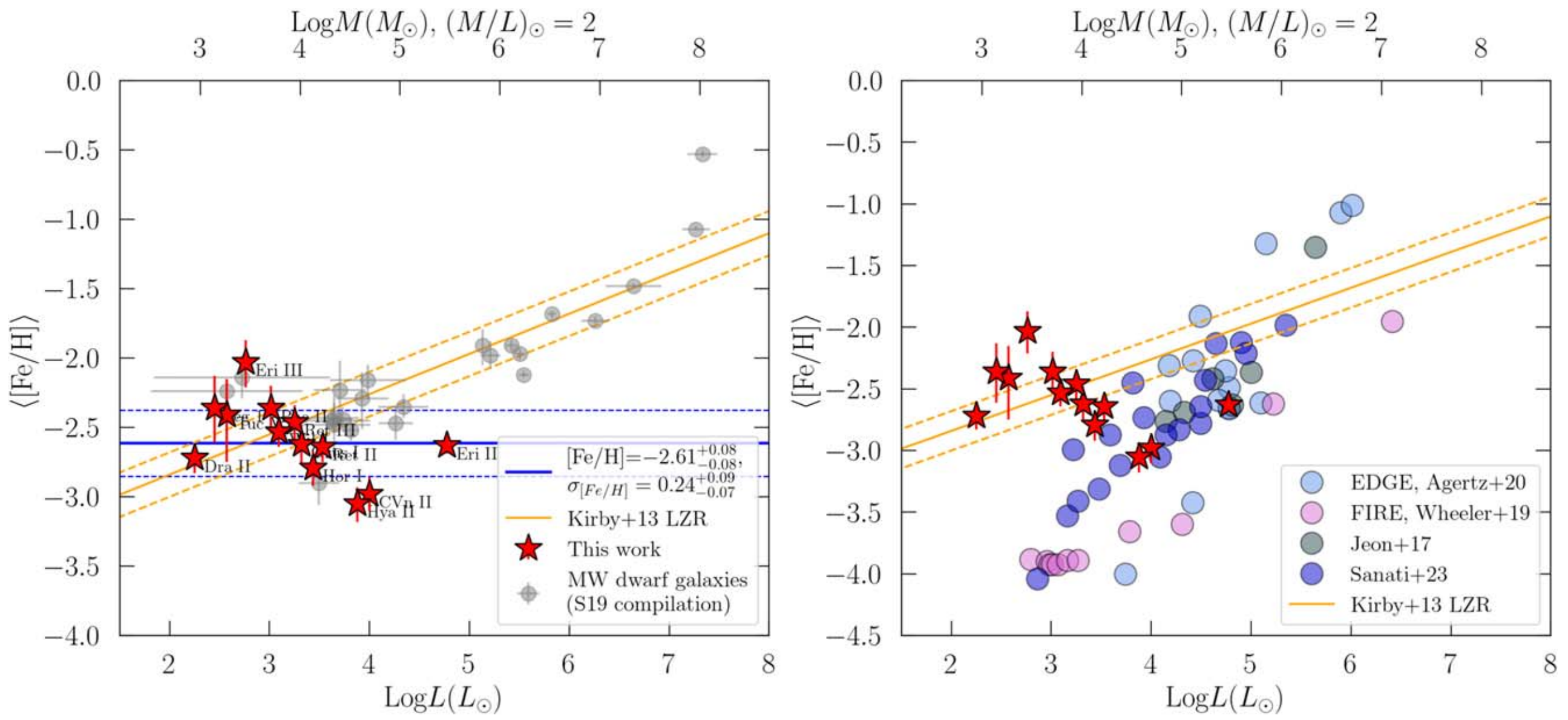

**Figure 12.** The mass–metallicity relation for dwarf galaxies, updated with results from this work. (Left) We add our ⟨[Fe/H]⟩ measurements to the ⟨[Fe/H]⟩ measurements of other MW dwarf galaxies, using the table compiled by Simon (2019). Our results show a clear floor in the mass–metallicity relation in the UFD regime. We characterize this putative floor using only our measurements, assuming that the floor can be described by a Gaussian mean and scatter, placing the mean of the floor at $[Fe/H] = -2.61 \pm 0.08$ dex. (Right) Our ⟨[Fe/H]⟩ measurements compared to those from select cosmological simulations. While simulations can broadly reproduce the mass–metallicity relation for the more luminous dwarf galaxies, they currently struggle to enrich less-luminous dwarf galaxies to the same level as we observe.

One notable discrepancy between current UFD simulations and the UFDs that we have observed is the impact of the environment. While the simulations quoted thus far evolve UFD halos in isolation, all of the UFDs in our sample are present-day satellites either of the MW or the Magellanic clouds. Environment can matter because the presence of a nearby, more massive host could provide pre-enriched gas and introduce additional metals into the UFD's internal ecosystem (e.g., Jeon et al. 2017). Additionally, reionization is thought to play a central role in truncating the star formation of UFDs (e.g., Brown et al. 2014), and a galaxy's environment becomes a proxy for uneven distance and exposure from reionizing sources (e.g., Dawoodbhoy et al. 2018).

On the other hand, studies of UFD infall times based on Gaia proper motions and simulation analogs suggest that the majority of them fell into the MW after forming all of their stars (Fillingham et al. 2019; Rodriguez Wimberly et al. 2019, Applebaum et al. 2021). In that case, perhaps the immediate MW environment would not be relevant for interpreting their present-day MDFs. Additionally, some galaxies within our sample are also satellites of the LMC prior to falling into the MW (e.g., Patel et al. 2020), and early studies already suggest that they may have experienced different chemical enrichment pathways and SFHs compared to MW satellites (Ji et al. 2020; Sacchi et al. 2021). Detailed orbital histories of LG satellites, and subsequently, a more complete account of LG assembly history at high redshifts, would also be constructive for disentangling competing physics that contribute to UFD formation.

In Figure 13, we present $\sigma_{[Fe/H]}$ of UFDs as a function of luminosity and in comparison with data from other LG dwarfs. Our data show that UFDs span a range of $\sigma_{[Fe/H]}$, from ∼0.3 to ∼0.7 dex. This empirical result was also hinted at in prior studies, and our data set enables its confirmation. The interpretation of UFD $\sigma_{[Fe/H]}$ is an active area of investigation, and potential physical mechanisms responsible include the stochasticity of chemical enrichment processes and metal mixing (e.g., Frebel et al. 2014; Emerick et al. 2020).

### 6.3. Prevalence of Metal-rich Stars

Our sample contains a higher fraction of stars that are more metal-rich than $[Fe/H] > -2.0$ than previously discovered in the literature. Here, we discuss some of the implications of and potential concerns with this discovery.

As discussed in Appendix A and Section 5, based on foreground modeling and the narrow FoV of our observations, we do not expect the foreground to be a major source of metal-rich stars in our sample. At the same time, we acknowledge the possibility that our sample may be affected by foreground in excess of our estimates, e.g., there is a nonzero probability that other yet-to-be-discovered MW substructures spatially overlap with our HST fields. To address this small, but nonzero, possible impact on our results, we take the very conservative approach of recomputing the MDFs with and without metal-rich stars for individual UFD cases where this may be a concern (e.g., they are outliers compared to the rest of the sample; see Section 5). However, because it seems unlikely that all metal-rich stars in all our UFDs are MW interlopers, we choose to be inclusive in our membership sample. Accordingly, we provide these metal-rich stars in Table 6 and welcome follow-up studies by the community to refine membership and undertake other studies of these stars (e.g., detailed abundance patterns).

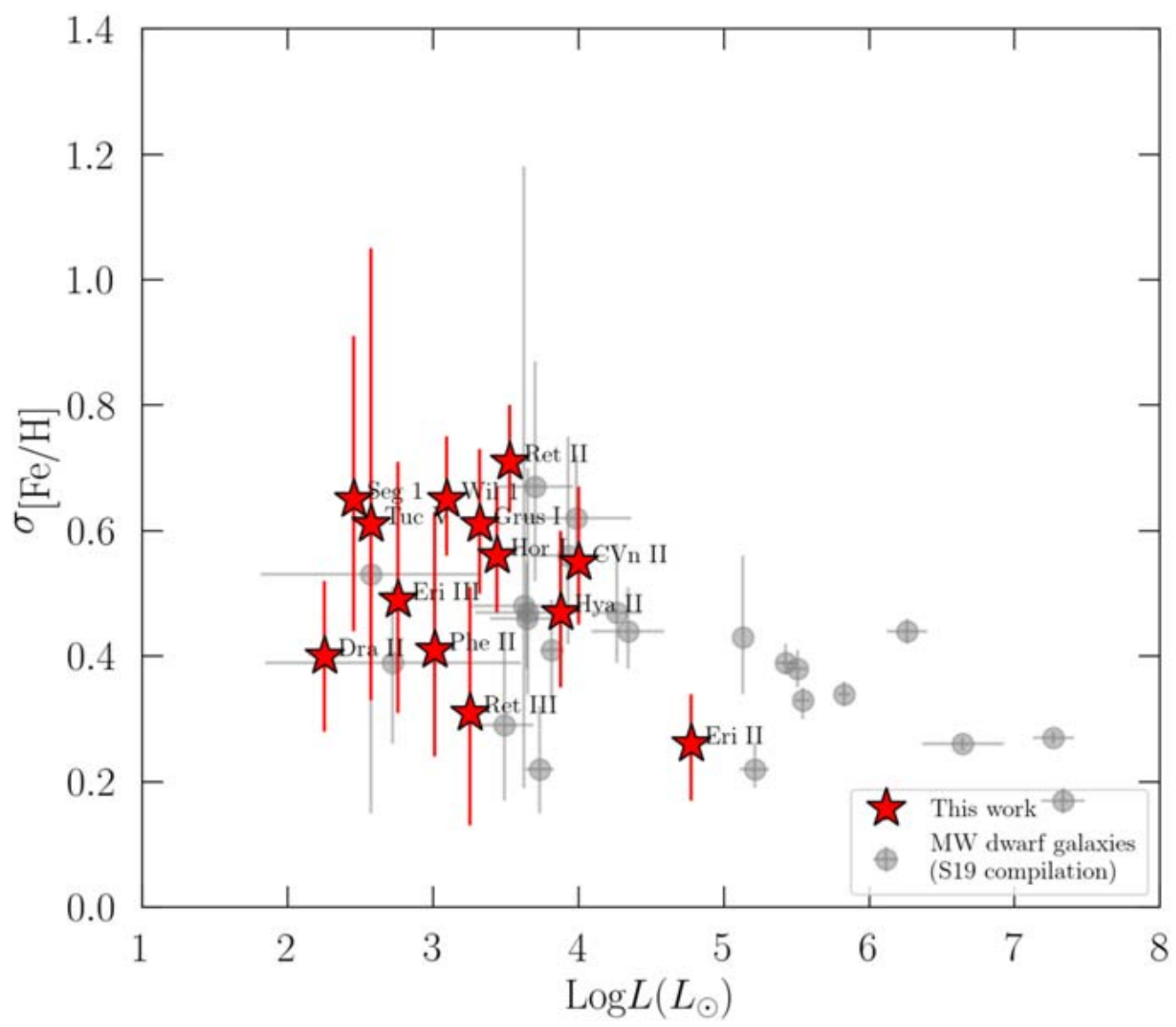

**Figure 13.** We compare our $\sigma_{[Fe/H]}$ measurements, made assuming a Gaussian characterization of the MDF, to the $\sigma_{[Fe/H]}$ measurements of other MW dwarf galaxies (compiled by Simon 2019). Our results confirm the large internal metallicity variations in these systems.

To date, the broad consensus for the abundance of metal-rich ($[Fe/H] > -2.0$) stars in UFDs is not well established. However, there also are known cases of spectroscopically confirmed metal-rich stars in UFDs (e.g., CVn II, Wil 1, and Seg 1). As a result, we also remark on the astrophysical implications of our findings. metal-rich stars are quite important for detailed interpretations of chemical evolution in UFDs: the shape of the metal-rich end of a galaxy's MDF is driven by the equilibrium between gas enrichment processes and accretion of pristine gas, as well as the rapidity of star formation truncation (e.g., Jenkins et al. 2021; Sandford et al. 2022). The detailed abundance patterns of metal-rich stars may also provide insight into enrichment mechanisms and/or the number of enrichment events that preceded their formation (e.g., Frebel & Norris 2015).

If the dwarf galaxy mass–metallicity relation could be extrapolated down to the UFD regime (Kirby et al. 2013), then the expectation prior to this work is that UFDs should indeed be composed primarily of metal-poor stars. However, as discussed in Section 6.2, our results have shown that the faintest UFDs deviate from this expectation. These observations suggest that different galaxy formation physics and enrichment processes dominate in the UFD regime, and additional theoretical investigations are needed to ascertain the extent to which UFDs can enrich stars beyond $[Fe/H] = -2.0$.

### *6.4. Metallicities for a Large Sample of Ancient Stars*

A main challenge in extremely metal-poor studies in the MW is a lack of precise ages. Extremely metal-poor stars are typically found in the field and age dating techniques rely on fitting single stars to isochrones, which can produce very precise, but inaccurate ages, or, in some cases, nuclear cosmochronology (i.e., isotopic age dating), which has large uncertainties (see, e.g., Boylan-Kolchin & Weisz 2021 for a detailed discussion).

While it would seem that lower metallicity stars in the MW formed at older ages, theory paints a more complicated picture. Cosmological simulations of the MW have already suggested that stellar age and metallicity are not monotonic relations (e.g., Starkenburg et al. 2017b; El-Badry et al. 2018). For example, these studies show that stars as metal-rich as $[Fe/H] \sim -1.0$ could have formed as early as within the first 1 Gyr of the Galaxy's lifetime alongside stars as metal-poor as $[Fe/H] \sim -4.0$. A star with $[Fe/H] \sim -2.5$ could have formed as long ago as 13 Gyr or as recently as 7 Gyr ago. As a result, due to the complexity of the MW's stellar populations (see, e.g., Grieco et al. 2012; Matteucci et al. 2019; Kerber et al. 2019; Savino et al. 2020 for corroborating observations), ages are challenging to infer based on metallicity alone.

In contrast, UFDs have well-constrained ages. By fitting the MSTO of deep optical CMDs, numerous studies have shown that UFDs that orbit the MW formed the majority of their stars during or prior to the epoch of reionization (e.g., Brown et al. 2014; Weisz et al. 2014; Gallart et al. 2021; Simon et al. 2021, 2023; Sacchi et al. 2021). Consequently, the metallicities and MDFs we measure all arise from stars that formed within the first billion years of the Universe. Chemical evolution modeling can provide insights into the physical mechanisms by which such small galaxies can create such large metallicity spreads so rapidly in the early Universe (e.g., Sandford et al. 2022; Alexander et al. 2023). We will pursue a similar investigation with our current sample in the next paper in this series.

### *6.5. Hierarchical Structure Formation*

The formation of larger galaxies from the hierarchical mergers of their smaller counterparts has been a long-standing tenet of Lambda cold dark matter cosmology (e.g., Searle & Zinn 1978; White & Rees 1978). An increasingly common scenario in the literature is that the observed signatures of accretion are dominated by larger galaxies (e.g., LMC-sized) over small UFD-like systems (e.g., Côté et al. 2000; Deason et al. 2015; Helmi et al. 2018; Belokurov et al. 2018; Naidu et al. 2020), though the discovery of numerous fainter streams and substructures (e.g., Li et al. 2018; Shipp et al. 2018; Li et al. 2022; Ibata et al. 2021) may provide a more granular perspective and reveal the contribution of fainter galaxies.

Figure 14, shows the distribution of mean metallicities of our satellites compared to the stellar streams recently characterized by S5 (Li et al. 2022) and Pristine (Martin et al. 2022). The mean metallicities of our galaxies are centered at $\langle[Fe/H]\rangle \sim -2.6$ and range from $\langle[Fe/H]\rangle = -3.0$ to $\langle[Fe/H]\rangle = -2.0$. In contrast, the streams observed by S5 and Pristine tend to be more enriched, and span a larger range of metallicities. For both comparison studies, the teams were unable to resolve $\sigma_{[Fe/H]}$ in their streams, suggesting that they were of GC origin. These preliminary comparisons suggest that the UFDs we observe are a distinct class of objects from the newer streams that are being discovered in the MW.

From a theoretical perspective, Brauer et al. (2022) explored the possibility of recovering completely tidally disrupted UFDs by searching for clustering in dynamical action space. Overall, they found that the prospects are slim because the signal of UFD remnants is weak compared to the background. These results affirm the emerging picture based on our metallicity measurements, that UFDs do not have similar properties as currently known MW streams and their progenitors. It is however possible that our current understanding of the UFD stream composition in the MW is limited by selection effects and the absence of available information in current stream-finding efforts. Brauer et al. (2022) note that additional chemical abundance information such as metallicity and/or *r*-process elements may improve the efficacy of such searches, in which case, the results of our study can assist in future efforts by establishing a baseline for the expected chemical profile of UFDs.

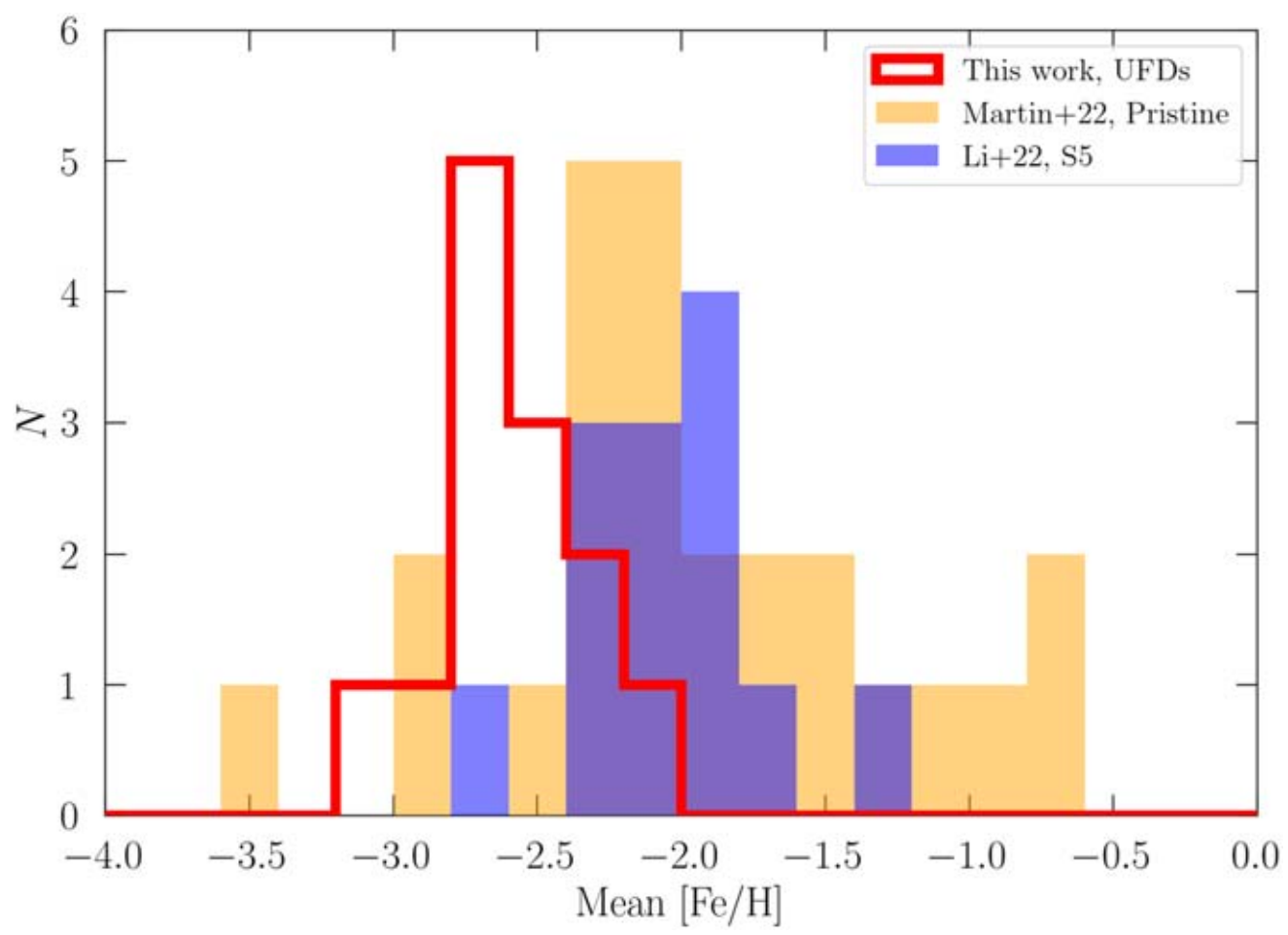

**Figure 14.** The mean metallicities of our UFDs compared to those of recent streams characterized by S5 (Li et al. 2022) and Pristine (Martin et al. 2022). The distributions are noticeably different, suggesting UFDs are not the main progenitors of the currently known MW stream population.

## 7. Conclusion

We present metallicity measurements of ∼500 stars across 13 UFDs, measured from HST narrowband imaging in the F395N filter. Our data set doubles the number of available metallicities by a factor of 5 among just the UFDs in our study, and doubles the number of available metallicities across all UFDs. We use these stellar metallicities to measure the MDFs of these 13 systems.

We summarize the key results found in our study:

1. HST F395N narrowband imaging can recover UFD MDFs to the same level of fidelity as numerous other metallicity measurement methods currently used by the community.
2. Our results are the largest homogeneous set of stellar metallicities measured in UFDs to date.
3. With this vastly expanded sample size, we are able to robustly resolve nonzero metallicity dispersions for all 13 of our targets. For Eri III, we confirm its status as a UFD (as opposed to a GC) for the first time.
4. The composite MDF of the UFDs has $\langle[\mathrm{Fe/H}]\rangle = -2.66 \pm 0.04$ dex and a dispersion of $0.56 \pm 0.03$. Individually, our UFDs span a range of $\langle[\mathrm{Fe/H}]\rangle$ from $\sim -3.0$ to $\sim -2.0$, and dispersions ranging from ∼0.3 to ∼0.7. With $\langle[\mathrm{Fe/H}]\rangle \sim -3.0$ as measured by our study, CVn II and Hya II are the most metal-poor UFDs known to date.
5. We identify stars on the extreme ends of the UFD MDFs ($[\mathrm{Fe/H}] < -3.0$ and $[\mathrm{Fe/H}] > -2.0$) that would be promising candidates for detailed spectroscopic follow-up studies to confirm their metallicities and origins. Respectively, these stars make up 24% and 19% of our sample. The extremely metal-poor star fraction we measure is larger than the literature value across all known UFDs of 14%, and the metal-rich star fraction is in agreement with the literature fraction of 19%.
6. We quantify the metallicity floor in the dwarf galaxy mass–metallicity relation as a distribution centered on $[\mathrm{Fe/H}] \sim -2.61 \pm 0.08$ with a dispersion of $0.24^{+0.09}_{-0.06}$ dex that ranges across three orders of luminosity in the UFD regime, from $10^2$– $10^5\, L_\odot$.
7. We provide the largest set of stellar metallicity measurements for a population of stars, which owing to HST SFH studies of UFDs, are known to have overwhelmingly formed within 1 Gyr of the Big Bang. This is in contrast to metal-poor stars in the MW, for which age uncertainties can be several gigayears owing to the complex formation history of the MW.
8. The mean metallicities of our UFD sample are different from the mean metallicities of many known streams in the MW, reinforcing the idea that surviving UFDs do not make up the vast majority of MW halo substructure that can be detected at present.
9. Our well-populated and homogeneous MDFs pave the way for detailed interpretation of the physics of UFD formation, with recent works like Sandford et al. (2022) providing an analytic framework for such studies moving forward.

Our study demonstrates the power of HST narrowband imaging to amass a large sample of standardized metallicity measurements and enable science cases in UFDs that have not been previously pursued due to sparse and heterogeneous data sets. This observational technique, as well as analogous techniques from the ground (e.g., Han et al. 2020; Chiti et al. 2020; Longeard et al. 2022) will continue to play a pivotal role in measuring metallicities of faint stars in distant galaxies that are anticipated to be uncovered by next-generation photometric surveys.

## Acknowledgments

We thank the anonymous referee whose comments enhanced the readability and clarity of this manuscript. We thank A. Dotter and C. Conroy for making $\alpha$-enhanced MIST models available for our analysis ahead of their release. We thank J. Simon for making the UFD compilation data used in Simon (2019) available for comparison to our results and for additional helpful comments on the manuscript. We thank C. Wheeler for making the simulations of UFDs from Wheeler et al. (2019) available for our discussion of the theoretical implications of our results. We thank Y. Revaz for providing the simulated data from Sanati et al. (2023) for comparison with our measurements. We thank A. Chiti, R. Naidu, and A. Frebel for helpful discussions that improved the interpretation of our results. We thank M. Jeon for providing the simulated UFDs from Jeon et al. (2017) for comparison with our work, for helpful conversations around simulation interpretation, and for hosting S.W.F. at Kyung Hee University, where a significant portion of this work was completed.

S.W.F. acknowledges support from a Paul & Daisy Soros Fellowship. S.W.F. and N.R.S. acknowledge support from the NSFGRFP under grants DGE 1752814 and DGE 2146752. S.W.F., D.R.W., M.B.K., A.S., and N.R.S. acknowledge support from HST-GO-15901, HST-GO-16226, and HST-GO-16159 from the Space Telescope Science Institute, which is operated by AURA, Inc., under NASA contract NAS5-26555. E.S. acknowledges funding through the VIDI grant "Pushing Galactic Archaeology to its limits" (project number VI.Vidi.193.093), which is funded by the Dutch Research Council (NWO).

This work made use of the Savio computational cluster provided by the Berkeley Research Computing Program at the University of California Berkeley. Some/all of the data presented in this paper were obtained from the Mikulski Archive for Space Telescopes (MAST) at the Space Telescope Science Institute. The specific observations analyzed can be accessed via doi:10.17909/cfn9-gt96. STScI is operated by the Association of Universities for Research in Astronomy, Inc., under NASA contract NAS5-26555. Support to MAST for these data is provided by the NASA Office of Space Science via grant NAG57584 and by other grants and contracts.

*Software:* DOLPHOT (Dolphin 2016, 2000), astropy (The Astropy Collaboration et al 2018), NumPy (Oliphant 2006), matplotlib (Hunter 2007), emcee (Foreman-Mackey et al. 2013), corner (Foreman-Mackey 2016), SciPy (Virtanen et al. 2020)

## Appendix A
## Foreground Contaminants

Our MDFs may suffer from some degree of MW foreground contamination. Unlike spectroscopy, we are unable to readily identify most contaminants (i.e., via kinematic selection). However, in contrast to spectroscopic studies, we intentionally target the inner regions of UFDs (i.e., within 1 $r_h$), for which the contamination fraction is expected to be lower than many ground-based surveys that extend to much larger areas.

To estimate the effect of foreground interlopers, we adopt a statistical approach. For each UFD, we use the TRILEGAL MW model (Vanhollebeke et al. 2009) to query all simulated stars within 0.5 deg$^2$ of the center of each galaxy. We query such a large on-sky area to adequately sample the distribution generated from the model. TRILEGAL outputs simulated stars in the filters and with known physical parameters (e.g., metallicity), which we use to evaluate the expected contamination fraction.

We estimate the expected number of foreground contaminants in our data by (i) using the ratio of model area to the WFC3 field of view and (ii) requiring the possible contaminants to fall into our member selection region on the CMD of each galaxy. The result for most galaxies is that we expect at most one foreground contaminant. The combination of the small WFC3 FoV, focus on the central regions of the galaxies, and CMD-based selection criteria ensure that we are predominantly including UFD member stars in our MDFs.

Figure 15 illustrates this process using the foreground model in the direction of Ret II. The left panel shows the TRILEGAL CMD of the queried foreground model within 0.5 deg$^2$, with the stars color coded by metallicity. The center panel shows the stars that remain from the same isochrone cut that we use to select Ret II members. The right panel illustrates the expected impact of foreground contamination when scaling the number of expected stars that pass the isochrone cut by the WFC3 FoV.

This is only an estimate, as in some cases, UFDs may have contamination in excess of the smooth TRILEGAL model (e.g., due to MW substructure in the case of Seg 1). In such cases, we are able to use some stellar kinematics to further clean our sample. We discuss these cases individually in Section 5. In general, the odds that our targeted HST observations fall onto another substructure with stars that satisfy all our selection criteria are generally quite small. Accordingly, we find that interlopers are unlikely to significantly alter our MDFs.

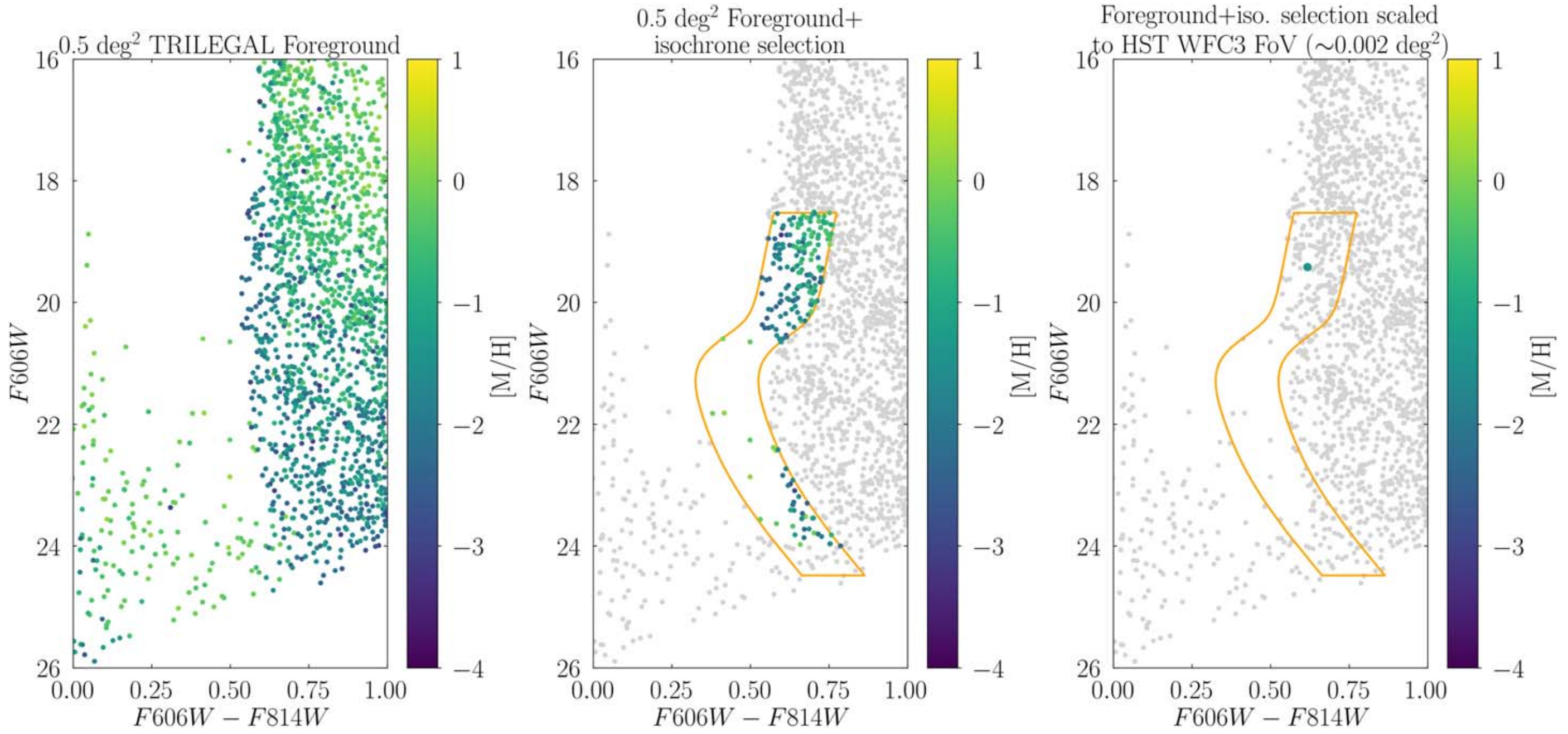

**Figure 15.** The potential impact of foreground contamination on our results illustrated using Ret II. (Left) CMD of the TRILEGAL foreground model (Vanhollebeke et al. 2009) by querying 0.5 deg$^2$ centered around Ret II, color coded by metallicity. Middle: remaining MW foreground stars (colored) from applying the same isochrone cut used to select Ret II members. (Right) Impact of foreground contamination when scaling down the number of expected interlopers by the WFC3 FoV. We typically expect ∼1 MW interloper in our final stellar sample. The nature of our observations and selection criteria combine to make foreground contamination minimal.

## Appendix B
## Systematic Uncertainties in Individual Star Metallicities

There are several plausible physics effects that could affect our metallicity estimates beyond what is discussed in the main paper. Here, we describe these effects and estimate their possible contribution to our metallicity and MDF determinations.

### B.1. Impact of α-Enhancements

UFDs are known to host stars with different degrees of $\alpha$-enhancement (Simon 2019). The effect on CaHK-based metallicities is modest, as shown in Fu et al. (2022). There, we demonstrated that using MIST models with different levels of $\alpha$-enhancements (0.4 versus 0.0 dex) shift the inferred stellar metallicity by no more than ~0.2 dex.

### B.2. Light-element Abundance Variations

Lighter element abundance variations can affect absorption around the CaHK lines. Notably, enhancements in C and N can introduce contaminating absorption into the near-UV region of the CaHK lines (e.g., Figure 2 from Starkenburg et al. 2017a).

C is an element of particular concern, as nearly half of known metal-poor ([Fe/H]$< -3$ stars in the MW show carbon enhancements (e.g., Frebel & Norris 2015). Carbon enhancement can introduce additional absorption around the CaHK lines, potentially polluting the CaHK narrowband measurement (see Figure 2 of Starkenburg et al. 2017a). The expected impact is that a C-enhanced star would have a higher inferred CaHK-based metallicity than its true metallicity (i.e., there would be more absorption in the F395N band). As Starkenburg et al. (2017a) show, modest carbon enhancements do not drastically affect the CaHK band. Quantifying the impact of carbon on every star in our sample is a complicated process and is beyond the scope of this paper. Instead, **we use** an example case to illustrate the impact of strong C-enhancement on our inferred metallicities and MDFs.

One star in our entire sample is known to be C-enhanced from high-resolution spectroscopy. This star, J100714+160154 in Seg 1, was determined by Frebel et al. (2014) to have [Fe/H] $= -1.42 \pm 0.24$, and, notably, a carbon enhancement of [C/Fe] $= +1.44 \pm 0.2$. In this paper, we find a value of [Fe/H] $= -1.86^{+0.11}_{-0.12}$(stat.) $\pm$ 0.3(syst.), indicating they are inconsistent within $1.2\sigma$. Though it is only a single example, it does show that even a very strong C-enhancement does not change the metallicity significantly beyond the bounds of our $1\sigma$ uncertainties.

For all other stars in our sample, we perform a coarser check on the impact of carbon. Since line blanketing from C-enhancement affects the near-UV spectral region covered by the F475W filter, we should expect that a C-enhanced star would appear redder in F475W–F814W than compared to expectations. In contrast, F606W–F814W colors should be comparatively unaffected because there are few carbon features in the F606W filter. We do not find a strong trend in our sample. Though it is likely that a modest fraction of our stars have some degree of C-enhancement, we believe it is unlikely that they are affecting our MDFs beyond the reported uncertainties.

### B.3. Binary Stars

At least one-third of low-mass MS stars are accompanied by a companion(s) (Duchêne & Kraus 2013). In our sample, Ret II, Wil 1, Seg 1, and Dra II are primarily characterized by stars on the lower MS (Figure 1), a place in which unresolved binaries have the largest photometric impacts. The broadness of their lower MS is suggestive of binary companions. Using the method detailed below, we estimate the impact of unresolved binaries on CaHK-based metallicities to be small, thus we do not a priori exclude any potential unresolved binaries from our sample.

We estimate the effects of binaries by calculating the synthetic photometry of single and binary stars. Specifically, we simulate a binary system with a primary mass of 0.7 $M_\odot$, a typical stellar mass in our sample, and accompanied by a lower-mass companion star of varying masses, down to the lowest mass limit available from the MIST isochrones (0.1 $M_\odot$). We assume that both stars have the same metallicity, and conduct this calculation for metallicities from [Fe/H] $= -1.0$ to $-4.0$. We also assume that light from both stars is visible, which means that the deviations from CaHK for a single-star system that we calculate are upper limits on the magnitude of the effect.

Figure 16 illustrates the impact of a binary companion on the CaHK color. The presence of an unresolved companion makes a star appear more metal-poor (i.e., bluer) in CaHK space. This effect is more pronounced for stars that are more metal-rich. For stars with [Fe/H]$> -1.5$, the impact of a binary companion can shift it away from the CaHK track by up to 0.1 mag, which translates into a metallicity shift of ~0.2 dex. For lower metallicity, the effect is ~0.1 dex. The assumptions in this calculation mean that these numbers describe the upper limit of the impact of binarity. The impact of unresolved binaries is modest compared to our overall uncertainty budgets, indicating that unresolved binaries are not a major uncertainty on our stellar metallicities.

### B.4. Uncertainties in Stellar Evolution Modeling

There are uncertainties intrinsic to stellar evolution modeling and the generation of stellar atmospheres that can affect the placement and curvature of CaHK color tracks. In Figure 17, we compare the MIST tracks used in our work to the tracks from BaSTI (Hidalgo et al. 2018; Pietrinferni et al. 2021) across the range of metallicities common to both models.

For RGB stars, the tracks are offset by up to 0.05 mag at F475W–F814W ~ 2.0, and <0.02 mag at F475W–F814W ~ 1.0. This translates to up to ~0.2 dex systematic difference in metallicity, and the difference diminishes at bluer colors. For MS stars, the tracks are offset by less than 0.02 mag between F475W–F814W ~ 0.8 and F475W–F814W ~ 1.5, which translates to at most 0.2 dex systematic difference in metallicity. These systematic differences are well within the statistical and systematic uncertainties we adopt for our measurements.

### B.5. Quantifying Systematic Uncertainties

While a detailed investigation of the physics contributing to CaHK metallicity uncertainties is beyond the scope of this work, we do attempt to quantify their impact. We begin by discussing the extremely metal-poor end of our measurements ([Fe/H]$< -3$). Toward lower metallicities, CaHK loses its discriminating power, as shown in the CaHK tracks in Figure 9. The extremely metal-poor star search from the Pristine survey using CaHK photometry has yielded success rates of $\sim$20% for stars with [Fe/H]$< -3$ (Youakim et al. 2017; Venn et al. 2020).

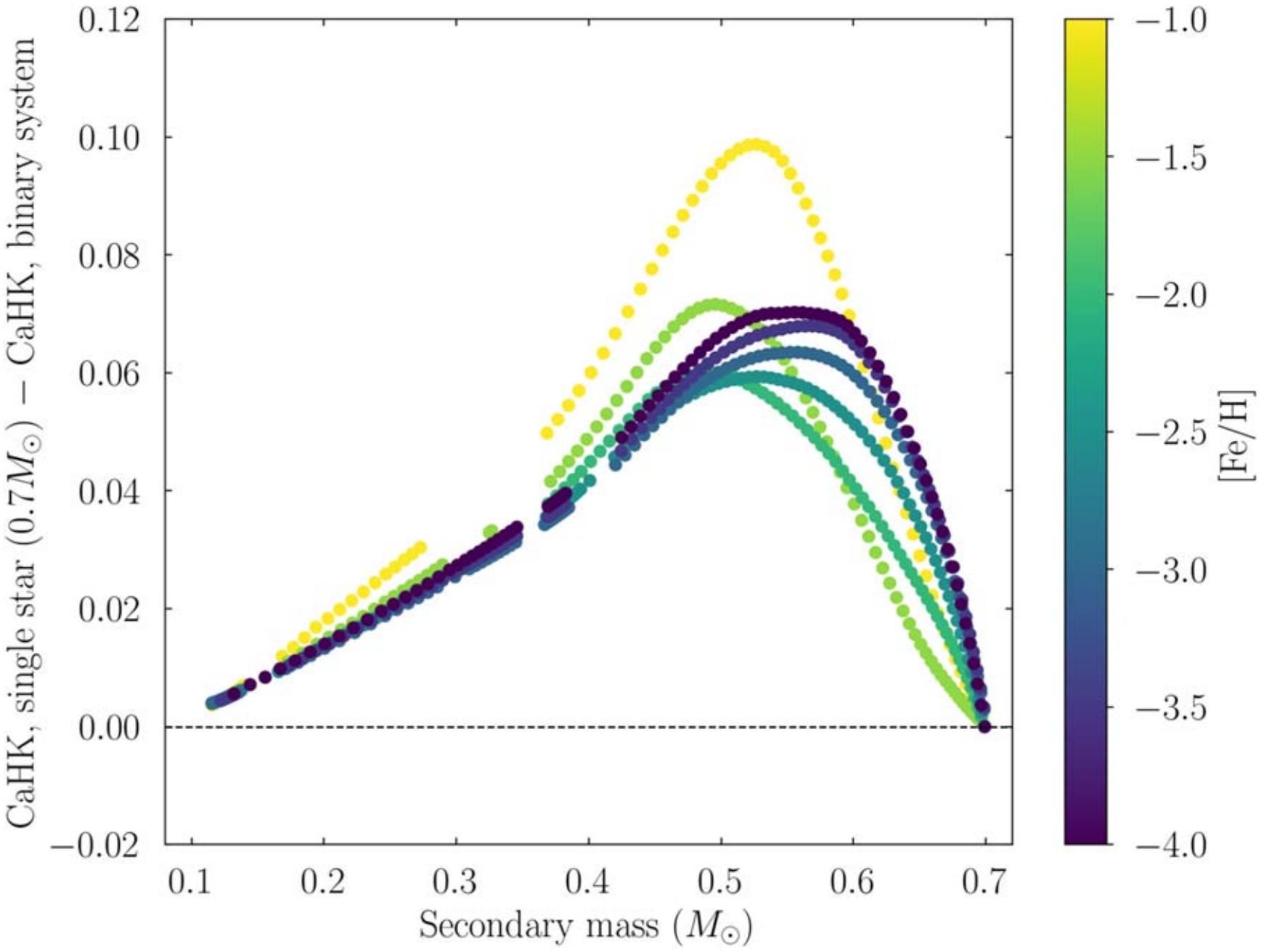

**Figure 16.** The effect of unresolved binaries on the CaHK color. The $y$-axis shows the difference in CaHK color between a single MS star at 0.7 $M_\odot$ and a binary system composed of a 0.7 $M_\odot$ MS star with companions from 0.1–0.7 $M_\odot$, for different metallicities spanned by our CaHK color grid. The impact of binary stars on our metallicities is small-to-modest compared to other sources of uncertainty.

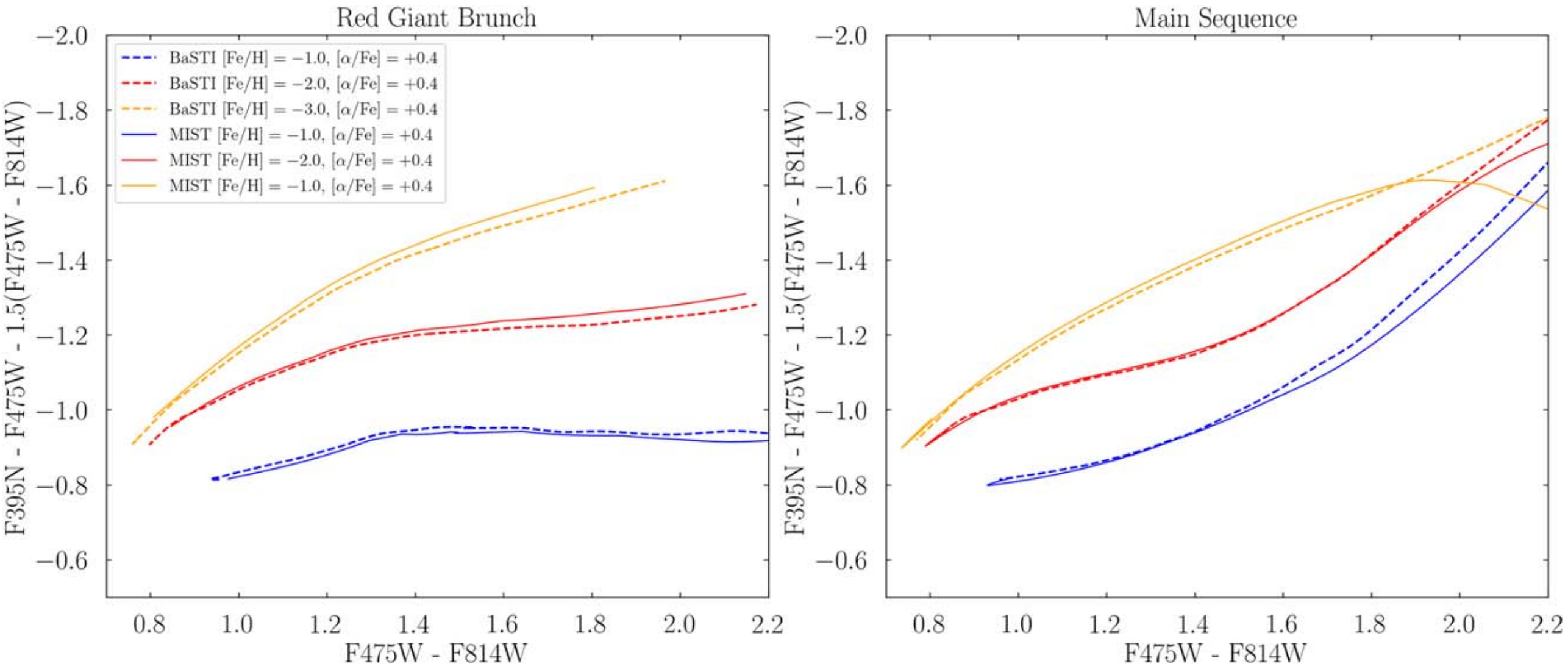

**Figure 17.** Comparing MIST and BaSTI stellar evolution models in CaHK space for $\alpha$-enhanced, 13 Gyr RGB (left) and MS (right) stars.

Additionally, there are known issues with the MIST/MESA isochrones for metal-poor stars.[21] Given that refining the models of metal-poor stars is an active area of research (Karovicova et al. 2020), we adopt a systematic error floor of $\sim$0.5 dex for all stars whose metallicity measurements are below −3.

For stars at more intermediate metallicities, we quantify the impact of systematics by attempting to recover the MDF of M92, a metal-poor GC with ⟨[Fe/H]⟩ on a par with those expected for UFDs (⟨[Fe/H]⟩ = −2.2) and no known metallicity dispersion, $[\mathrm{Ca/Fe}] = 0.10 \pm 0.05$ with a dispersion consistent with zero,[22] $[\mathrm{Mg/Fe}] = 0.14 \pm 0.04$ with a dispersion of $0.22^{+0.03}_{-0.02}$, and $[\mathrm{C/Fe}] < 0$ (Mészáros et al. 2015). We retrieve archival F395N narrowband imaging for M92 taken by HST program GO-11729 (PI: Holtzman) and reduce it alongside short (∼10s) F475W and F814W archival exposures from HST programs GO-10505 (PI: Gallart) and GO-12116 (PI: Dalcanton).

[21] For stars at lower metallicities, temperatures inferred from isochrone mapping tended to be hotter by up to $\Delta T_{\mathrm{eff}} = +500$ K compared to spectroscopic methods (e.g., Monty et al. 2020; Kielty et al. 2021).

[22] Using the data from Mészáros et al. (2015), we compute the mean and dispersion for additional elements following the procedure in Section 3.2.

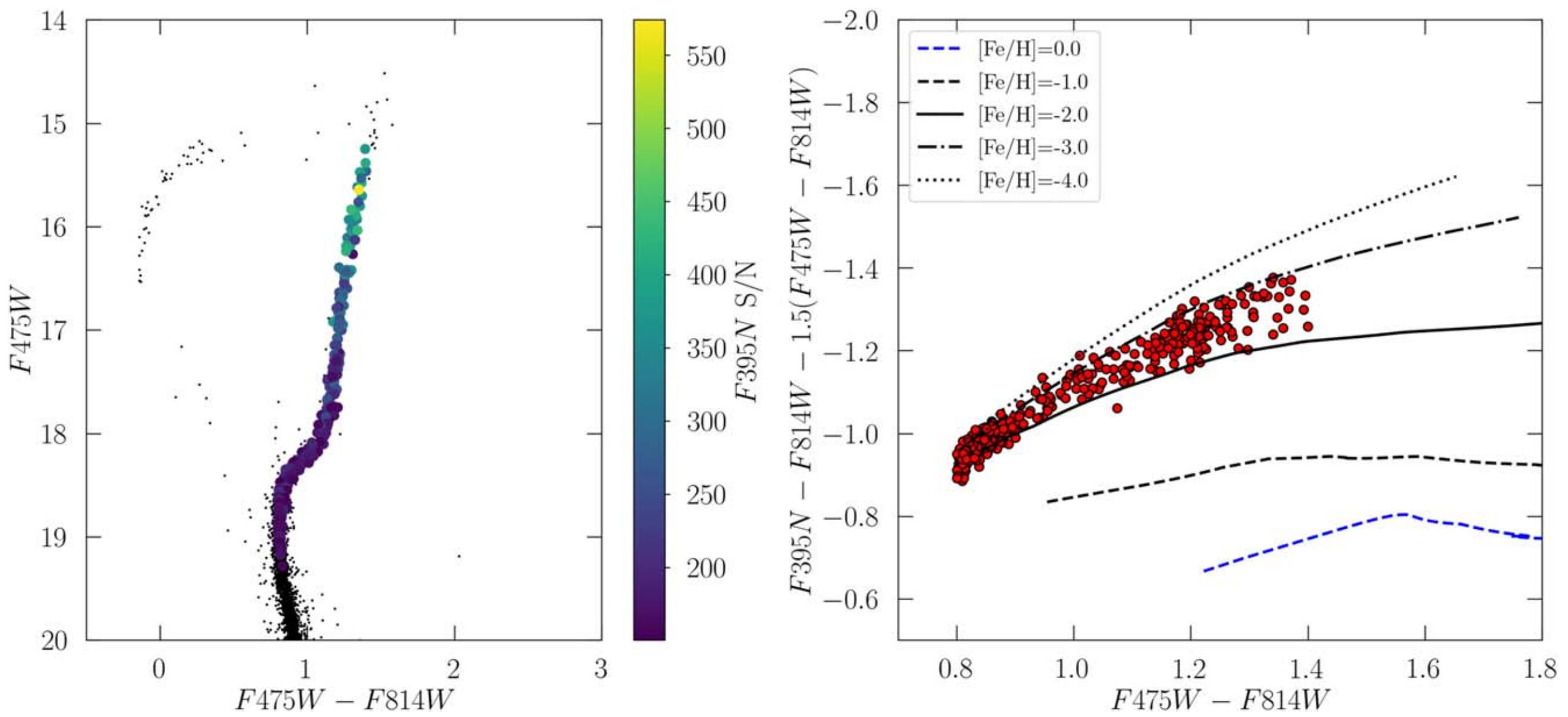

**Figure 18.** HST data of the M92 GC, a monometallic population, that we use to determine systematic uncertainties for our metallicity measurements. (Left) CMD of M92 in F475W–F814W. (Right) The position of M92 members on the CaHK diagram. The majority of these stars lie between the [Fe/H]= −3 and [Fe/H]= −2 tracks.

From the resulting catalog, we select stars with F395N S/N >100. The resulting M92 data begin below the horizontal branch and extend down to the MSTO. We present the data in Figure 18. The stars do not follow a perfectly narrow track, but they lie overwhelmingly between the [Fe/H]= −3 and [Fe/H]= −2 tracks. With our S/N selection criteria, any scatter observed in the placement of M92 members in CaHK should be driven largely by abundance variations and other modeling uncertainties.

Using this data, we aim to find the minimum individual metallicity measurement uncertainty that would result in the inferred metallicity dispersion of M92 being zero. As part of this exercise, we split the samples by stellar evolution phase, which also functions as a proxy for temperature: stars on the MSTO are hotter than their counterparts on the RGB. For hotter stars, the CaHK monometallic tracks begin to converge and make it more difficult to discern metallicities.

For each star, we assign it a metallicity based on its location on a grid of interpolated MIST $[\alpha/\mathrm{Fe}] = +0.20$ models.[23] We then assign a measurement uncertainty to each star and infer the metallicity dispersion of M92 following the procedure outlined in Section 3.2. For stars on the lower RGB, we find that a minimum measurement uncertainty of 0.2 dex is necessary to reduce the metallicity dispersion of M92 to be consistent with zero. For stars on the MSTO, a minimum measurement uncertainty of 0.3 dex is necessary.

### *B.6. Impact of Systematic Offset on Our Conclusions*

In Appendix C, we make direct comparisons to literature measurements where available, but the interpretation of discrepancies and offsets is ultimately complicated by small sample sizes and a diversity of metallicity inference methods. Here, we consider the robustness of the key results of our paper to systematic shifts in metallicity. These shifts can be caused, for example, by using different stellar evolution models, scaling our measurements to different solar abundances, and/or assuming different levels of $\alpha$-enhancement, all of which are not standardized among literature measurements.

If an metal-poor offset was applied to our measurements:

1. The overall level of the luminosity–metallicity relation for UFDs would be lowered across our sample, but the break in the LZR would remain. For the mean metallicities to begin approaching agreement with simulations, an offset of 0.5 dex at minimum would be required. Such an offset would bring our results out of agreement with literature values.
2. We would expect fewer metal-rich stars in our sample, bringing the metal-rich star fraction out of agreement with the literature.
3. Our distribution of mean metallicities will deviate further from the MW stream population from Martin et al. (2022) and Li et al. (2022), and it is unlikely that the range of UFD mean metallicities will broaden enough to be comparable to results from stream searches.

If an metal-rich offset was applied to our measurements:

1. The overall level of the luminosity–metallicity relation for UFDs would be raised across our sample, thereby exacerbating the problem of under-enriching UFDs in simulations.
2. Accordingly, we would expect more metal-rich stars in our sample, leading to tension with spectroscopic results.
3. Our distribution of mean metallicities will fall closer to the peak of the MW population of streams from Martin et al. (2022) and Li et al. (2022), but it is still unlikely that the range of UFD mean metallicities will broaden enough to be comparable to results from stream searches.

Overall, these offsets would not affect the main results of the paper.

[23] Although we use $[\alpha/\mathrm{Fe}] = +0.20$ models for this exercise, they look very similar to $[\alpha/\mathrm{Fe}] = +0.40$ models in CaHK color space.

# Appendix C
# Comparisons to Literature Methods

In Figure 19, we compare metallicity measurements from this work to stars with metallicities in the literature derived from different techniques.

The upper left panel of Figure 19 shows a comparison between our metallicities and those derived using the Carrera et al. (2013) CaT equivalent widths calibration. The galaxies whose data are used for this comparison are Eri II (Magellan/IMACS, Li et al. 2017) and Grus I (Magellan/IMACS, Chiti et al. 2022). This comparison shows about a 0.3 dex systematic offset and agrees reasonably well (within $\sim 1.5\sigma$). Uncertainties between the two methods are generally comparable in size. The Carrera et al. (2013) calibration is anchored by GC and open clusters for metallicities between −2.33 and +0.47, and below that metallicity range, the calibration was anchored by known metal-poor galactic field stars known at the time of writing. This calibration sample includes both solar-scaled and $\alpha$-enhanced stars, and though not factored into the calibration, it is acknowledged in the paper that the CaT equivalent widths, and therefore the inferred metallicity, have a secondary dependence on Ca abundances. The scatter in the comparison may be due to differences in assumed $\alpha$ abundances as well as additional systematic effects.

The upper right panel of Figure 19 shows a comparison of our metallicities to those derived using spectral synthesis methods of medium-resolution spectra. The galaxies represented in this panel are CVn II (Keck/DEIMOS, Kirby et al. 2013, later analyzed by Vargas et al. 2013), Hya II (Keck/DEIMOS, Kirby et al. 2015), and Ret III and Phe II (VLT/FLAMES, Fritz et al. 2019). The metallicity measurements from Kirby et al. (2013) and Kirby et al. (2015) tend to be more metal-rich than our measurements (falling above the 1:1 line) by ∼0.6 dex, though the overall samples agree within $1\sigma$.

On average, the measurements of Fritz et al. (2019) tend to be more metal-poor than our measurements by ∼0.5 dex (i.e., they fall below the 1:1 line). Where the Kirby et al. studies do not assume $\alpha$-enhancement, the Fritz et al. (2019) measurements assume $[\alpha/\mathrm{Fe}] = +0.5$, which is slightly enhanced relative to the models that we use. In CaHK color space, we found that additional $\alpha$-enhancement makes a star appear more metal-poor, and this may account for some of the discrepancies that we observe. Some of the technical details in Fritz et al. (2019) are not entirely clear to us, making it a little challenging to determine what other factors might lead to the observed discrepancies.

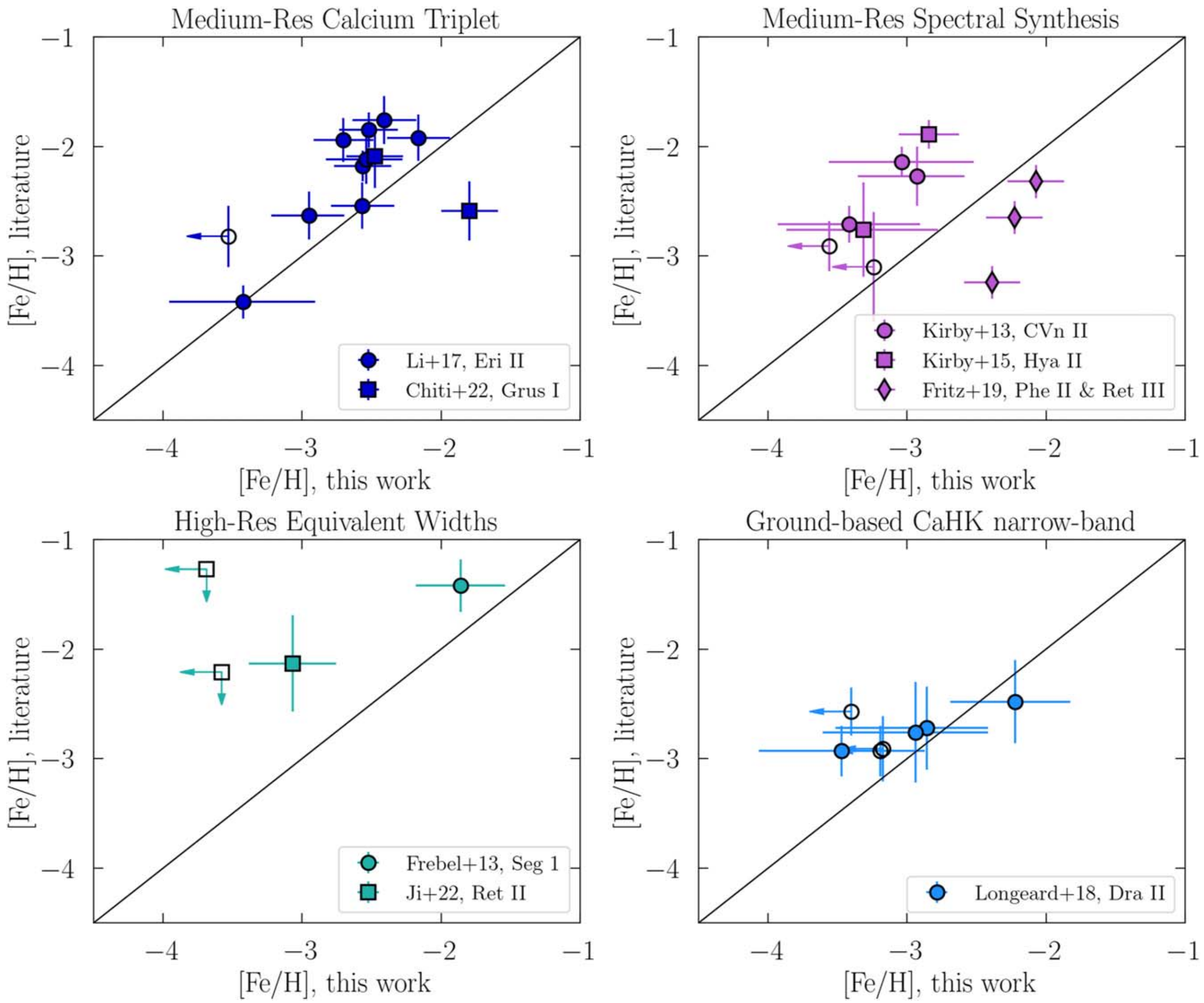

**Figure 19.** Comparisons between our measurements and those from the literature, between different measurement methods. The vast majority of literature metallicity measurements come from medium-resolution studies, particularly those leveraging the CaT metallicity calibration. The ground-based CaHK narrowband measurements are from the Pristine study of Dra II (Longeard et al. 2018).

The lower left panel of Figure 19 shows a comparison of our metallicities to those derived using equivalent widths of high-resolution spectroscopy. The galaxies in this sample are Seg 1 (Magellan/MIKE, Frebel et al. 2014) and Ret II (Magellan/M2FS and FLAMES/GIRAFFE, Ji et al. 2023). As discussed in previous sections, our metallicity measurement for the star that we have in common with Frebel et al. (2014) is in good agreement at [Fe/H]$\sim -1.5$. The remaining stars in this panel are from Ji et al. (2023). For the single star that is not an upper limit, the two methods are in agreement at $\sim 1.5\sigma$. The spectra of this star suggest [Ca/Fe]$= -0.25 \pm 0.48$ and an unknown $\alpha$-enhancement. Knowledge of the latter could bring them into better agreement. Ji et al. (2023) were only able to place upper limits on the metallicities of the other two stars, and our metallicities are consistent with these upper limits.

Finally, the bottom-right panel of Figure 19 shows ground-based CaHK Pristine measurements from Dra II (Longeard et al. 2018) compared to our CaHK measurements. The level of agreement is good, given the uncertainties. Though our uncertainties are much larger than Longeard et al. (2018), this is because we adopt a more conservative treatment of systematic uncertainties, as described in Appendix B.5. Our random uncertainties for these stars are 0.3 dex, which are comparable to what (Longeard et al. 2018) report.

Due to calibration limits, Longeard et al. (2018) also placed a metallicity floor on their measurements at [Fe/H]$= -3.0$. Because we use synthetic model grids that extend to [Fe/H] $= -4.0$, we are able to report more metal-poor values, as opposed to upper limits. Though limited in sample size, this comparison affirms the metal-poor nature of Dra II and demonstrates reasonable consistency between ground-based and space-based CaHK narrowband metallicities.

## Appendix D
## Metallicity Measurements as a Function of Stellar Evolutionary Phase and Brightness

As a diagnostic of the reliability of our measurements, we investigate whether there are systematic effects in metallicity measurements as a function of stellar evolutionary phase and/or apparent brightness that may be affecting our mean and dispersion measurements. We present all of the measurements that we use for the Gaussian MDF inference fit (see Section 3.2 for the procedure on their treatment) in Figure 20.

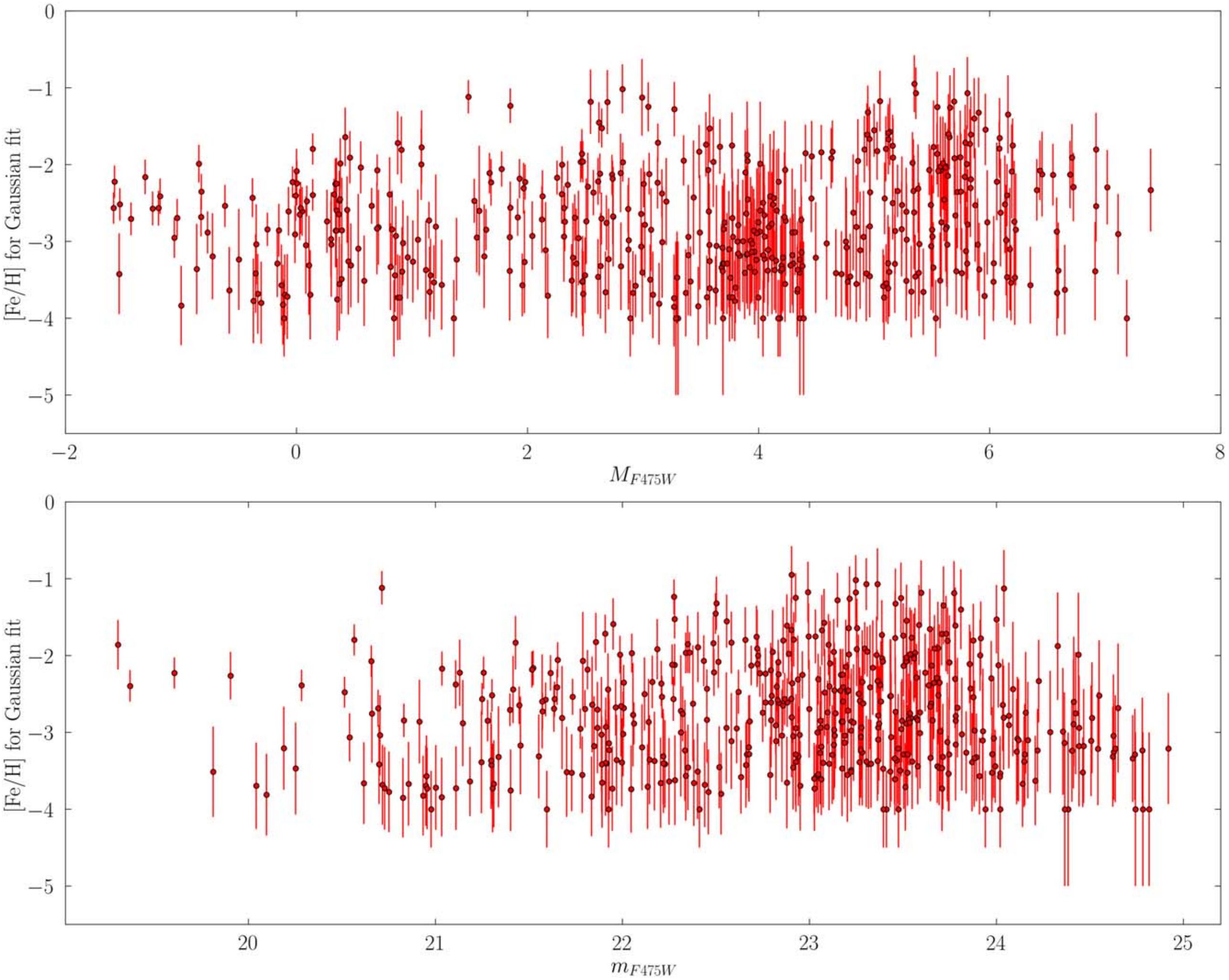

**Figure 20.** Stellar metallicities used for fitting the mean and dispersion (Section 3.2) of our MDFs as a function of brightness across our entire sample of galaxies. The measurements shown here are specifically for the process of fitting the MDFs, and may not fully reflect the final reported measurements (e.g., asymmetric uncertainties, upper limit constraints). The uncertainties shown here also take into account the systematic uncertainties we adopt as discussed in Appendix B.5, and statistical uncertainties for our brighter stars are often smaller. (Top panel) Metallicity measurements as a function of absolute magnitude, where absolute magnitude is a stand-in for the stellar evolutionary phase. (Bottom panel) Metallicity measurements as a function of apparent magnitude.

In the top panel, we plot our measurements as a function of absolute magnitude, where absolute magnitude is a proxy for the stellar evolutionary phase. There is no apparent systematic trend in metallicity as a function of absolute magnitude across our sample, and the larger scatter for MS stars can be attributed to their larger systematic uncertainties as determined in Appendix B.5, as well as the lower S/N of the fainter stars. Similarly, there is no apparent systematic trend in metallicity as a function of apparent magnitude, and the larger scatter in metallicity measurements for fainter stars is in line with expectations for measurements made using lower S/N data.

## Appendix E
## Additional MDF Summary Statistics

To assess deviations from Gaussianity, in this section, we present the 16th, median, and 84th percentile measurements for our MDFs, as well as the skew and kurtosis. We compute these quantities by first assuming Gaussian uncertainties on our individual measurements following the procedure outlined in Section 3.2 to generate 10,000 Monte Carlo realizations of our MDF, and then deriving summary statistics and their uncertainties from these realizations. In all cases, the median we recover is in $1\sigma$ agreement with the Gaussian mean. For the cases of Wil 1, Ret II, and Dra II, we resolve a kurtosis above at least the $2\sigma$ level, suggesting that their MDFs have weaker tails compared to a standard Gaussian. In all other cases, we are unable to resolve significant deviations from Gaussianity. We present these measurements in Table 3.

## Appendix F
## Tables of Metallicity Measurements

In Table 4, we present metallicity measurements of all of our stars. In Table 5, we present stars that we designate as extremely metal-poor ([Fe/H] $-2$). In Table 6, we present metal-rich stars ([Fe/H] $> -2$).

**Table 3**
Additional MDF Summary Statistics

| Galaxy | $M_v$ (mag) | $N_{\rm stars}$ | 16th Percentile MDF (dex) | Median (dex) | 84th Percentile (dex) | Skew | Kurtosis |
|---|---|---|---|---|---|---|---|
| Eridanus II | −7.1 | 75 | $-3.50^{+0.12}_{-0.12}$ | $-2.70^{+0.07}_{-0.07}$ | $-2.12^{+0.08}_{-0.08}$ | $-0.3^{+0.2}_{-0.2}$ | $-0.1^{+0.5}_{-0.4}$ |
| Canes Venatici II | −5.2 | 40 | $-3.88^{+0.15}_{-0.15}$ | $-3.09^{+0.13}_{-0.14}$ | $-2.23^{+0.13}_{-0.13}$ | $0.3^{+0.3}_{-0.3}$ | $0.0^{+0.8}_{-0.5}$ |
| Hydra II | −4.9 | 30 | $-3.93^{+0.17}_{-0.18}$ | $-3.16^{+0.14}_{-0.15}$ | $-2.46^{+0.14}_{-0.13}$ | $0.1^{+0.4}_{-0.5}$ | $0.2^{+1.1}_{-0.7}$ |
| Reticulum II | −4.0 | 77 | $-3.73^{+0.12}_{-0.12}$ | $-2.74^{+0.12}_{-0.12}$ | $-1.61^{+0.12}_{-0.11}$ | $0.1^{+0.2}_{-0.2}$ | $-0.8^{+0.2}_{-0.2}$ |
| Horologium I | −3.8 | 40 | $-3.72^{+0.16}_{-0.17}$ | $-2.88^{+0.14}_{-0.14}$ | $-2.06^{+0.15}_{-0.14}$ | $0.1^{+0.3}_{-0.3}$ | $-0.4^{+0.5}_{-0.3}$ |
| Grus I | −3.5 | 36 | $-3.66^{+0.20}_{-0.20}$ | $-2.80^{+0.15}_{-0.16}$ | $-1.86^{+0.17}_{-0.17}$ | $0.0^{+0.3}_{-0.4}$ | $-0.3^{+0.7}_{-0.4}$ |
| Reticulum III | −3.3 | 18 | $-3.46^{+0.26}_{-0.28}$ | $-2.49^{+0.14}_{-0.15}$ | $-1.94^{+0.17}_{-0.14}$ | $-0.4^{+0.4}_{-0.4}$ | $-0.5^{+0.7}_{-0.5}$ |
| Willman 1 | −2.9 | 68 | $-3.64^{+0.14}_{-0.13}$ | $-2.55^{+0.13}_{-0.13}$ | $-1.57^{+0.12}_{-0.11}$ | $-0.1^{+0.2}_{-0.2}$ | $-0.8^{+0.3}_{-0.2}$ |
| Phoenix II | −2.7 | 10 | $-3.01^{+0.23}_{-0.27}$ | $-2.39^{+0.14}_{-0.14}$ | $-1.99^{+0.16}_{-0.15}$ | $-0.4^{+0.6}_{-0.6}$ | $-0.3^{+1.0}_{-0.6}$ |
| Eridanus III | −2.1 | 13 | $-2.77^{+0.28}_{-0.38}$ | $-2.05^{+0.14}_{-0.14}$ | $-1.84^{+0.18}_{-0.16}$ | $-0.6^{+0.4}_{-0.4}$ | $-0.2^{+1.0}_{-0.6}$ |
| Tucana V | −1.6 | 6 | $-3.15^{+0.30}_{-0.44}$ | $-2.43^{+0.18}_{-0.16}$ | $-1.98^{+0.17}_{-0.15}$ | $-0.5^{+0.5}_{-0.5}$ | $-0.8^{+0.8}_{-0.6}$ |
| Segue 1 | −1.3 | 12 | $-3.30^{+0.29}_{-0.30}$ | $-2.29^{+0.20}_{-0.22}$ | $-1.69^{+0.17}_{-0.17}$ | $-0.6^{+0.4}_{-0.4}$ | $-0.7^{+0.9}_{-0.5}$ |
| Draco II | −0.8 | 38 | $-3.68^{+0.17}_{-0.17}$ | $-2.75^{+0.13}_{-0.13}$ | $-2.04^{+0.12}_{-0.12}$ | $-0.3^{+0.2}_{-0.3}$ | $-0.7^{+0.5}_{-0.3}$ |

**Note.** Additional summary statistics for our MDFs. We list the 16th, 50th, and 84th percentiles, as well as the skew and kurtosis measured from our MDFs.

**Table 4**
Metallicity Measurements of All Stars

| UFD | Star | R.A. (deg) | Decl. (deg) | F814W (mag) | F606W (mag) | F475W (mag) | F395N (mag) | *VI* (mag) | *CaHK* (mag) | [Fe/H] (dex) |
|---|---|---|---|---|---|---|---|---|---|---|
| Eri II | 0 | 56.099955 | −43.545433 | 19.311 ± 0.000 | 20.292 ± 0.001 | 21.248 ± 0.003 | 22.634 ± 0.033 | 1.937 ± 0.003 | −1.520 ± 0.033 | $-2.56^{+0.04}_{-0.05}$ (stat.) ± 0.2 (syst.) |
| Eri II | 1 | 56.111058 | −43.523014 | 19.397 ± 0.001 | 20.351 ± 0.001 | 21.301 ± 0.002 | 22.692 ± 0.048 | 1.904 ± 0.002 | −1.465 ± 0.048 | $-2.52^{+0.07}_{-0.06}$ (stat.) ± 0.2 (syst.) |
| Eri II | 2 | 56.114817 | −43.526851 | 19.423 ± 0.001 | 20.354 ± 0.001 | 21.257 ± 0.002 | 22.680 ± 0.034 | 1.834 ± 0.002 | −1.328 ± 0.034 | $-2.22^{+0.06}_{-0.07}$ (stat.) ± 0.2 (syst.) |
| Eri II | 3 | 56.123845 | −43.529984 | 19.526 ± 0.001 | 20.444 ± 0.001 | 21.298 ± 0.002 | 22.246 ± 0.029 | 1.772 ± 0.002 | −1.710 ± 0.029 | $-3.42^{+0.19}_{-0.14}$ (stat.) ± 0.2 (syst.) |
| Eri II | 4 | 56.086611 | −43.541012 | 19.548 ± 0.001 | 20.498 ± 0.001 | 21.398 ± 0.003 | 22.672 ± 0.044 | 1.850 ± 0.003 | −1.501 ± 0.044 | $-2.71^{+0.07}_{-0.08}$ (stat.) ± 0.2 (syst.) |
| Eri II | 5 | 56.088979 | −43.505809 | 19.660 ± 0.001 | 20.577 ± 0.001 | 21.522 ± 0.003 | 23.008 ± 0.030 | 1.862 ± 0.003 | −1.307 ± 0.030 | $-2.16^{+0.11}_{-0.11}$ (stat.) ± 0.2 (syst.) |
| Eri II | 6 | 56.084220 | −43.536364 | 19.729 ± 0.001 | 20.674 ± 0.001 | 21.588 ± 0.004 | 22.902 ± 0.047 | 1.859 ± 0.004 | −1.475 ± 0.047 | $-2.57^{+0.09}_{-0.10}$ (stat.) ± 0.2 (syst.) |
| Eri II | 7 | 56.060979 | −43.526336 | 19.866 ± 0.001 | 20.758 ± 0.001 | 21.637 ± 0.004 | 22.861 ± 0.030 | 1.771 ± 0.004 | −1.433 ± 0.030 | $-2.57^{+0.11}_{-0.12}$ (stat.) ± 0.2 (syst.) |
| Eri II | 8 | 56.110605 | −43.545527 | 19.888 ± 0.001 | 20.770 ± 0.001 | 21.650 ± 0.003 | 22.898 ± 0.034 | 1.762 ± 0.003 | −1.395 ± 0.034 | $-2.41^{+0.11}_{-0.12}$ (stat.) ± 0.2 (syst.) |
| Eri II | 9 | 56.075786 | −43.519993 | 20.126 ± 0.001 | 20.969 ± 0.001 | 21.775 ± 0.003 | 22.740 ± 0.034 | 1.649 ± 0.003 | −1.508 ± 0.034 | $-2.95^{+0.18}_{-0.16}$ (stat.) ± 0.2 (syst.) |

**Note.** Metallicity measurements for all stars in our sample. Statistical uncertainties on the metallicity measurements originate from photometric uncertainties, and systematic uncertainties are assigned following the investigation in Appendix B.5.

(This table is available in its entirety in machine-readable form.)

**Table 5**
Extremely Metal-poor Star Candidates

| UFD | Star | R.A. (deg) | Decl. (deg) | F814W (mag) | F606W (mag) | F475W (mag) | F395N (mag) | *VI* (mag) | CaHK (mag) | [Fe/H] (dex) |
|---|---|---|---|---|---|---|---|---|---|---|
| Eri II | 3 | 56.123845 | −43.529984 | 19.526 ± 0.001 | 20.444 ± 0.001 | 21.298 ± 0.002 | 22.246 ± 0.029 | 1.772 ± 0.002 | −1.710 ± 0.029 | $-3.42^{+0.19}_{-0.14}$ (stat.) ± 0.5 (syst.) |
| Eri II | 12 | 56.096139 | −43.523451 | 20.193 ± 0.001 | 21.038 ± 0.001 | 21.835 ± 0.003 | 22.640 ± 0.048 | 1.642 ± 0.003 | −1.658 ± 0.048 | $< -3.53$ |
| Eri II | 13 | 56.085980 | −43.552250 | 20.368 ± 0.001 | 21.193 ± 0.001 | 21.969 ± 0.003 | 22.798 ± 0.041 | 1.601 ± 0.003 | −1.573 ± 0.041 | $-3.36^{+0.38}_{-0.25}$ (stat.) ± 0.5 (syst.) |
| Eri II | 17 | 56.084077 | −43.509971 | 20.484 ± 0.001 | 21.308 ± 0.001 | 22.104 ± 0.003 | 23.007 ± 0.042 | 1.620 ± 0.003 | −1.527 ± 0.042 | $-3.20^{+0.28}_{-0.23}$ (stat.) ± 0.5 (syst.) |
| Eri II | 18 | 56.078614 | −43.541732 | 20.651 ± 0.001 | 21.468 ± 0.001 | 22.249 ± 0.004 | 23.067 ± 0.055 | 1.598 ± 0.004 | −1.579 ± 0.055 | $-3.64^{+0.23}_{-0.29}$ (stat.) ± 0.5 (syst.) |
| Eri II | 20 | 56.114148 | −43.547555 | 20.791 ± 0.001 | 21.587 ± 0.001 | 22.334 ± 0.004 | 23.168 ± 0.035 | 1.543 ± 0.004 | −1.480 ± 0.035 | $-3.23^{+0.41}_{-0.45}$ (stat.) ± 0.5 (syst.) |
| Eri II | 22 | 56.070042 | −43.529300 | 21.053 ± 0.001 | 21.798 ± 0.001 | 22.460 ± 0.003 | 22.979 ± 0.044 | 1.407 ± 0.003 | −1.592 ± 0.044 | $< -3.24$ |
| Eri II | 23 | 56.076370 | −43.538978 | 21.064 ± 0.001 | 21.848 ± 0.002 | 22.527 ± 0.005 | 23.117 ± 0.063 | 1.463 ± 0.005 | −1.605 ± 0.063 | $< -3.35$ |
| Eri II | 25 | 56.100176 | −43.539428 | 21.215 ± 0.001 | 21.990 ± 0.002 | 22.666 ± 0.005 | 23.371 ± 0.046 | 1.451 ± 0.005 | −1.472 ± 0.046 | $-3.29^{+0.38}_{-0.41}$ (stat.) ± 0.5 (syst.) |
| Eri II | 32 | 56.092274 | −43.523353 | 21.449 ± 0.001 | 22.198 ± 0.002 | 22.913 ± 0.005 | 23.675 ± 0.062 | 1.464 ± 0.005 | −1.434 ± 0.062 | $-3.05^{+0.50}_{-0.28}$ (stat.) ± 0.5 (syst.) |

**Note.** Extremely metal-poor ([Fe/H]$< -3.0$) star candidates identified in our work.

(This table is available in its entirety in machine-readable form.)

**Table 6**
Metal-rich ([Fe/H] > −2.0) Stars

| UFD | Star | R.A. (deg) | Decl. (deg) | F814W (mag) | F606W (mag) | F475W (mag) | F395N (mag) | *VI* (mag) | CaHK (mag) | [Fe/H] (dex) |
|---|---|---|---|---|---|---|---|---|---|---|
| Eri II | 11 | 56.105741 | −43.529386 | 20.129 ± 0.001 | 20.954 ± 0.001 | 21.987 ± 0.003 | 23.548 ± 0.063 | 1.858 ± 0.003 | −1.226 ± 0.063 | $-1.99^{+0.14}_{-0.15}$ (stat.) ± 0.2 (syst.) |
| Eri II | 48 | 56.106782 | −43.547699 | 21.819 ± 0.002 | 22.560 ± 0.002 | 23.211 ± 0.006 | 24.147 ± 0.100 | 1.392 ± 0.006 | −1.152 ± 0.100 | $-1.99^{+0.27}_{-0.25}$ (stat.) ± 0.2 (syst.) |
| Eri II | 49 | 56.102513 | −43.533665 | 21.817 ± 0.002 | 22.565 ± 0.002 | 23.254 ± 0.006 | 24.338 ± 0.107 | 1.437 ± 0.006 | −1.072 ± 0.107 | $-1.64^{+0.27}_{-0.37}$ (stat.) ± 0.2 (syst.) |
| Eri II | 53 | 56.095432 | −43.523738 | 21.883 ± 0.002 | 22.611 ± 0.002 | 23.295 ± 0.006 | 24.267 ± 0.103 | 1.412 ± 0.006 | −1.146 ± 0.103 | $-1.91^{+0.26}_{-0.29}$ (stat.) ± 0.2 (syst.) |
| Eri II | 63 | 56.091969 | −43.509073 | 22.349 ± 0.002 | 23.058 ± 0.003 | 23.707 ± 0.008 | 24.673 ± 0.102 | 1.358 ± 0.008 | −1.071 ± 0.102 | $-1.72^{+0.32}_{-0.40}$ (stat.) ± 0.2 (syst.) |
| Eri II | 64 | 56.078675 | −43.556979 | 22.408 ± 0.002 | 23.116 ± 0.003 | 23.739 ± 0.009 | 24.647 ± 0.108 | 1.331 ± 0.009 | −1.089 ± 0.108 | $-1.81^{+0.38}_{-0.39}$ (stat.) ± 0.2 (syst.) |
| Eri II | 68 | 56.089733 | −43.506056 | 22.603 ± 0.002 | 23.310 ± 0.004 | 23.915 ± 0.008 | 24.814 ± 0.102 | 1.312 ± 0.008 | −1.069 ± 0.102 | $-1.78^{+0.39}_{-0.48}$ (stat.) ± 0.2 (syst.) |
| Eri II | 69 | 56.113460 | −43.540911 | 22.595 ± 0.002 | 23.296 ± 0.003 | 23.912 ± 0.008 | 24.771 ± 0.090 | 1.317 ± 0.008 | −1.116 ± 0.090 | $-1.99^{+0.41}_{-0.43}$ (stat.) ± 0.2 (syst.) |
| CVn II | 21 | 194.302909 | 34.324379 | 22.284 ± 0.003 | 22.902 ± 0.003 | 23.597 ± 0.029 | 24.701 ± 0.083 | 1.313 ± 0.029 | −0.866 ± 0.088 | $-1.18^{+0.34}_{-0.40}$ (stat.) ± 0.2 (syst.) |
| CVn II | 26 | 194.310078 | 34.320700 | 22.754 ± 0.004 | 23.354 ± 0.004 | 24.041 ± 0.030 | 25.150 ± 0.103 | 1.287 ± 0.030 | −0.822 ± 0.107 | $-1.13^{+0.44}_{-0.48}$ (stat.) ± 0.2 (syst.) |

**Note.** Metal-rich ([Fe/H] > −2.0) stars identified in our work.

(This table is available in its entirety in machine-readable form.)

## ORCID iDs

Sal Wanying Fu https://orcid.org/0000-0003-2990-0830
Daniel R. Weisz https://orcid.org/0000-0002-6442-6030
Nicolas Martin https://orcid.org/0000-0002-1349-202X
Alessandro Savino https://orcid.org/0000-0002-1445-4877
Michael Boylan-Kolchin https://orcid.org/0000-0002-9604-343X
Patrick Côté https://orcid.org/0000-0003-1184-8114
Andrew E. Dolphin https://orcid.org/0000-0001-8416-4093
Alexander P. Ji https://orcid.org/0000-0002-4863-8842
Ekta Patel https://orcid.org/0000-0002-9820-1219
Nathan R. Sandford https://orcid.org/0000-0002-7393-3595

## References

Agertz, O., Pontzen, A., Read, J. I., et al. 2020, MNRAS, 491, 1656
Aguado, D. S., Youakim, K., González Hernández, J. I., et al. 2019, MNRAS, 490, 2241
Alexander, R. K., Vincenzo, F., Ji, A. P., et al. 2023, MNRAS, 522, 5415
An, D., Beers, T. C., Johnson, J. A., et al. 2013, ApJ, 763, 65
Andrews, B. H., Weinberg, D. H., Schönrich, R., & Johnson, J. A. 2017, ApJ, 835, 224
Applebaum, E., Brooks, A. M., Christensen, C. R., et al. 2021, ApJ, 906, 96
Applebaum, E., Brooks, A. M., Quinn, T. R., & Christensen, C. R. 2020, MNRAS, 492, 8
Asplund, M., Grevesse, N., Sauval, A. J., & Scott, P. 2009, ARA&A, 47, 481
Baumgardt, H., Faller, J., Meinhold, N., McGovern-Greco, C., & Hilker, M. 2022, MNRAS, 510, 3531
Bechtol, K., Drlica-Wagner, A., Balbinot, E., et al. 2015, ApJ, 807, 50
Belokurov, V., Erkal, D., Evans, N. W., Koposov, S. E., & Deason, A. J. 2018, MNRAS, 478, 611
Belokurov, V., Zucker, D. B., Evans, N. W., et al. 2007, ApJ, 654, 897
Boylan-Kolchin, M., & Weisz, D. R. 2021, MNRAS, 505, 2764
Boylan-Kolchin, M., Weisz, D. R., Bullock, J. S., & Cooper, M. C. 2016, MNRAS, 462, L51
Brauer, K., Andales, H. D., Ji, A. P., et al. 2022, ApJ, 937, 14
Brown, T. M., Tumlinson, J., Geha, M., et al. 2014, ApJ, 796, 91
Cantu, S. A., Pace, A. B., Marshall, J., et al. 2021, ApJ, 916, 81
Carigi, L., Hernandez, X., & Gilmore, G. 2002, MNRAS, 334, 117
Carrera, R., Pancino, E., Gallart, C., & del Pino, A. 2013, MNRAS, 434, 1681
Cerny, W., Martínez-Vázquez, C. E., Drlica-Wagner, A., et al. 2023, ApJ, 953, 1
Cerny, W., Pace, A. B., Drlica-Wagner, A., et al. 2021, ApJ, 910, 18
Chiti, A., Frebel, A., Jerjen, H., Kim, D., & Norris, J. E. 2020, ApJ, 891, 8
Chiti, A., Simon, J. D., Frebel, A., et al. 2022, ApJ, 939, 41
Choi, J., Dotter, A., Conroy, C., et al. 2016, ApJ, 823, 102
Conn, B. C., Jerjen, H., Kim, D., & Schirmer, M. 2018, ApJ, 852, 68
Conroy, C., Naidu, R. P., Zaritsky, D., et al. 2019, ApJ, 887, 237
Côté, P., Marzke, R. O., West, M. J., & Minniti, D. 2000, ApJ, 533, 869
Crnojević, D., Sand, D. J., Zaritsky, D., et al. 2016, ApJL, 824, L14
Dawoodbhoy, T., Shapiro, P. R., Ocvirk, P., et al. 2018, MNRAS, 480, 1740
Deason, A. J., Belokurov, V., & Weisz, D. R. 2015, MNRAS, 448, L77
Dolphin, A., 2016 DOLPHOT: Stellar Photometry, Astrophysics Source Code Library, ascl:1608.013
Dolphin, A. E. 2000, PASP, 112, 1383
Dolphin, A. E. 2002, MNRAS, 332, 91
Dotter, A. 2016, ApJS, 222, 8
Dotter, A., Chaboyer, B., Jevremović, D., et al. 2008, ApJS, 178, 89
Drlica-Wagner, A., Bechtol, K., Rykoff, E. S., et al. 2015, ApJ, 813, 109
Duchêne, G., & Kraus, A. 2013, ARA&A, 51, 269
El-Badry, K., Bland-Hawthorn, J., Wetzel, A., et al. 2018, MNRAS, 480, 652
Emerick, A., Bryan, G. L., & Mac Low, M.-M. 2020, ApJ, 890, 155
Fillingham, S. P., Cooper, M. C., Kelley, T., et al. 2019, arXiv:1906.04180
Foreman-Mackey, D. 2016, JOSS, 1, 24
Foreman-Mackey, D., Hogg, D. W., Lang, D., & Goodman, J. 2013, PASP, 125, 306
Frebel, A., & Norris, J. E. 2015, ARA&A, 53, 631
Frebel, A., Simon, J. D., & Kirby, E. N. 2014, ApJ, 786, 74
Fritz, T. K., Carrera, R., Battaglia, G., & Taibi, S. 2019, A&A, 623, A129
Fu, S. W., Simon, J. D., Shetrone, M., et al. 2018, ApJ, 866, 42
Fu, S. W., Weisz, D. R., Starkenburg, E., et al. 2022, ApJ, 925, 6
Gallart, C., Monelli, M., Ruiz-Lara, T., et al. 2021, ApJ, 909, 192
Gandhi, P. J., Wetzel, A., Hopkins, P. F., et al. 2022, MNRAS, 516, 1941
Geha, M., Willman, B., Simon, J. D., et al. 2009, ApJ, 692, 1464
Gelman, A., & Rubin, D. B. 1992, StaSc, 7, 457
Greco, C., Dall'Ora, M., Clementini, G., et al. 2008, ApJL, 675, L73
Grevesse, N., & Sauval, A. J. 1998, SSRv, 85, 161
Grieco, V., Matteucci, F., Pipino, A., & Cescutti, G. 2012, A&A, 548, A60
Grillmair, C. J. 2014, in IAU Symp. 298, Setting the scene for Gaia and LAMOST, ed. S. Feltzing et al. (Cambridge: Cambridge Univ. Press), 405
Han, S.-I., Kim, H.-S., Yoon, S.-J., et al. 2020, ApJS, 247, 7
Helmi, A., Babusiaux, C., Koppelman, H. H., et al. 2018, Natur, 563, 85
Hidalgo, S. L., Pietrinferni, A., Cassisi, S., et al. 2018, ApJ, 856, 125
Hogg, D. W., Bovy, J., & Lang, D. 2010, arXiv:1008.4686
Homma, D., Chiba, M., Komiyama, Y., et al. 2019, PASJ, 71, 94
Homma, D., Chiba, M., Okamoto, S., et al. 2018, PASJ, 70, S18
Hunter, J. D. 2007, CSE, 9, 90
Ibata, R., Malhan, K., Martin, N., et al. 2021, ApJ, 914, 123
Ivezić, Ž., Sesar, B., Jurić, M., et al. 2008, ApJ, 684, 287
Jenkins, S., Li, T. S., Pace, A. B., et al. 2021, ApJ, 920, 92
Jeon, M., Besla, G., & Bromm, V. 2017, ApJ, 848, 85
Jeon, M., & Bromm, V. 2019, MNRAS, 485, 5939
Jerjen, H., Conn, B., Kim, D., & Schirmer, M. 2018, arXiv:1809.02259
Ji, A. P., Frebel, A., Chiti, A., & Simon, J. D. 2016, Natur, 531, 610
Ji, A. P., Li, T. S., Simon, J. D., et al. 2020, ApJ, 889, 27
Ji, A. P., Simon, J. D., Frebel, A., Venn, K. A., & Hansen, T. T. 2019, ApJ, 870, 83
Ji, A. P., Simon, J. D., Roederer, I. U., et al. 2023, AJ, 165, 100
Karovicova, I., White, T. R., Nordlander, T., et al. 2020, A&A, 640, A25
Kerber, L. O., Libralato, M., Souza, S. O., et al. 2019, MNRAS, 484, 5530
Kielty, C. L., Venn, K. A., Sestito, F., et al. 2021, MNRAS, 506, 1438
Kirby, E. N., Cohen, J. G., Guhathakurta, P., et al. 2013, ApJ, 779, 102
Kirby, E. N., Lanfranchi, G. A., Simon, J. D., Cohen, J. G., & Guhathakurta, P. 2011, ApJ, 727, 78
Kirby, E. N., Simon, J. D., & Cohen, J. G. 2015, ApJ, 810, 56
Koposov, S. E., Belokurov, V., Torrealba, G., & Evans, N. W. 2015a, ApJ, 805, 130
Koposov, S. E., Casey, A. R., Belokurov, V., et al. 2015b, ApJ, 811, 62
Laevens, B. P. M., Martin, N. F., Bernard, E. J., et al. 2015, ApJ, 813, 44
Lanfranchi, G. A., Matteucci, F., & Cescutti, G. 2008, A&A, 481, 635
Li, T. S., Ji, A. P., Pace, A. B., et al. 2022, ApJ, 928, 30
Li, T. S., Simon, J. D., Drlica-Wagner, A., et al. 2017, ApJ, 838, 8
Li, T. S., Simon, J. D., Kuehn, K., et al. 2018, ApJ, 866, 22
Longeard, N., Jablonka, P., Arentsen, A., et al. 2022, MNRAS, 516, 2348
Longeard, N., Martin, N., Ibata, R. A., et al. 2021, MNRAS, 503, 2754
Longeard, N., Martin, N., Starkenburg, E., et al. 2018, MNRAS, 480, 2609
Longeard, N., Martin, N., Starkenburg, E., et al. 2020, MNRAS, 491, 356
Manning, E. M., & Cole, A. A. 2017, MNRAS, 471, 4194
Martin, N. F., Ibata, R. A., Starkenburg, E., et al. 2022, MNRAS, 516, 5331
Martin, N. F., Nidever, D. L., Besla, G., et al. 2015, ApJL, 804, L5
Martínez-Vázquez, C. E., Monelli, M., Cassisi, S., et al. 2021, MNRAS, 508, 1064
Matteucci, F., Grisoni, V., Spitoni, E., et al. 2019, MNRAS, 487, 5363
Mészáros, S., Martell, S. L., Shetrone, M., et al. 2015, AJ, 149, 153
Monty, S., Venn, K. A., Lane, J. M. M., Lokhorst, D., & Yong, D. 2020, MNRAS, 497, 1236
Muñoz, R. R., Côté, P., Santana, F. A., et al. 2018, ApJ, 860, 66
Muratov, A. L., Kereš, D., Faucher-Giguère, C.-A., et al. 2015, MNRAS, 454, 2691
Mutlu-Pakdil, B., Sand, D. J., Carlin, J. L., et al. 2018, ApJ, 863, 25
Mutlu-Pakdil, B., Sand, D. J., Crnojević, D., et al. 2021, ApJ, 918, 88
Nagasawa, D. Q., Marshall, J. L., Li, T. S., et al. 2018, ApJ, 852, 99
Naidu, R. P., Conroy, C., Bonaca, A., et al. 2020, ApJ, 901, 48
Norris, J. E., Gilmore, G., Wyse, R. F. G., Yong, D., & Frebel, A. 2010, ApJL, 722, L104
Oliphant, T. E. 2006, A Guide to NumPy, Vol. 1 (USA: Trelgol Publishing)
Patel, E., Kallivayalil, N., Garavito-Camargo, N., et al. 2020, ApJ, 893, 121
Pietrinferni, A., Hidalgo, S., Cassisi, S., et al. 2021, ApJ, 908, 102
Prgomet, M., Rey, M. P., Andersson, E. P., et al. 2022, MNRAS, 513, 2326
Revaz, Y., & Jablonka, P. 2018, A&A, 616, A96
Rodriguez Wimberly, M. K., Cooper, M. C., Fillingham, S. P., et al. 2019, MNRAS, 483, 4031
Roederer, I. U., Mateo, M., Bailey, J. I. I., et al. 2016, AJ, 151, 82
Sacchi, E., Richstein, H., Kallivayalil, N., et al. 2021, ApJL, 920, L19
Sanati, M., Jeanquartier, F., Revaz, Y., & Jablonka, P. 2023, A&A, 669, A94
Sand, D. J., Strader, J., Willman, B., et al. 2012, ApJ, 756, 79
Sandford, N. R., Weinberg, D. H., Weisz, D. R., & Fu, S. W. 2022, arXiv:2210.17045

Sandford, N. R., Weisz, D. R., & Ting, Y.-S. 2023, ApJ, 267, 18
Savino, A., Koch, A., Prudil, Z., Kunder, A., & Smolec, R. 2020, A&A, 641, A96
Schlegel, D. J., Finkbeiner, D. P., & Davis, M. 1998, ApJ, 500, 525
Searle, L., & Zinn, R. 1978, ApJ, 225, 357
Shipp, N., Drlica-Wagner, A., Balbinot, E., et al. 2018, ApJ, 862, 114
Simon, J. D. 2019, ARA&A, 57, 375
Simon, J. D., Brown, T. M., Drlica-Wagner, A., et al. 2021, ApJ, 908, 18
Simon, J. D., Brown, T. M., Mutlu-Pakdil, B., et al. 2023, ApJ, 944, 43
Simon, J. D., Drlica-Wagner, A., Li, T. S., et al. 2015, ApJ, 808, 95
Simon, J. D., & Geha, M. 2007, ApJ, 670, 313
Simon, J. D., Geha, M., Minor, Q. E., et al. 2011, ApJ, 733, 46
Simon, J. D., Li, T. S., Erkal, D., et al. 2020, ApJ, 892, 137
Smith, S. E. T., Jensen, J., Roediger, J., et al. 2023, AJ, 166, 76
Starkenburg, E., Martin, N., Youakim, K., et al. 2017a, MNRAS, 471, 2587
Starkenburg, E., Oman, K. A., Navarro, J. F., et al. 2017b, MNRAS, 465, 2212
The Astropy Collaboration, Price-Whelan, A. M., Sipőcz, B. M., et al. 2018, ApJ, 156, 123
Tolstoy, E., Hill, V., & Tosi, M. 2009, ARA&A, 47, 371
Vanhollebeke, E., Groenewegen, M. A. T., & Girardi, L. 2009, A&A, 498, 95
Vargas, L. C., Geha, M., Kirby, E. N., & Simon, J. D. 2013, ApJ, 767, 134
Vargas, L. C., Geha, M. C., & Tollerud, E. J. 2014, ApJ, 790, 73
Venn, K. A., Kielty, C. L., Sestito, F., et al. 2020, MNRAS, 492, 3241
Virtanen, P., Gommers, R., Oliphant, T. E., et al. 2020, NatMe, 17, 261
Vivas, A. K., Olsen, K., Blum, R., et al. 2016, AJ, 151, 118
Walker, M. G., Mateo, M., Olszewski, E. W., et al. 2006, AJ, 131, 2114
Walker, M. G., Mateo, M., Olszewski, E. W., et al. 2015, ApJ, 808, 108
Walker, M. G., Mateo, M., Olszewski, E. W., et al. 2016, ApJ, 819, 53
Weinberg, D. H., Andrews, B. H., & Freudenburg, J. 2017, ApJ, 837, 183
Weisz, D. R., & Boylan-Kolchin, M. 2017, MNRAS, 469, L83
Weisz, D. R., Dolphin, A. E., Skillman, E. D., et al. 2014, ApJ, 789, 147
Wheeler, C., Hopkins, P. F., Pace, A. B., et al. 2019, MNRAS, 490, 4447
White, S. D. M., & Rees, M. J. 1978, MNRAS, 183, 341
Willman, B., Blanton, M. R., West, A. A., et al. 2005, AJ, 129, 2692
Willman, B., Geha, M., Strader, J., et al. 2011, AJ, 142, 128
Willman, B., Masjedi, M., Hogg, D. W., et al. 2006, arXiv:astro-ph/0603486
Willman, B., & Strader, J. 2012, AJ, 144, 76
Youakim, K., Starkenburg, E., Aguado, D. S., et al. 2017, MNRAS, 472, 2963